\documentclass{nature}

\usepackage{comment}

\usepackage{amsmath}
\usepackage[dvipsnames]{xcolor}
\usepackage{soul}
\usepackage{graphicx}
\usepackage{amssymb}
\usepackage{pifont}
\usepackage{longtable}
\usepackage{multirow}
\usepackage{arydshln}
\usepackage{hyperref}
\usepackage{caption}

\newcommand{\cmark}{\ding{51}}%
\newcommand{\xmark}{\ding{55}}%

\usepackage[switch, modulo]{lineno}
\modulolinenumbers[1]

\newcommand{\kepler}{{\it Kepler}}

\newcommand{\gaia}{{\it Gaia}}

\newcommand{\cofiam}{{\tt cofiam}}
\newcommand{\polyam}{{\tt polyam}}
\newcommand{\local}{{\tt local}}
\newcommand{\gp}{{\tt gp}}

\newcommand{\python}{{\tt Python}}

\newcommand{\multi}{{\sc MultiNest}}
\newcommand{\luna}{{\tt LUNA}}
\newcommand{\isochrones}{{\tt isochrones}}
\newcommand{\keplerports}{{\tt KeplerPORTS}}
\newcommand{\starini}{{\tt star.ini}}
\newcommand{\vespa}{{\tt vespa}}
\newcommand{\radecpix}{{\tt raDec2Pix}}

\newcommand{\kicname}{KIC-7906827}
\newcommand{\kicnameb}{KIC-7906827.01}

\newcommand{\kepx}{Kepler-1708}
\newcommand{\kepxb}{Kepler-1708 b}

\newcommand{\farcs}{\mbox{\ensuremath{.\!\!^{\prime\prime}}}}%

\newcommand{\wwwbryson}{\href{https://github.com/stevepur/Kepler-RaDex2Pix}{this URL}}
\newcommand{\wwwneaburke}{\href{https://exoplanetarchive.ipac.caltech.edu/docs/Kepler_completeness_reliability.html}{this URL}}
\newcommand{\wwwkepports}{\href{https://github.com/nasa/KeplerPORTs}{this URL}}
\newcommand{\wwwcoolworlds}{\href{https://doi.org/10.5061/dryad.18931zcz9}{this URL}}

\title{An Exomoon Survey of 70 Cool Giant Exoplanets and the New Candidate Kepler-1708 b-i}

\author{David~Kipping$^{1}\star$,
Steve~Bryson$^{2}$,
Chris~Burke$^{3}$,
Jessie~Christiansen$^{4}$,
Kevin~Hardegree-Ullman$^{4}$,
Billy~Quarles$^{5}$,
Brad~Hansen$^{6}$,
Judit~Szul\'agyi$^{7}$,
Alex~Teachey$^{8}$
}

\begin{document}

\maketitle

\begin{affiliations}
 \item Department of Astronomy, Columbia University, 550 W 120th St., New York, NY 10027, USA
 \item NASA Ames Research Center, Moffett Field, CA 94035, USA
 \item Department of Physics and Kavli Institute for Astrophysics and Space Research, Massachusetts Institute of Technology, Cambridge, MA 02139, USA
 \item Caltech/IPAC-NASA Exoplanet Science Institute, 770 S. Wilson Ave, Pasadena, CA 91106, USA
 \item Department of Physics, Astronomy, Geosciences and Engineering Technology, Valdosta State University, Valdosta GA, 31698, USA
 \item Mani Bhaumik Institute for Theoretical Physics, Department of Physics and Astronomy, UCLA, Los Angeles, CA 90095, USA
 \item Institute for Particle Physics \& Astrophysics, ETH Zurich, Wolfgang-Pauli-Str. 27, 8093 Z\"urich, Switzerland
 \item Institute of Astronomy and Astrophysics, Academia Sinica, Taipei 10617, Taiwan
\end{affiliations}

\begin{abstract}
Exomoons represent a crucial missing puzzle piece in our efforts to understand
extrasolar planetary systems. To address this deficiency, we here describe an
exomoon survey of 70 cool, giant transiting exoplanet candidates found by
\kepler. We identify only one which exhibits a moon-like signal that passes a
battery of vetting tests: \kepxb. We show that \kepxb\ is a
statistically validated Jupiter-sized planet
orbiting a Sun-like quiescent star at ${\sim}1.6$\,AU. The signal of the exomoon candidate, \kepxb-i, is a 4.8-sigma effect and is persistent across different instrumental detrending methods, with a $1$\% false-positive probability via injection-recovery. \kepxb-i is ${\sim}2.6$ Earth radii and is located in an approximately coplanar orbit at
${\sim}12$ planetary radii from its ${\sim}1.6$\,AU Jupiter-sized host. Future
observations will be necessary to validate or reject the candidate.
\end{abstract}

\newpage

In the last three decades, more than 4000 planets around stars other than the
Sun, exoplanets, have been discovered. These worlds display remarkable
diversity, from highly eccentric Jupiters\cite{hd80606b} to compact, coplanar
systems of terrestrial planets\cite{trappist1}. In an effort to understand the
formation and evolution of such systems, more detailed knowledge about their
environment and properties is sought\cite{morbidelli2016} - such as the
existence and nature of potential satellites\cite{heller2014}. Given
the abundance of moons in our Solar System, it's reasonable to presume that
exomoons will reside around some exoplanets - which has motivated efforts to
detect them\cite{brown2001,hek1}.

One of the most promising strategies for seeking exomoons focusses on
transiting planets\cite{sartoretti1999,szabo2006,kipping2009a}; worlds which
periodically eclipse their star and make up the majority of the discovered
exoplanets. However, the observational bias of transit surveys\cite{beatty2008}
leads to an underrepresentation of long-period, cool planets - precisely the
type of planet where moons are thought most likely due to dynamical
considerations\cite{namouni2010,barnes2002}. Nevertheless, a small sample of
long-period planetary candidates was discovered by \kepler\cite{wang2015,
uehara2016,dfm2016,wheeler2019,kawahara2019} - worlds with orbits greater than
that of the Earth around the Sun. The Jupiter-sized planets amongst these are
of particular interest, as satellite formation is thought to be a natural
outcome of how such planets form\cite{canup2006}.

To date, very little is known about the prevalence and properties of exomoons.
Initial surveys largely focussed on planets interior to 1\,AU\cite{hek5},
since this was broadly the only sample available at the time. Around
these relatively close-in planets, large moons appear uncommon, with the
abundance of Galilean-like satellite systems measured to be ${<}38$\% to 95\%
confidence\cite{hek6}. However, amongst the longest periods of these worlds,
the ${\sim}1$\,AU Jupiter-sized planet Kepler-1625b was reported to exhibit
a timing variation and transit signature consistent with a large
Neptune-sized/mass moon using Hubble Space Telescope (HST)
photometry\cite{k1625}. Both of these were independently recovered in one
study\cite{heller2019}, but only one (the timing) in
another\cite{kreidberg2020} - shown later to be possibly due to higher
systematics in their photometric reduction\cite{looseends}. Much like
hot-Jupiters, such large moons were not widely anticipated in the literature.
However, subsequent theoretical work has shown that the candidate exomoon could
form through a capture scenario\cite{hansen2019} or a massive
circumplanetary disk\cite{moraes2020}.

With no published exomoon surveys for $\gtrsim1$\,AU planets, and the
intriguing hint of Kepler-1625b-i, the aforementioned \kepler\ sample of
long-period giant planetary candidates represents one of the most promising
unturned stones. To address this, we here present a survey of \kepler's cool,
gas giants.

\section*{Results}
\label{sec:results}


We first curated known long-period transiting planets discovered by
\kepler\ from the literature, selecting any object with a reported radius
within a factor of two of Jupiter's, and with either i) a period
$>400$\,days, ii) an equilibrium temperature $<300$\,K, or iii) an instellation
less than that of Earth (see Methods). After removing any
object listed as a false-positive or with less than two available transits,
our sample comprised of 73 cool giants.

After the analysis and removal of long-term systematic trends in the
\kepler\ photometry, three targets were rejected as being of unacceptably poor
quality. Light curve detrending was performed using four different algorithms
applied to two photometric reductions, with the results cross-compared and
averaged (``method marginalised''), to ensure a robust correction against
algorithmic choices (see Methods).


For the 70 cool giants remaining, we fit several light curve models to each
detrended photometric time series. Using these fits, we applied a battery of
initial tests to check for the presence of exomoons. These are described in
detail in the Methods, and include 1) the Bayes factor of a photodynamical
planet-moon model must be favoured over a planet-only model by at least a
factor of 10 (i.e. ``strong'' evidence\cite{kass1995}), 2) light curve is
consistent with a planet on a near-circular orbit (as high eccentricities
diminish the suitability for moons\cite{domingos2006,gong2013}), and 3) if
more than two transits are available, the object should exhibit transit timing
variations. For the eccentricity test, it was necessary to derive fundamental
stellar properties for each host star, which was achieved through an isochrone
analysis (see Methods).

\begin{figure}
\centering
\includegraphics[angle=0, width=16.0cm]{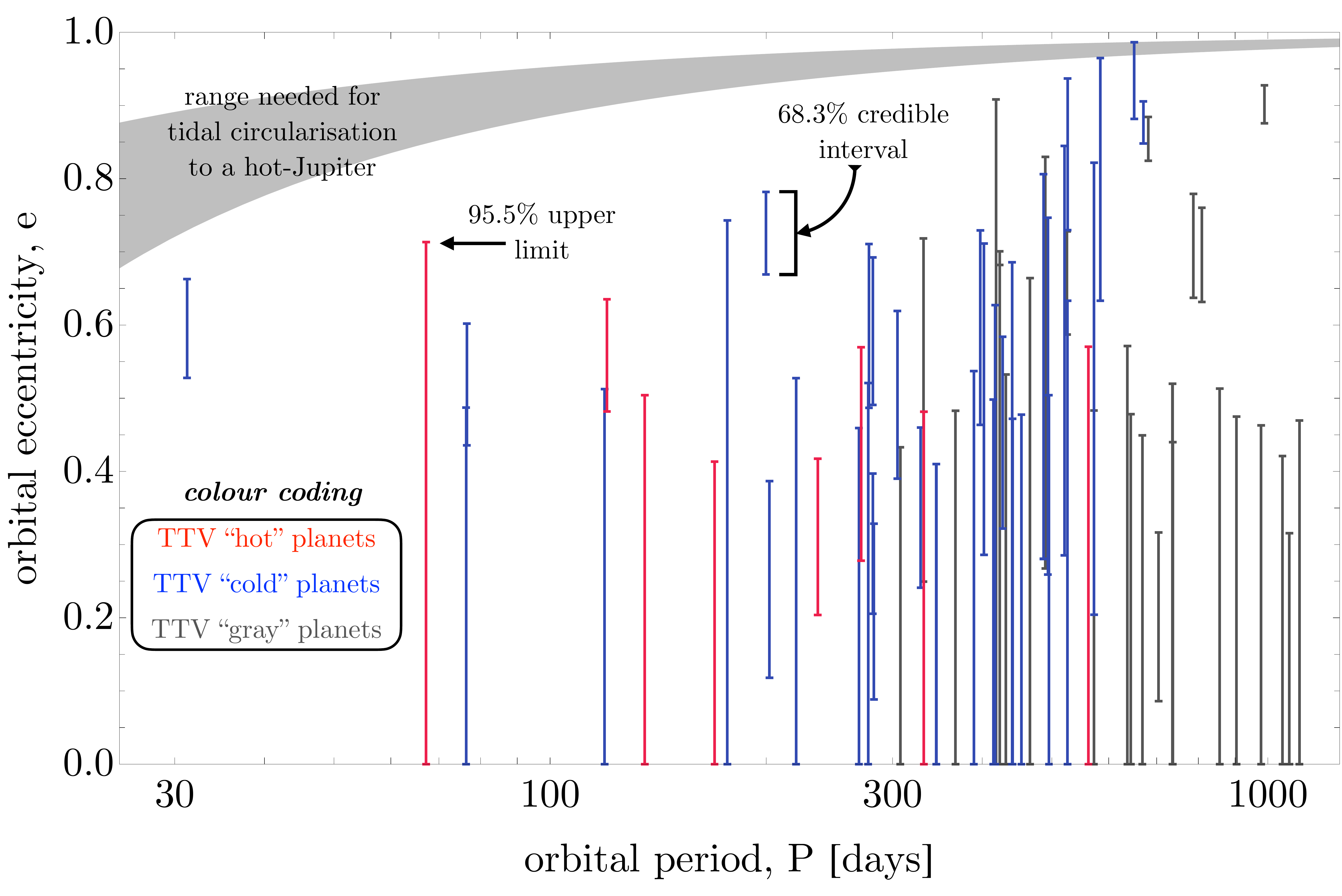}
\caption{\label{fig:e_vs_P}
\textbf{Orbital properties of the 70 cool giants.}
A comparison of the derived orbital eccentricities from this work (y-axis) versus the
orbital periods (previously known) for our planetary candidates sample. We use colour
(see legend) to depict the inferred absence/presence of transit timing variations
(TTVs).
}
\end{figure}

Although our primary goal is to search for exomoons, these tests provide some
novel dynamical insights that we highlight here. First, we find no clear
correlation between planetary candidates which exhibit transit timing
variations and those with eccentric orbits (see Fig.~\ref{fig:e_vs_P}).
Further, the eccentricity distribution does not appear sufficiently
extreme to explain the origin of hot-Jupiters through tidal circularisation
theory\cite{dawson2015}. However, we do see tentative evidence that the
eccentricity distribution comprises of more than one component (Extended Data
Fig.~\ref{fig:e_distribution}), indicative of multiple evolutionary
paths.

From the initial trio of tests, 11 planetary candidates passed these criteria
and were thus considered further (Supplementary Table~1). We
emphasise that this does not mean that none of the other targets host moons,
indeed some of our moon fits may have recovered genuine moon signals. However,
in each of these, there is at least one reason why the signal is weaker than
that expected of a faultless detection. By rejecting these, we follow a
conservative approach of only tolerating signals with zero reasons for concern.


We next applied three additional tests to the surviving 11 (see
Methods). Specifically, the light curves were re-fit with another planet-moon
model but one which permits unphysical parameter values, such as negative
radius moon (inverted transits) and unphysically large bulk densities for the
planet and moon. This allowed us to fairly evaluate the preference of the
models for 4) a non-zero moon mass, 5) a non-zero moon radius and 6) a positive
moon radius.


Only three objects survive these additional checks:
KIC-8681125.01\cite{kawahara2019}, KIC-5351250.06 (also known as
Kepler-150f\cite{schmitt2017}) and KIC-79068275.01\cite{kawahara2019}.
At this point, we turned to more detailed vetting tailored to each object.
In general, our goal is to identify if there is any basis to eliminate the
objects as a possible exomoon, and we work through various tests in an effort
to accomplish this. As soon as a test is failed, the object is rejected
from further consideration, regardless of how significant other aspects of
the exomoon-like signal may have been.


Inspection of the best-fitting planet-moon model to KIC-8681125.01 (Extended
Data Fig.~\ref{fig:KIC868_lcs}) revealed immediate cause for skepticism. No
distinct moon-like features are observed with the signal dominated by a transit
depth change between the two available epochs of $3590_{-130}^{+160}$\,ppm to
$3030_{-110}^{+140}$\,ppm. Since the two epochs are separated by six \kepler\
quarters, the spacecraft has physically rolled into a distinct position,
meaning that unknown background stars can contribute differently between the
two epochs. No high resolution imaging exists for this source, but \gaia\ lists
the closest companion 11\farcs8 away and 1.7 magnitudes fainter, which is
somewhat too faint and distant to easily explain the required dilution. Sources
interior to \gaia's arcsecond resolving power\cite{zieglergaia} remain
possible, as $(6.3\pm0.9)$\% of single-planetary candidate \kepler\ stars have
companions within 2\farcs0\cite{ziegler2018}. However, as discussed in the
Methods, this possibility also does not easily explain the depth change and an
instrumental effect could ultimately be responsible too.

To evaluate the fitness of the blend model, we re-fit the light curve with the
planet-only model but modifying such that the second epoch is diluted by some
factor, $\gamma$, from which we obtain $\gamma=1.188\pm0.037$. This model
yields the highest marginal likelihood score of all models tried, implying a
Bayes factor for the blend model of $6.8$. Taking all of these points together,
questions certainly remain about the cause of the depth change, but we consider
it unlikely it is ultimately driven by an exomoon, given both the alternative
possibilities and the nature of the signal.


Turning to Kepler-150f, inspection of the planet-moon fit reveals in-transit
morphological differences between the two available epochs. Unlike
KIC-8681125.01, this is not characterised by a simple depth change, but rather
complex morphology within the second event particularly (Extended Data
Fig.~\ref{fig:KIC535_lcs}). This raises concern that the signal is
spurious and caused by Kepler-150f passing over dark star spots
on the stellar surface, as has been reported previously\cite{ojeda2012}.

As the star rotates, spots cause the brightness to periodically change. If the
second transit was afflicted by spots, one might expect it to coincide with a
flux minimum in these rotational modulations\cite{ojeda2012}, when one is
observing the spot-covered face of the star - and this is indeed the case here.
Further, the star is known to be rotationally active with a
reported\cite{mcquillan2014} periodicity of 17.6\,days and amplitude of
10.9\,mmag. This far exceeds the depth of Kepler-150f's transit
(${\sim}1.5$\,mmag), indicating that the spot covered area of the star is
much larger than the planet itself and thus large transit distortions can
occur. Finally, we show in Methods that a modification to the planet-only
model that includes two spot crossings (using a simplified prescription) yields
a $\Delta\chi^2=9.2$ improved fit to the transits, versus the planet-moon
model, despite using the same number of free parameters. On this basis, we
conclude that this is most likely spot-driven activity rather than an exomoon
signature.



Lastly, we turn to \kicnameb. As with the other two, only two transits were
available given the long-period of $P=737.1$\,days. The Bayes factor of the
planet-moon model against the planet-only is 11.9, formally passing our
threshold of 10 (``strong evidence'' on the Kass \& Raftery scale
\cite{kass1995}). Inspection of the maximum likelihood moon fit, shown in
Fig.~\ref{fig:KIC790_lcs}, reveals that the signal is driven by an unexpected
decrease in brightness on the shoulder of preceding the first planetary
transit, as well as a corresponding increase in brightness preceding the egress
of that same event. The time interval between these two anomalies is
approximately equal to the duration of the planetary transit, which is
consistent with that expected for an exomoon\cite{luna2011}.
The second transit shows more marginal evidence for a
similar effect. The planet-moon model is able to well-explain these features,
indicative of an exomoon on a fairly compact orbit, in order to explain the
close proximity of the anomalies to the main transit. In a raw $\chi^2$ sense,
the inclusion of the exomoon leads to a $\Delta\chi^2=23.2$ improved fit,
indicating a 4.8-$\sigma$ effect. This does not penalise the model for its
extra complexity though; but that is accounted for in the previously mentioned
Bayes factor calculation of 11.9.


Our first concern was whether these undulations could be a spurious product of
the detrending process. Inspection of the individual method based detrendings,
rather than the method marginalisation, shows that the anomalies are present in
all detrendings (Extended Data Fig.~\ref{fig:lcdetrendings}), and further the
planet-moon fit from the method marginalised light curve is always a closer
match than that of planet-only fit (see Methods). Thus, the moon-like signal
appears robust against detrending choices.


Unlike KIC-8682235.01, transit 1's pre-ingress dip cannot be a starspot
crossing, since it occurs before the planet even enters the stellar
disk. It's also unlikely to be caused by an unseen contaminant star - this
would require such a star to coincidentally undergo a transit at nearly
precisely the same instant as the unrelated source star (although we
investigate this possibility shortly). Additionally, unlike
KIC-8682235, the two epochs are separated by an integer number of 4-quarters,
going from Q8 to Q16, thus placing the star on the same detector module in each
quarter, and indeed the same optimal aperture is used. Thus, any
difference between the two epochs cannot be caused by a contaminant being
present in one epoch but not the other.


Detailed inspection of the pixel light curves shows that the pixels of
highest planetary transit SNR coincide with the highest flux region, as
expected. We also applied this to the pixel location of the moon signature,
by evaluating the $\Delta\chi^2$ between the planet-moon and planet-only
model in each pixel, with \local\ detrending of the pixel light curves
and normalisation. This test was used for Kepler-90g in 2014 indicating that
a hypothesised exomoon was a false-positive\cite{kepler90g}, possibly
caused by a Sudden Pixel Sensitivity Dropout (SPSD)
event\cite{christiansen2013}. In contrast, we find here (see Methods) that
the candidate moon's SNR is co-located with the planetary signal, consistent
with a genuine signal (Extended Data Fig.~\ref{fig:kic790_pixels}).


Analysis of the flux weighted centroids reveals a small shift of
$\{-0.52\pm0.06,+0.62\pm0.05\}$\,millipixels in the $\{X,Y\}$ directions during
the two transits of \kicnameb. This can sometimes indicate that the transit
occurs on a different star than that assumed, potentially killing \kicnameb\ as
a bona-fide planet, but it can also simply occur because of nearby stars within
the aperture\cite{bryson2013}. A detailed centroid analysis (see Methods) shows
that amongst the known nearby stars in the \gaia\ catalog, only \kicname\ could
plausibly be the host of the transit signal. The shift is indeed broadly
consistent with that expected as a result of the known stars and the estimated
blend probability was calculated to be $2.6 \times 10^{-6}$.


Although the centroids indicate that the signal is a real planet, other
information (such as the transit light curve shape) can also be used to assess
this hypothesis. Accordingly, we used the \vespa\ package\cite{morton2016}
to rigorously calculate a statistical probability of planet-hood (see Methods).
From this, we estimate the false-positive probability (FPP) to be 0.024\%,
substantially below the 1\% threshold typically used to define a ``validated''
exoplanet\cite{morton2016} - we thus refer to the planet as \kepxb\ henceforth.


With \kepxb\ validated, we return to the exomoon signal. A formal assumption in
the light curve fits is that the noise is described by an independent normal
distribution. Time-correlated noise would render this assumption invalid and
could introduce deviations into the photometry to such a degree that the
planet-moon model fits are spuriously favoured - a false-positive. The act of
detrending the photometry attenuates this possibility but residual correlated
noise could still persist. Although we see no evidence for this (see Methods),
it cannot be excluded and to some degree will always be present in real world
conditions. We thus performed an injection-recovery exercise where we
injected the planet-only template for \kepxb\ into the \kicname\ photometry at
random times away from the real events and performed the same battery of tests
to see how often we would erroneously claim an exomoon. By using real light
curves, any time-correlated noise structure associated with the source is
properly accounted for.


Computational constraints limit us to 200 such fake systems, and amongst these
injections we find two cases where we would spuriously claim an exomoon
(see Methods), and thus the FPP for the exomoon signal is
$1.0_{-1.0}^{+0.7}$\%. If the signal is indeed not from time-correlated noise,
the most likely astrophysical false-positive is an unseen second transiting
planet, for which we find the probability is $\lesssim 1$\%
(Fig.~\ref{fig:moonfpp}).

\section*{Discussion}
\label{sec:discussion}

From a survey of 70 cool giant exoplanets, we find no compelling evidence for
an exomoon around any, bar one: \kicnameb/\kepxb. That candidate is presently
uncertain, with an estimated FPP of being an astrophysical signal of
${\sim}1$\% and a $\lesssim1$\% probability of a previously undetected
transiting planet causing such a signal. One detection from a sample of 70
and a 1\% FPP naively appears consistent with zero moons, but it's also fully
consistent with one real signal with the actual odds being dependent upon the
underlying (and unknown) occurrence rate of large exomoons (see Methods). In
short, we can find no grounds to reject \kepxb-i as an exomoon candidate at
this time, but urge both caution and further observations.

Our photodynamical model predicts a planetary mass $<4.6$\,$M_J$ [2\,$\sigma$],
corresponding to a predicted radial velocity (RV) amplitude of $<98$\,m/s. As a
faint source ($K_P=15.8$), RV detection would be a major challenge. TTVs are
generally expected and could be observed with future transits. Although we only
have photodynamically derived upper limits on the planet and satellite masses,
we can forecast them based upon their radii\cite{chen2017} to predict that the
TTV amplitude has a 95\% confidence range between 1.2 to 77.0 minutes. Future
observations with HST, JWST or PLATO could seek these TTVs or repeated
moon transits ($\simeq500$\,ppm).

\kepxb-i joins Kepler-1625b-i\cite{teachey2018} as another example of an
unexpectantly large exomoon candidate - echoing the surprise that hot-Jupiters
discoveries elicited in the mid-1990s\cite{mayor1995}. The basic properties are
listed in Table~\ref{tab:system}, and can be summarised as being a mini-Neptune
moon orbiting approximately 12 planetary radii around a Jupiter-sized planet,
which itself orbits a Sun-like star at 1.6\,AU. Compared to Kepler-1625b-i, the
moon candidate is substantially smaller, on a tighter orbit and more
consistent with a coplanar geometry. Although the reality of Kepler-1625b-i
remains unclear\cite{kreidberg2019}, the existence of this second candidate
challenges us to consider the origins of such large moons.

We first consider the moon's possible orbital migration from tidal
interactions with \kepxb. We evolved a constant time lag tidal
model\cite{hut1981} using the system parameters from our posterior
samples\cite{quarles2020}. The tidal model adopts Jupiter-like parameters for
the tidal Love number $k_{2J}$\cite{ni2018}, moment of inertia\cite{ni2018},
and time lag\cite{leconte2010}. We evolved the models over 10\,Gyr
($\simeq99.5$\% confidence upper limit on stellar age), assuming
that the moon forms in-situ at twice the Roche limit and that the planet has a
initial spin period of 5-10 hours. Over this timescale, the moon begins well
beyond the co-rotation radius so that it slowly migrates outwards. Over the
full 10\,Gyr simulation, the moon migrates to ${\sim}20$\,$R_P$, which is both
well within the Hill stability limit (${\sim}250$\,$R_P$) and
consistent with our favoured solution of ${\sim}12$\,$R_P$.

The fact that this candidate can plausibly migrate outwards via tides blurs the
distinction between formation scenarios, as any model that produces a massive
moon on a compact orbit can match the observations. There are several broad
scenarios for moon formation: planet-planet collisions, formation of moons
within gaseous circumplanetary disks (e.g. the Galilean moons), or
direct capture - either by tidal dissipation or pulldown during the growth of
the planet. For a gaseous planet, the first scenario is unlikely to produce a
debris disk massive enough to form a moon this large. The moon is also at
the extreme end of the mass range produced by primordial disks in the
traditional core collapse picture of giant planet formation\cite{szulagyi2017,
cilibrasi2018,inderbitzi2020}, but is easier in the case where planets form by
disk instability\cite{shabram2013,szulagyi2017}. Such models also naturally
produce moons on low inclination orbits. Direct capture by tidal dissipation is
also possible, although the parameter range for capture without merger is
limited. Pulldown capture can produce large moons within ${\sim}10$ Jupiter
radii, with a wide range of inclinations depending on the timescale for
planetary growth. Low inclinations such as those observed here argue for a
slower envelope growth\cite{hansen2019}.

Together then, the formation and properties of a moon like this certainly
challenge conventional thinking, but plausible mechanisms have been previously
proposed. Ultimately, the reality of supermoons like \kepxb-i and
Kepler-1625b-i will require follow-up transit photometry as both their
nature and supporting evidence demand appropriate skepticism at this time.

\clearpage

\bibliographystyle{naturemag}
\newcounter{firstbib}

\clearpage


\begin{addendum} 
\item[Author Correspondence]
All correspondence regarding this work should be directed to D. Kipping.
\item
D.K. thanks donors Mark Sloan, Laura Sanborn, Douglas Daughaday, Andrew Jones, Marc Lijoi, Elena West, Tristan Zajonc, Chuck Wolfred, Lasse Skov, Alex de Vaal, Mark Elliott, Methven Forbes, Stephen Lee, Zachary Danielson, Chad Souter, Marcus Gillette, Tina Jeffcoat,  Jason Rockett, Scott Hannum, Tom Donkin, Andrew Schoen, Jacob Black, Reza Ramezankhani, Steven Marks, Philip Masterson \& Gary Canterbury.
D.K. acknowledges support from NASA Grant \#80NSSC21K0960.
J.Sz. acknowledges the financial support from the European Research Council (ERC) under the European Union’s Horizon 2020 research and innovation programme (Grant agreement \#948467).
Analysis was carried out in part on the NASA Supercomputer PLEIADES (Grant \#HEC-SMD-17- 1386), provided by the NASA High-End Computing (HEC) Program through the NASA Advanced Supercomputing (NAS) Division at Ames Research Center.
This paper includes data collected by the \kepler\ Mission. Funding for the \kepler\ Mission is provided by the NASA Science Mission directorate.
This work has made use of data from the European Space Agency (ESA) mission
\gaia\ (\href{https://www.cosmos.esa.int/gaia}{https://www.cosmos.esa.int/gaia}),
processed by the \gaia\ Data Processing and Analysis Consortium (DPAC, \href{https://www.cosmos.esa.int/web/gaia/dpac/consortium}{https://www.cosmos.esa.int/web/gaia/dpac/consortium}). Funding for the DPAC has been provided by national institutions, in particular
the institutions participating in the \gaia\ Multilateral Agreement.

\item[Author contributions] 
D.K. performed the data reduction, analysis and interpretation and wrote the majority of the text. S.B. performed the centroid analysis and interpretation. C.B. performed the \keplerports\ analysis. J.S. and K.H.U. performed the \vespa\ validation. B.Q., B.H. \& J.S. wrote the formation and evolutionary interpretation sections. A.T. consulted with the authors on false-positive calculations and signal morphologies.

\clearpage
\item[Author Information] Reprints and permissions information is available at www.nature.com/reprints. Correspondence and requests for materials should be addressed to D.K.~(email: dkipping@astro.columbia.edu).

\item[Competing Interests] The authors declare that they have no competing interests.

\end{addendum}


\newpage

\newpage
\begin{figure}
\centering
\includegraphics[angle=0, width=16.8cm]{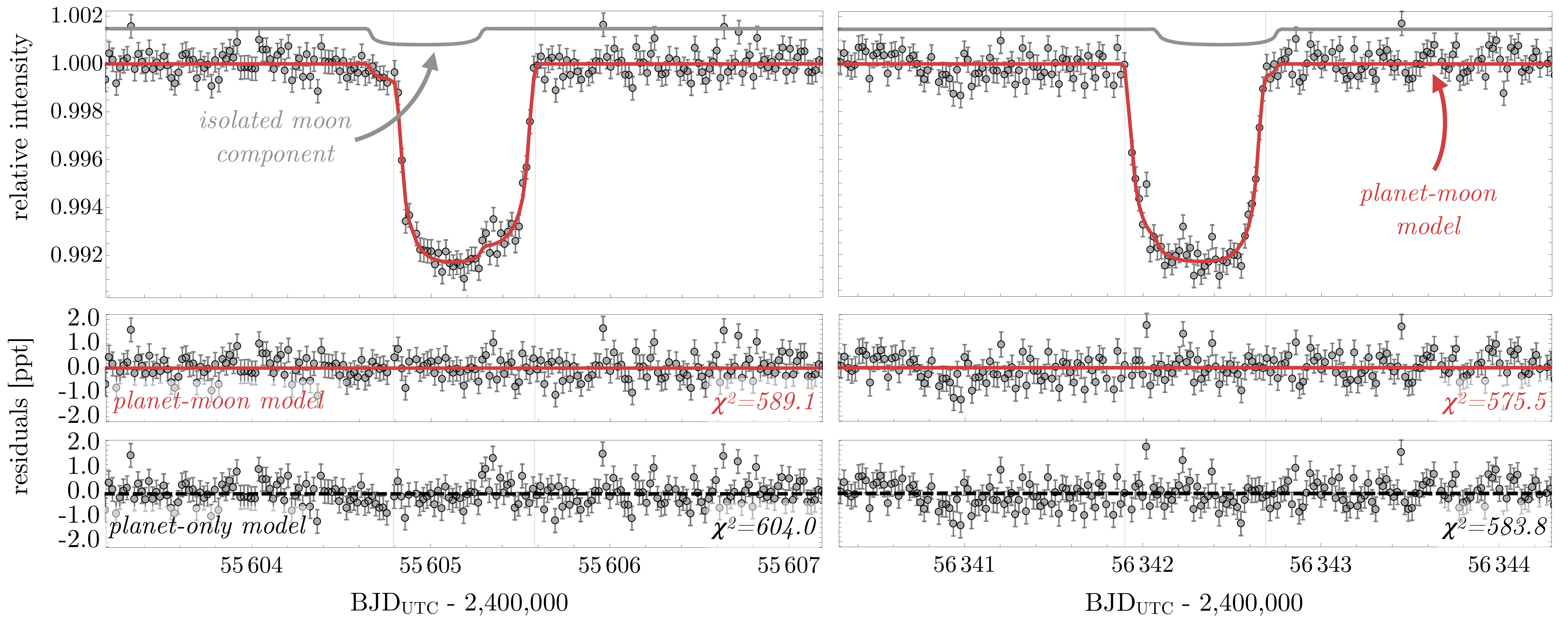}
\caption{\label{fig:KIC790_lcs}
\textbf{Transit light curves of \kepxb.}
Left/right column shows the first/second transit epoch, with the maximum likelihood planet-moon model overlaid in solid red. The grey line above shows the contribution of the moon in isolation. Lower panels show the residuals between the planet-moon model and the data, as well as the planet-only model.
}
\end{figure}

\newpage
\begin{figure}
\centering
\includegraphics[angle=0, width=16.0cm]{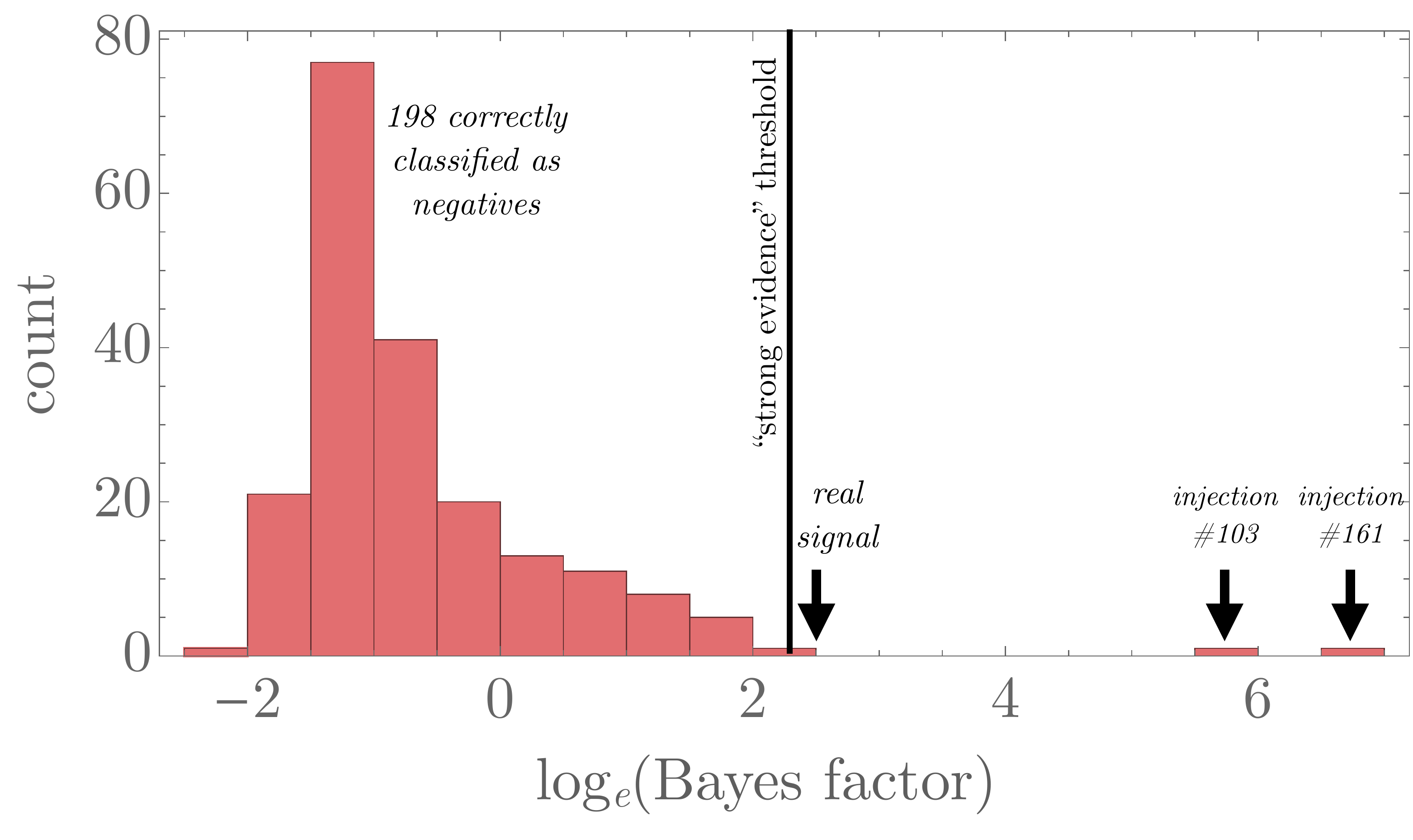}
\caption{\label{fig:moonfpp}
  \textbf{False-positive probability calculation for \kepxb-i.}
    Histogram of the log Bayes factor between a planet-moon and planet-only
	model from 200 fake planet-only signals injected into the light curve. Two
	signals pass the threshold $(=\log_e 10)$ and have positive-radii, indicating
	a 1\% FPP.
  }
\end{figure}

\newpage
\begin{longtable}{llc}
\hline
Parameter & Description & Value \\
\hline
$R_P/R_{\star}$ & Planet-star radius ratio & $0.0818_{-0.0010}^{+0.0011}$ \\
$P_P$\,[days] & Planet's orbital period & $737.1131_{-0.0077}^{+0.00146}$ \\
$b_P$ & Impact parameter of the planet & $<0.37$ [2\,$\sigma$] \\
$a_P/R_{\star}$ & Planet-star radius ratio & $317.9_{-8.4}^{+5.2}$ \\
$R_P$\,[$R_{\oplus}$] & Planetary radius & $9.96_{-0.59}^{+0.60}$ \\
$M_P$\,[$M_J$] & Planetary mass & $<4.6$ [2\,$\sigma$] \\
$a_P$\,[AU] & Planet's orbital radius & $1.64_{-0.10}^{+0.10}$ \\
$e_P$ & Orbital eccentricity of the planet & $<0.40$ [2\,$\sigma$] \\
$S_{P}(=S_{S})$\,[$S_{\oplus}$] & Instellation c.f. Earth & $0.561_{-0.068}^{+0.074}$ \\
\hline 
$P_S$\,[days] & Satellite's orbital period & $4.6_{-1.8}^{+3.1}$ \\
$a_{SP}/R_P$ & Satellite's orbital radius in planetary radii & $11.7_{-2.2}^{+3.9}$ \\
$i_S$'\,[$^{\circ}$] & Satellite's inclination c.f. planetary orbit & $9_{-45}^{+38}$ \\
$\Omega_S$\,[$^{\circ}$] & Satellite's longitude of the ascending node & $6_{-150}^{+140}$ \\
$R_S/R_P$ & Satellite-planet radius ratio & $0.263_{-0.042}^{+0.040}$ \\
$M_S/M_P$ & Satellite-planet mass ratio & $<0.11$ [2\,$\sigma$] \\
$\mathrm{TTV}$\,[mins] & Predicted TTV amplitude & $<41$ [2\,$\sigma$] \\
$R_S$\,[$R_{\oplus}$] & Satellite radius & $2.61_{-0.43}^{+0.42}$ \\
$M_S$\,[$M_{\oplus}$] & Satellite mass & $<37$ [2\,$\sigma$] \\
\hline
\caption{System parameters for \kepxb.
}
\label{tab:system} 
\end{longtable}



\clearpage

\begin{methods}

\subsection{Target selection.}
\label{sub:targets}


The focus of this work is the population of cool, giant transiting exoplanets
observed by \kepler. The most comprehensive catalog of \kepler\ transiting
planet candidates comes from the NASA Exoplanet Archive (NEA\cite{akeson2013})
and thus we began by downloading this catalog at the
start of this investigation (March 27th 2018). Not all \kepler\ Objects of
Interest (KOIs) in the catalog are viable planet candidates though, and so we
applied a cut to remove any objects which have been dispositioned as a likely
``FALSE POSITIVE'' by the NEA.

Both the term ``cool'' and ``giant'' are somewhat subjective and thus require
a clear definition for the purposes of target selection. For ``giant'', we
elected to use a cut of $\hat{R_P} > 0.5$\,$R_J$, where $\hat{R_P}$ is the
most probable radius value listed in the NEA. This choice is motivated to
minimise the number of sub-Neptunes that make it into our sample, and thus
focus on Jupiter-sized worlds.

The definition of ``cool'' is again subjective, but here we are primarily
interested in planets in cooler environments than that of the Earth.
Using a simple instellation or temperature cut alone is inadequate
though, as these values depend upon the stellar parameters which have been
subject to substantial revision over time\cite{mathur2017}. On this basis,
they may not be reliable in isolation as a means of capturing all of the cool
giants. Instead, we apply three different definitions of cool and accept giant
planets that satisfy any of the three. These criteria are
$\hat{S_P} < 1$\,$S_{\oplus}$,
or $\hat{T_P} < 300$\,K,
or $\hat{P_P} > 400$\,days
- where $S_P$ is the planetary instellation, $T_P$ is the blackbody equilibrium
temperature of the planet, and $P_P$ is the orbital period.


Using these cuts, 48 KOIs were identified from the NEA. However, we noted
that a subset of these had suspiciously large radii. We thus applied an
additional cut to remove any with best reported planetary radii in excess of
2 Jupiter radii (removing KIC-3644071.01, KIC-6426592.01, KIC-6443093.01,
KIC-9025662.01, KIC-9011955.01, KIC-8240617.01 \& KIC-8868364.01).
This left us with 41 unique planetary candidates. Of particular note is
KIC-5437945, which possesses two cool giants associated with a single source.

At this point, we introduce another cut: that at least two transits have been
observed by \kepler. Without the two transits, the orbital period cannot
be precisely measured and this in turn makes it impossible to measure the
eccentricity of the planets photometrically - a test we will depend on later
in our exomoon analysis. Of the 41 cool giants from the NEA, 5 were found to
only exhibit a single transit in the available \kepler\ data (KIC-2162635.01,
KIC-3230491.01, KIC-3962440.01, KIC-11342550.01, KIC-11709124.02) and thus were
excluded. This leaves us with 36 cool giants.


Although the NEA is the most complete catalog available, long-period planets
are more challenging to find than their shorter-period counter-parts and
thus numerous independent studies have identified long-period exoplanetary
candidates that were not present in the NEA. In particular, we identified
several additional studies\cite{wang2015,uehara2016,dfm2016,wheeler2019}
that we inspected in an effort to locate any additional possible planets
missed thus far. In what follows, we attempt to
apply the same filters as before, such as exhibiting at least two transits, but
note that in some cases the radius and instellation values had not been
computed by the original authors.

From Wang and colleagues\cite{wang2015}, KIC-8012732, KIC-9413313 \&
KIC-11465813 were flagged as exhibiting 3-4 visible transits and have long
periods (431\,d, 440\,d \& 671\,d) and large radii (9.8\,$R_J$ and
13.8\,$R_J$). KIC-5437945 and KIC-7619236 also satisfy the criteria but are
already included in the NEA catalog. KIC-5652983 is long-period and large but
has been argued to be a likely false-positive\cite{wang2015} due to the
observation of large RV variations. Amongst the 2-transit cases, KIC-5732155,
KIC-6191521 and KIC-10255705 are also added, which leads to 6 new cool giants
from this sample.

From Uehara and colleagues\cite{uehara2016}, only KIC-10460629 exhibits at
least two transits separated by 525\,days. Although the radius is not computed,
the grazing, deep (2\%) transit is plausibly a giant around a diminutive star
and thus was included in what follows.

From Foreman-Mackey and colleagues\cite{dfm2016}, only three of the candidates
have more than two transits KIC-3239945, KIC-8410697 and KIC-8800954. However,
KIC-3239945 is already found in our NEA catalog and KIC-8800954 has a radius of
0.39\,$R_J$, making it too small for our sample. Accordingly, we only add
KIC-8410697, of period 1047\,days and 0.70\,$R_J$.

From Wheeler \& Kipping\cite{wheeler2019}, we pick an extra transiter of 3\%
depth but unreported radius (KIC-8508736) with a period of 681\,days.

Combining these targets to the NEA sample yielded 45 cool giants.


As our study progressed, spanning multiple years, the possibility of new
cool giants detections appearing in the NEA grew. To address this, we re-ran
our NEA search June 12th 2019, which identified 24 new cool giants. Of these, 6
were single transit systems (KIC-2162635, KIC-3230491, KIC-3644071,
KIC-3962440, KIC-11342550, KIC-11709142) and were thus removed. In addition,
upon querying KIC-8308347 on MAST, we noted it had been flagged as an
``Eclipsing Binary: Likely False Positive'' under Condition Flag and thus
rejected it in what follows. These new inclusions raise our total number of
cool giants to 62.


As a final addition to our catalog, a recently published
study\cite{kawahara2019} (although quite far into our own analysis) presented
23 long-period \kepler\ planetary candidates. Of these, 14 were suitable for
our study and all but one were double-transit systems. However, three of these
were already included in our sample by this point (KIC-3756801, KIC-9663113,
KIC-10460629), meaning that the \cite{kawahara2019} sample added 11 cool giants
to our ensemble. Of the new objects, KIC-5351250 is of particular note since it
represents the fifth planetary candidate to the Kepler-150
system\cite{schmitt2017}. Together, this brings our final catalog of cool
giants up to 73 planetary candidates. Three of these were found to exhibit
unacceptably high correlated noise structure in the light curves and were
thus rejected, as is described in later in this Methods section. The
remaining 70 are listed in Supplementary Table~1.

\subsection{Data preparation.}
\label{sub:data}


For each target, the light curve files were downloaded from the Mikulski
Archive for Space Space Telescopes (MAST), using primarily the long-cadence
(LC) data but short-cadence (SC) where available. However, given the
long-period nature of our transiting bodies, the value of SC is considerably
less than typical transiters\cite{binning2010}. In all cases, the data were
processed as part of the 25th and final data release issued by the \kepler\
science team\cite{thompson2018}, dubbed DR25 hereafter.


For all light curves, we trimmed any points with an error flag equal to
anything other than zero - thus removing points known to be afflicted by
effects such as reaction wheel zero crossings\cite{thompson2016}. Additional
outliers (e.g. unidentified cosmic ray hits\cite{morris2012}) were
removed independently for the ``Simple Aperture Photometry'' (SAP) and
``Pre-search Data Conditioning'' (PDC\cite{smith2012}) light curves, by
flagging points more than 3-$\sigma$ deviant from moving median of bandwidth 20
LC cadences.


\kepler\ light curves exhibit modulations in intensity due to a myriad of
effects. Ultimately, the short-term modulations corresponding to a transit are
of central interest to this study, but longer term variability is also present
and introduces sizeable trends that require correction. Such variability could
originate from the instrument (e.g. focus drift\cite{jenkins2010}) or the
parent star (e.g. rotational modulations\cite{walker2007}). In what
follows, we describe our approach for detrending these effects.

As a brief aside, we note that short-term variability on the same timescale as
the transits can also be present (e.g. pulsations in evolved
stars\cite{kallinger2016}) and is generally much more difficult to remove
since it is not separable in the frequency-domain. Consequently, attempts to
remove such noise come at grave risk of distorting a transit signal of
interest. Given that our primary objective is to look for exomoons, which
manifest as small undulations on this timescale, it was considered an
unacceptable risk to attempt to remove such short-term variability - since such
efforts may in fact introduce false-positives into the time series. Instead,
the philosophy in what follows is to use statistical tests to identify light
curves corrupted by such noise and simply discard them. This naturally comes at
the expense of increasing our false-negative rate, since such systems are not
even analysed further.

For each cool giant target, we detrend the light curves of the individual
transit epochs individually, rather than imposing that the noise in one quarter
need be representative of others. This is largely motivated by the fact that
the spacecraft rolls every quarter and thus sources appear on different silicon
with different optimal apertures, blend contaminations and CCD behaviours. In
addition, we adopt the approach to detrend each transit epoch eight different
ways. The reasoning here is that, although we generally consider each of the
eight different methods to be fairly accurate (else we would not be
using them), we cannot guarantee that any of them are going to work in every
situation. From experience, peculiarities in particular light curves can
interact with detrending algorithms in unanticipated ways, leading to anything
from a complete failure to a subtle residual trend. Fundamentally, any claim we
make about the presence of an exomoon needs to be robust against choices made
in this detrending stage and a path to achieving that is to simply use multiple
detrending methods and compare.

The details of the different detrending algorithms used are presented shortly,
but once in hand they are combined into a single data product (per transit
epoch) known as a ``method marginalised'' light curve\cite{k1625}.
In this work, we generate such light curves by simply taking the median of
the multiple detrended intensities at each time stamp. The formal uncertainty
on each photometric data point is also inflated by adding it in quadrature to
1.4286 multiplied by the median absolute deviation (MAD) between the methods.
Median statistics are used throughout here to mitigate the influence of a
failed detrending(s). In this way, we increase the robustness of our light
products against detrending choices and also inflate the errors to propagate
the uncertainty in the detrending procedure itself.

As an additional safeguard against poorly detrended light curves, we compute
two light curve statistics to measure their Gaussianity. If any of the
eight light curves fail this test, they are rejected prior to the method
marginalisation procedure. For the first test, we bin the light curves
(after removing the transits) into ever larger bins and compute the standard
deviation versus bin size, against which we then fit a linear slope in log-log
space. For such a plot, the slope should be minus one-half, reflecting the
behaviour of Poisson counting of independent measures. However, time-correlated
noise structure will lead to a shallower slope that can be used to flag such
problematic sources\cite{carter2009}. We thus generate 1000 light curves
of precisely the same time sampling and pure Gaussian noise and measure their
slopes in this way. This allows us to construct a distribution of expected
slope values. If the real slope deviates from the Monte Carlo experiments with
a $p$-value exceeding 2\,$\sigma$, the light curve is flagged as non-Gaussian.

For the second test, we compute the Durbin-Watson\cite{durbin1950} statistic
of the unbinned light curves (after removing the transits). This is essentially
a test for autocorrelation at the timescale of the data's cadence, where
uncorrelated time series should yield a score of 2. As before, we test for
non-Gaussian cases by generating 1000 fake Gaussian light curves at the same
time sampling and scoring their Durbin-Watson metrics. If the real light curve
is deviant from this distribution by more than 2-$\sigma$, the light curve is
rejected.

The above describes how we combine eight light curves detrended independently,
but we have yet to describe how these eight light curves are generated in the
first place - which we turn to in what follows. In total, four different
detrending algorithms are used, which are then applied to the SAP and PDC data
to give eight light curves. The four algorithms are described in what follows.


\textbf{\cofiam:}
Cosine Filtering with Autocorrelation Minimisation (\cofiam) builds upon the
cosine filtering approach previously developed for
\textit{CoRoT}\cite{mazeh2010} data. Cosine filtering is attractive because it
behaves in a predictable manner in the frequency-domain, unlike the other
methods used here which leak power across frequency space. Fourier
decomposition of the transit morphology reveals dominant power at the timescale
of the transit duration and higher frequencies\cite{waldmann2012}. Thus, by
only removing frequencies substantially lower than this, one can ensure that
the morphology of the transit is not distorted by the process of detrending
itself. On the other hand, cosine filtering is problematic in that one could
regress a very large number of cosines to the data. Much like fitting
high order polynomials, predictions from such model become unstable at high
order. In our case, we train on the out-of-transit data (in fact the entire
quarter) and interpolate the model into the transit window, thus introducing
the possibility of high order instabilities here.

This is where our implementation deviates from that used for
\textit{CoRoT}\cite{mazeh2010}, in order to account for this effect. We detrend
the light curve up to 30 different ways, in each case choosing a different
number of cosine components to include. The simplest model is a single cosine
of frequency given by twice the baseline of available observations (thus looks
like a quadratic trend) - known as the basic frequency. At each step, we add
another cosine term of higher frequency to the function (equal to a harmonic
of the basic frequency), train the updated model, detrend the light curve and
compute statistics concerning the quality of the detrending. We continue up to
30 harmonics, or until we hit 1.5 times the reported transit duration. From the
30 options, we pick the one which leads to the most uncorrelated light curve -
as measured from the Durbin-Watson statistic evaluated on the data surrounding
(but not including) the transit (specifically to within six transit durations
either side). This local data is then exported with the data further away from
transit trimmed at this point. We direct the reader to our previous
paper\cite{hek2} for more details on this approach, including the underlying
formulae used.

\textbf{\polyam:}
Polynomial detrending with Autocorrelation Minimisation (\polyam) is similar
to the above except that the basis function is changed from a series of
harmonic cosines to polynomials. As before, 30 different possible maximum
polynomials orders are attempted from 1st- to 30th-order. And, as before,
for each epoch the least autocorrelated light curve is selected as the
accepted detrending on a transit-by-transit basis.

\textbf{\local:}
The next approach again uses polynomials, and again up to 30th order, but
this time the final accepted polynomial order is the which leads to the lowest
Bayesian Information Criterion\cite{schwarz1978} as computed on the
data directly surrounding the transit (specifically to within six transit
durations). This is arguably the simplest of the four algorithms attempted
and is a fairly typical strategy in the analysis of short-period
transiters\cite{sandford2017}.

\textbf{\gp:}
Finally, we implemented a Gaussian Process (\gp) regression to the light curve.
As with all of the methods above, the transits are masked during the regression
by using the best available ephemeris. We implemented the regression using a
squared exponential kernel where the hyper-parameters (e.g. length scale) are
optimised for each epoch independently. For consistency, we only export the
data that is within six transit durations of the transit, although technically
the entire segment ($\pm0.5$ orbital periods of each transit) is detrended.


As a final note, recall that during the method marginalisation process, we
perform checks for identifying detrended light curves which do not conform to
Gaussian behaviour. If all eight detrendings of a given epoch fail these tests,
then there will be no accepted light curves to combine and thus the epoch is
dropped. In some cases then, this can reduce the number of available transits
(post detrending) to less than two - thereby making it fail our basic criteria
of presenting two epochs or more. We found this happened for three objects in
our sample, KIC-10255705, KIC-11513366 and KIC-6309307. This removal of these
three targets reduces our sample size from 73 to 70 (which are the ones listed
in Supplementary Table~1).

\subsection{Isochrone analysis.}
\label{sub:isochrones}

The 70 transiting planet candidates are almost all associated with distinct
stars to one another, with the exception of one pair associated with
KIC-5437945, leading to 69 unique stars. In order to derive physical dimensions
for the planetary candidates, it is necessary to first derive stellar
parameters. This is accomplished using an isochrone analysis, which compares
observable quantities associated with a star against a grid of stellar models,
assuming different masses, radii, ages, etc. In this way, best matching
solutions can be inferred in a Bayesian framework to derive posterior
distributions for the stellar properties.

Given that our star are broadly FGK-type, we elected to use the Dartmouth
stellar isochrones models\cite{dotter2008} to describe these stars. For
each target, we took the Gaia DR2 parallax\cite{luri2018}, the \kepler\
bandpass apparent magnitude, and the stellar atmosphere properties reported
in the \kepler\ DR25 catalog\cite{mathur2017}, and appended them to a
\starini\ file along with their associated errors. These were then fed into
the \isochrones\ package\cite{morton2015} to obtain \textit{a-posteriori}
fundamental stellar parameters, including $\rho_{\star}$. These fundamental
parameters are reported in Supplementary Table~2 and were used
later in our analysis for deriving planet/moon radii/masses.

Due to particular interest of the target \kicname, we updated our isochrone
analysis to include the \gaia\ DR3 parallax when it became available
(reducing the parallax uncertainty by 25\%). We also elected to use the
stellar atmosphere properties from the transit detection
paper\cite{kawahara2019} ($T_{\mathrm{eff}} = 6157_{-202}^{+231}$\,K,
$\log g = 4.37_{-0.05}^{+0.04}$, Fe/H $= 0.0_{-0.2}^{+0.2}$) rather than
the DR25 catalog\cite{mathur2017} ($T_{\mathrm{eff}} = 5977\pm176$\,K,
$\log g = 4.39 \pm 0.12$, Fe/H $= -0.08 \pm 0.26$), although we note that these
values are clearly very similar. It is using these revised choices that the
system parameters listed in Table~\ref{tab:system} correspond to,
for which the associated fundamental stellar parameters are
$M_{\star} = 1.088\pm0.072$\,$M_{\odot}$,
$R_{\star} = 1.117\pm0.064$\,$R_{\odot}$,
$\log_{10}(\mathrm{A}\,[\mathrm{yr}]) = 9.50\pm0.31$,
$\log_{10}(L_{\star}\,[L_{\odot}]) = 0.182\pm0.082$,
$d = 1712 \pm 75$\,pc and
$\log_{10}(\rho_{\star}\,[\mathrm{g}\,\mathrm{cm}^{-3}]) = 0.042\pm0.065$.

We note that these are not the same fundamental stellar parameters for
\kicname\ listed in Supplementary Table~2, which come from the
original \kepler\ DR25 and \gaia\ DR2 inputs. Once again though, we note that
there is very little difference between the two, with
$M_{\star} = 1.056\pm0.067$\,$M_{\odot}$,
$R_{\star} = 1.098\pm0.095$\,$R_{\odot}$,
$\log_{10}(\mathrm{A}\,[\mathrm{yr}]) = 9.61\pm0.33$,
$\log_{10}(L_{\star}\,[L_{\odot}]) = 0.140\pm0.087$,
$d = 1750 \pm 100$\,pc and
$\log_{10}(\rho_{\star}\,[\mathrm{g}\,\mathrm{cm}^{-3}]) = 0.05\pm0.11$.
As a final point of comparison, both sets of values are in good
agreement with the independent analysis (using Gaia DR2) of
Berger and colleagues\cite{berger2020}, who find
$M_{\star} = 1.061_{-0.079}^{+0.073}$\,$M_{\odot}$,
$R_{\star} = 1.141_{-0.066}^{+0.073}$\,$R_{\odot}$,
$\log_{10}(L_{\star}\,[L_{\odot}]) = 0.140\pm0.087$ and
$d = 1640 \pm 100$\,pc.

\subsection{Light curve fits.}
\label{sub:fits}


For planets exhibiting three or more transits, at least three different light
curve models, or hypotheses, were proposed to explain the data. The first is
model $\mathcal{P}$, which represents the null hypothesis of a transiting
planet orbiting its star on a strictly Keplerian orbit. In this case, the
Mandel-Agol\cite{mandel2002} light curve algorithm is used. The second
hypothesis, model $\mathcal{T}$, expands upon the first by adding transit
timing variations (TTVs). This is formally accounted for using same algorithm
again but allowing each transit epoch to have a unique time of transit minimum,
$\tau_i$. The third model considered is that of a planet-moon system,
$\mathcal{M}$, generated using the \luna\ photodynamic
algorithm\cite{luna2011}. For planets exhibiting just two transits, TTVs cannot
be distinguished from a linear ephemeris and thus model $\mathcal{T}$ was not
used.

In each model, the limb darkening of the star is modelled with a
quadratic limb darkening law using the $q_1$-$q_2$
re-parameterisation\cite{q1q2}. Since the majority of light curves are
long-cadence, the potentially significant light curve smearing effect is
accounted for by employing the numerical re-sampling method\cite{binning2010}
(with $N_{\mathrm{resamp.}}=30$). Finally, contaminated light from nearby
sources is tabulated in the \kepler\ {\tt fits} files as ``CROWDSAP'' and this
value is used in a blend correction to each quarter's light curve using a
previously published method\cite{nightside}. We also note that the models
formally assume circular orbits although elliptical planets are almost
perfectly described by these models too\cite{kipping2008}, with the exception
that the fitted stellar density will be skewed away from the true
value\cite{map2012}. Exomoon orbits are also treated as circular which is
justified on the basis of the expected rapid circularisation
timescales\cite{porter2011}.


Regressions were executed using the multimodal nested sampling algorithm
\multi\cite{feroz2009} using 4000 live points. The advantage of using
\multi\ over conventional MCMC methods is the ability to sample disparate
modes and efficiently evaluate the marginal likelihood of the proposed
hypotheses, which is used later in Bayesian model selection.

For model $\mathcal{P}$, seven parameters fully describe the light curve model
and thus are the free parameters in these fits. These are:
i) $P$, the orbital period of the planet
ii) $\tau$, the time of transit minimum
iii) $p$, the ratio of radii between the planet and the star
iv) $b$, the impact parameter of the planetary transit
v) $\rho_{\star}$, the mean density of the host star
vi) $q_1$, the first limb darkening coefficient
vii) $q_2$, the second limb darkening coefficient.
Uniform priors are adopted for all except $\rho_{\star,\mathrm{circ}}$
for which we use a log-uniform between $10^{-3}$\,g\,cm$^{-3}$ and
$10^{+3}$\,g\,cm$^{-3}$.

For model $\mathcal{T}$, we have 5+$N$ parameters, where $N$ is the number of
available transit epochs. The first five terms are the same as that of model
$\mathcal{P}$ except for $P$ and $\tau$. The extra $N$ terms are the individual
times of transit minimum for each epoch.

Finally, for model $\mathcal{M}$, we have 14 free parameters. The first seven
are simply the same as model $\mathcal{P}$ but the latter seven describe the
exomoon. Specifically, these are
i) $P_S$, the orbital period of the satellite
ii) $a_{SP}/R_P$, the planet-satellite semi-major axis in units of the
planetary radius
iii) $R_{SP}$, the ratio of radii between the satellite and the planet
iv) $M_{SP}$, the ratio of masses between the satellite and the planet
v) $\phi_S$, the orbital phase of the satellite at the instant of planet-star
inferior conjunction during the reference epoch
vi) $\cos(i_s)$, cosine of the satellite's orbital inclination angle, relative
to the star-planet orbital plane
vii) $\Omega_S$, longitude of the ascending node of the satellite's orbit,
relative to the star-planet orbital plane. As before, uniform priors are
adopted for all with the exception of $P_S$ which uses a log-uniform prior from
75\,minutes to the period corresponding to one Hill radius. The semi-major axis
of the satellite has a uniform prior from 2 to 100 planetary radii.

For all models, a normal likelihood function is adopted. The only addition made
to this is that we applied a likelihood penalty to model $\mathcal{M}$ which
explored unphysical parameter combinations. Specifically, we calculate the
satellite and planet density using previously published
expressions\cite{weighing} and reject any samples for which
$\rho_S > 20$\,g\,cm$^{-3}$ or $\rho_P > 150$\,g\,cm$^{-3}$ or
$\rho_P < 0.5$\,g\,cm$^{-3}$, in an effort to keep the sampler in the region
of physically plausible solutions. Additionally, solutions where
the satellite period exceeds 93.09\% of the Hill sphere are unstable even
for retrograde orbits\cite{domingos2006}, and are thus rejected.

\subsection{Initial checks for exomoon candidacy.}
\label{sub:vetting1}


The primary objective of this work is to search for new possible exomoon
candidates amongst \kepler's cool giant sample. One of the first observable
effects predicted for exomoons come from transit timing variations (TTVs)
imparted onto the planet by the moon's gravitational
influence\cite{sartoretti1999}. Thus, a basic criterion one might suggest is
that TTVs should be present. However, for 25 of our targets, only two
transits were available and thus TTVs cannot be inferred. For the remainder, we
apply a statistical test for TTVs as described later.

The other observational consequence of an exomoon is that its shadow can
eclipse either the star or planet (i.e. a syzygy), leading to additional
changes in flux on-top of the conventional transit signature\cite{luna2011}.
These changes can occur in- or out-of-transit and impart complex shapes.
Yet more, the limited number of transits available in our sample means that
stacking approaches\cite{simon2012,ose2014} will not be statistically valid.
However, our photodynamical planet-moon model (model $\mathcal{M}$) computed
using \luna\ does fully account for the moon's subtle influence on the light
curve. Since \multi\ provides marginal likelihoods, we can evaluate the Bayes
factor between models $\mathcal{P}$ and $\mathcal{M}$, whilst correctly
penalising the moon model for its greater complexity, to determine the
statistical evidence for a moon. We thus demand that the Bayes factor formally
favours the planet-moon model over the planet model to be considered further.

Finally, we require that the planet has an orbit that is consistent with a
circular path. Elliptical orbits can be produced through planet-planet
scattering\cite{carrera2019}, which is expected to strip
exomoons\cite{gong2013}. Even if the eccentricity is produced through some
other effect, the dynamical region of stability is severely truncated by
non-zero eccentricity\cite{domingos2006}. Thus, although an elliptical orbit
does not prohibit exomoons, we consider it \textit{a-priori} improbable and
thus reject any planets exhibiting eccentric orbits. 

If these 2(+1) criteria are satisfied, the object is promoted for further
checks, which we refer to as secondary tests (discussed in the next
subsection). We describe the details of the initial tests in the following
paragraphs and highlight that the results are listed in Supplementary
Table~1.


For the eccentricity test, we require some formal criterion to evaluate if
a planet has an eccentric orbit or not. Since we assume a circular orbit
in all of our light curve models, then the derived mean stellar density
will be offset from the true value if the orbit is in fact
eccentric\cite{investigations2010}. If one has an independent and unbiased
measure of the mean stellar density in hand, this offset can be detected and
thus used to constrain the orbital eccentricity\cite{dawson2012}.
Again, this philosophy here is to minimise the exomoon FPR at the
expense of the FNR.

Our eccentricity test thus begins by taking the \textit{a-posteriori}
parameter samples from either model $\mathcal{P}$ or $\mathcal{T}$ -
specifically we default to $\mathcal{P}$ unless we conclude the system
is ``TTV hot'' as defined by our TTV test described later. Naturally, for
two-transit planets we always use model $\mathcal{P}$.

Next, we need an independent measure of the true stellar density and here that
comes from an isochrone analysis. This is described in a dedicated section
later and results are summarised in Supplementary Table~2.

To determine an eccentricity posterior for each planet, we begin with the
asterodensity profiling relationship\cite{investigations2010}, which relates
the light curve derived stellar density (under the assumption of a circular 
orbit), $\rho_{\star,\mathrm{circ}}$, to the true value, $\rho_{\star}$:

\begin{align}
\rho_{\star,\mathrm{circ}} &\simeq \rho_{\star} \Psi,
\end{align}

where

\begin{align}
\Psi \equiv \frac{(1 + e\sin\omega)^3}{(1-e^2)^{3/2}}.
\label{eqn:Psi}
\end{align}

Since $\rho_{\star,\mathrm{circ}}$ and $\rho_{\star}$ are inferred
independently, we construct a $\Psi$ posterior by simply drawing random samples
from our light curve derived density and dividing them by random samples from
the isochrone density. This $\Psi$ posterior does not trivially lead to an
eccentricity posterior sadly because of the joint dependence on $\omega$, the
argument of periastron. Thus, we need to again sample the parameter space.
To achieve this, we used kernel density estimation (KDE) on the $\log\Psi$
posterior with a Gaussian kernel and a bandwidth optimised for using least
squares cross validation. The KDE distribution was then used as a log-likelihood
function for the purposes of an MCMC exploration in $\{e,\omega\}$ parameter
space.

Since transiting planets are more likely to be eccentric, \textit{a-priori}, as
a result of geometric bias\cite{barnes2007,burke2008}, it is necessary to
account for this selection bias during the inference. This in turn requires a
prior for the eccentricity distribution which cannot be strictly uniform to
avoid infinities\cite{kipping2014}. We thus assume that $\mathrm{Pr}(e)$, the
prior on eccentricity, is a Beta distribution with $\alpha=1$ and $\beta=3$,
broadly matching the long-period radial velocity population\cite{beta2013}. The
selection effect inherent to the transit method is then accounted for using the
joint prior, $\mathrm{Pr}(e,\omega|\mathrm{transiting})$ as derived for
eccentric planets\cite{kipping2014}.

We then sampled the $\{e,\omega\}$ parameter volume $110,000$ times, burning
out the first 10,000 steps. Note, that MCMC sampling suffers from biases at
boundary conditions, such as $e>0$, and this can lead to an artificial positive
skew in eccentricity\cite{ford2006}. This can be overcome by
re-parameterising\cite{anderson2011} to $\sqrt{e}\sin\omega$ and
$\sqrt{e}\cos\omega$, which we use here. Once the eccentricity posterior has
been evaluated, we next perform Bayesian model selection by evaluating the
Savage-Dickey ratio\cite{dickey1971}. This simply evaluates the posterior
density at $e=0$ versus the prior, where the ratio provides a direct estimate
of the Bayes factor of an eccentric versus circular orbit (in the case of
nested models such as here). In our case, any instance where the posterior
density is less than the prior at $e=0$ is hereby labelled as ``eccentric'',
otherwise ``circular''. The prior density is analytic and thus trivial to
evaluate at zero\cite{kipping2014} but for the posterior density we apply a
KDE to the posterior to evaluate the density at zero. Because of the boundary
condition at $e=0$, we mirror the posterior samples around zero and combine
them with the originals, and then apply a Gaussian KDE to the combined sample.
The density at zero is then equal to twice the density of this KDE at zero,
as a result of the doubling of the sample volume.

Thus far, the eccentricity test described above was applied to either the
planet-only or planet-with-TTVs light curve model, depending on whether
we classified the planet as ``TTV hot'' (see next paragraphs). However, we
also repeated this a second time applied to the planet-moon posteriors from
model $\mathcal{M}$. If the planet appears incompatible with a circular
orbit only after the moon component is introduced, we mark this as
\xmark$^{\dagger}$ symbol in Supplementary Table~1 and the
object is not considered further as a viable moon candidate. Such cases
essentially mean that the required moon solution demands a light curve shape
which is inconsistent with the derived stellar density unless eccentricity or
blending is introduced\cite{AP2014}.

We emphasise that transits planets preferring eccentric orbit
solutions are identified via the existence of a ``photoeccentric effect'',
which describes an apparent tension between the light curve derived
stellar density and an independently inferred value\cite{dawson2012}.
However, blends and starspots can also cause substantial tension,
and both would lead to an elevated risk of exomoon false-positives
motivating their exclusion. Finally, although we exclude these systems
in this study, that does not mean they are necessarily devoid of moons,
any more than hot-Jupiters are necessarily devoid of moons. But in both
cases, physical arguments suggest they are not the most suitable
environment. And so, although we elect to avoid such systems in this
study, efforts by other teams to survey such objects are by no means
futile and we encourage such work.


For the TTVs, as noted earlier, 25 of our targets have only two
transits available and thus cannot be tested. This is because TTVs are
defined as an excursion away from a linear ephemeris fit, but a linear
ephemeris model (governed by two free parameters) will always provide a
perfect fit to two arbitrary transit times (two data points). For the other
cases, we can search for TTVs as an indication for an exomoon, as well
as providing some novel insights about the propensity of cool giants to
exhibit TTVs more broadly.

Testing for TTVs through periodogram searches is impractical for the vast
majority of our sample. This is because 51/72 of our planets have have three
transits or less and thus will offer just three data points for a regression.
For a sinusoidal TTV, the simplest periodic function, five unknown parameters
describe the ephemeris (the period and reference time of transit minimum, as
well as three sinusoid parameters - period, phase, amplitude). Even in a
grid-search periodogram, which removes one parameter - TTV period, we still
have less data than free parameters. 
Note that if one possesses four data points, the system becomes constrained but
fits are typically ``perfect'', although in such cases one can apply
regularisation techniques on the amplitude term to make progress e.g. as has
been done for Kepler-1625b\cite{undersampled}.

Instead of trying to seek a \textit{periodic} TTV, we simply ask whether there
is evidence for a TTV. To this end, we follow earlier work\cite{sixexomoons}
and apply their first test which addressed this question. This takes the
maximum likelihood light curve fits of models $\mathcal{P}$ and $\mathcal{T}$
and compares their log-likelihood through a Bayesian Information Criterion
(BIC) evaluation\cite{schwarz1978}. By working with the light curves directly,
rather than derived products such as marginalised transit times, we are able
to extract as much information from the light curve as possible. Any planet
with a BIC preference for model $\mathcal{T}$ is labelled as ``TTV hot'',
else ``TTV cold'' unless only two transits exist in which we use ``TTV grey''.

\subsection{Secondary checks for exomoon candidacy.}
\label{sub:vetting2}


If a planet passes the basic checks described in the last section, we apply
additional checks to evaluate the plausibility of an exomoon. In total, 11 of
the 70 cool giants satisfy this criteria. First, we require that the planet's
eccentricity, as determined from model $\mathcal{M}$, also favours a circular
path. Following the same method described in the last section, we
find that all 11 indeed appear consistent with circular after applying this
test.

Next, we regressed a new moon model to the data, model $\mathcal{X}$, which
is identical to model $\mathcal{M}$, except that negative and zero radius moons
are permitted. Negative radii correspond to inverted transits are simply
implemented by flipping the signals. Zero radii moons are formally forbidden
in model $\mathcal{M}$ since we impose the density constraint that
$\rho_S<20$\,g\,cm$^{-3}$ and a zero-radius moon has infinite density. Thus,
to enable this we relax this condition by simply commenting out this check in
our code. Using the posteriors of model $\mathcal{X}$, we apply three
statistical additional tests to the 11 objects.

The first of these, which could be labeled test \#4 by this point, is that
we computed a Savage-Dickey ratio at the location of $M_S/M_P=0$ to
evaluate the statistical evidence in favour of a non-zero exomoon mass. If
the system has three or more transits, we imposed the constraint that the 
Bayes factor from this calculation must exceed 10 in favour of a finite
mass (i.e. ``strong'' evidence\cite{kass1995}). Next, test \#5,
we computed a Savage-Dickey ratio at the location of $R_S/R_P=0$ and demand
that, for all objects, the Bayes factor preference for a non-zero radius
exceeds 10. Finally, test \#6, we count up how many of the $R_S/R_P$ samples were
negative versus positive and demand that the positive:negative ratio exceeds
5. This final test catches the possibility that $R_S/R_P$ is
offset from zero but has substantial weight in the unphysical negative radius
regime.

After applying these cuts, three objects emerged as possible candidates:
KIC-8681125, KIC-7906827 and KIC-5351250 (aka Kepler-150).

\subsection{Vetting of KIC-8681125.01.}
\label{sub:kic868}

Advanced vetting of the KIC-8681125.01 planetary moon candidate begins by
visual inspection of the transit light curve fits to better understand
what type of moon signal is seemingly detected. As shown in Extended Data
Fig.~\ref{fig:KIC868_lcs}, the signal is remarkable for featuring no
clear moon-like transit. Instead, the main difference observed is a
transit depth change from $3590_{-130}^{+160}$\,ppm to
$3030_{-110}^{+140}$\,ppm. This is explained by the moon-model by placing
the moon transit on-top of the planetary transit in the first epoch but
then the moon avoids transiting the star altogether in the next epoch.

This situation was immediately suspicious and appeared somewhat convoluted
and fine-tuned, particularly when one compares to typical planet-moon
models generated in simulation work\cite{luna2011}. One possibility
is that a nearby contaminant source is more prominently included within
the aperture of the second epoch than the first, thus diluting the
second's transit depth. However, if the source landed on the same
silicon with the same aperture used each time, this would clearly be
excluded as a possibility.

To investigate this, we used the \kepler\ target pixel files to inspect
the photometry at the pixel level. Since the first epoch occurs in
quarter 10, but the second in quarter 16, the spacecraft has rolled
into a distinct position between the epochs (every four quarters it
returns to the same position). As a result, the source is on different
silicon between the two epochs. However, KIC-8681125 is located near
the centre of the entire detector array, within module 13, and thus
ends up remaining within this module even after the roll, since the
roll is itself uses an axis with an origin close to the centre of
the detector array. Despite this, it does indeed end up on different
silicon moving from quadrant 4 to quadrant 2 between the two epochs.

The optimal aperture used by the \kepler\ pipeline is also quite
distinct between the epochs, as shown in Supplementary
Fig.~\ref{fig:KIC868_pixels}. Epoch 1 has a simple 2x2 square pixel centred on
the source, but epoch 2 use a ``+'' shaped aperture with an extra pixel
included in one corner. In total, 6 pixels are used in the second aperture,
thus increasing the chance of a contaminant falling within the
aperture. On this basis, we consider that the hypothesis of a contaminant
driving the depth change as being highly plausible.

To investigate further, we fit the light curve with a model which
was identical to the planet-only model except for the fact the second
epoch had a unique blend factor associated with it, $\gamma$. The
maximum likelihood of this fit did not exceed the moon model, but it
led to a major $\Delta\chi^2 = 50$ improvement over the planet-only model.
Since the model only requires one extra parameter over the planet-only
model, whereas the planet-moon model needs 7, the blend model outperforms
the moon model in terms of the marginal likelihood. As a result, it
is formally the preferred model by a Bayes factor of 6.8.

Whilst the blend hypothesis seems to naturally resolve this system then,
we highlight that problems still remain with this idea.
Unfortunately, no high resolution imaging has been previously obtained
but \textit{Gaia} can resolve sources greater than 1-2 arcseconds away.
The closest source (source id 2127184090671914880) is 11\farcs8 away
and 1.7 magnitudes fainter. Given the pixel scale of \kepler\ of
4\farcs0, this is likely too far away to explain the relatively large
depth change, as well being somewhat fainter than expected to explain
the depth change. Another possibility is that an unseen source resides
closer within approximately one arcsecond of the source, evading
\textit{Gaia}. However, this is also not satisfactory as the contaminant
should then be sufficiently close as to be included in both epoch
apertures. The contaminant hypothesis is thus challenged by the lack
of an obvious known source.

We also considered the possibility that the star may be covered in
spots, and between the two epochs the spot coverage varies to manifest
the depth change. However, high spot coverage appears incompatible
with the \kepler\ photometry which is relatively flat. To explore this,
we ran a Lomb-Scargle periodogram on each quarter and find the amplitude
is consistently below 200\,ppm (see Supplementary
Fig.~\ref{fig:KIC868_activity}).

Other possibilities still remain, such as uncorrected stray-light
video cross-talk, for example, but it will be difficult to make
further progress in the absence of high resolution imaging, which we
encourage at this time. However, given our generally conservative
approach of seeking reasons to throw away moon candidates rather than
keep them, sufficient reason for skepticism exists about this object
that we do not consider it further as an exomoon candidate.

\subsection{Vetting of KIC-5351250.06.}
\label{sub:kic535}

In vetting the planetary moon candidate of KIC-5351250.06/Kepler-150f, we begin
by noting that the star is unusually active amongst the sample considered. This
is apparent from simple inspection of the light curves but has been also
previously reported\cite{mcquillan2014} as a rotationally active star with a
periodicity of 17.6\,days and amplitude of 10.9\,mmag (approximately 1\%).
Since the transit depth of Kepler-150f is ${\sim}1.5$\,mmag, this implies that
a larger area of the stellar surface is covered by spots than the sky-projected
area of the planetary disk. Accordingly, it is quite possible for the planet to
cross over one or more spots during the transit and induce upward flux
undulations\cite{ojeda2012} that mimic the signature of star-planet-moon
syzygies\cite{luna2011}.

If the spots are much colder than photosphere, then the spot crossings can be
up to the entire transit depth. In practice, this is somewhat rare for even the
most active stars\cite{beky2014}, requiring both a very cold spot and a perfect
alignment of the spot and planetary transit chord\cite{spotrod}. On this basis,
we proceeded with caution given the enhanced possibility of false-positives.

To investigate further, we ran a Lomb-Scargle periodogram of the PDC \kepler\
data, quarter-by-quarter. As shown in Supplementary
Fig.~\ref{fig:KIC535_activity}, we confirm the ${\sim}1$\% level activity
reported previously\cite{mcquillan2014} and note that the activity seems
greater in Q12 (corresponding to the second transit epoch of Kepler-150f)
versus Q5 (the first epoch). This indicates that spots are more likely to
corrupt the second transit than the first.

Inspection of the transit light curve itself, shown in Extended Data
Fig.~\ref{fig:KIC535_lcs}, reveals an apparent transit depth change from
epoch 1 to 2, going from $1350_{-190}^{+260}$\,ppm to $1100_{-100}^{+90}$\,ppm.
Closer inspection reveals that the trough of the second transit is not
uniformly higher, but rather bounces up and down sporadically - consistent with
the behaviour expect for spot-crossings\cite{spotrod}. Given that this transit
coincides with an episode of high activity, this begins to cast doubt on the
reality of the exomoon signal.

To go further, we fit just the first epoch in isolation with a planet-only
model and then used its maximum likelihood solution as ``template'' for
adding starspots to for the second transit. If the planet is passing
over a spotty, dark patch - as we hypothesise - then the transit will
also be diluted in depth because it is only now blocking out a relatively
dim region of the star's total intensity\cite{spotrod}. Thus, the second
transit is modified in two ways: 1) the addition of a dilution factor,
$\gamma$, and 2) the inclusion of $N$ spot crossing events. Since we are
not particularly interested in the properties of the spots themselves,
just whether they can fit the light curve better than a moon, we adopt a
simple heuristic model for the crossings. Specifically, we add on a
Gaussian of width $\sigma$, amplitude $A$ and central time $\mathbb{T}$;
thus meaning we have three parameters per spot.

In total, we regressed four different versions of this model to the second
epoch:
i) no spot crossings but a contamination factor (1 extra parameter),
ii) 1 spot crossing and the contamination factor (4 extra parameters),
iii) 2 spot crossings and the contamination factor (7 extra parameters),
iv) 3 spot crossings and the contamination factor (10 extra parameters).
Since the planet-only model has 7 native parameters, then the final model
includes 17 variables altogether. The results of these fits are shown in
Extended Data Fig.~\ref{fig:KIC535_lcs}, along with the fits from the
planet-only and planet-moon models

For the planet-only and planet-moon models, we have been thus far comparing
models using the Bayesian evidence. However, here, we seek an alternative
model selection method. To see why, consider that in the case of the
planet-moon and planet-only models, the model parameters are have physical
meaning and thus have well-defined parameter limits. For example, the moon's
orbital period is bounded by the inner Roche limit and the outer
Hill sphere. In contrast, our heuristic model has no clear bounds on the
parameters of interest. Thus, one could just increase the widths of the
priors somewhat arbitrarily, which would then dilute the Bayesian evidences.
Accordingly, the model selection results become highly subjective when
using marginal likelihoods for heuristic models, and we instead prefer to
use a model selection metric than compares the maximum likelihood solutions,
for which there is no sensitivity to prior widths.

The two most commonly used maximum likelihood model comparison metrics are
the Bayesian Information Criterion, or BIC\cite{schwarz1978}, and the
Akaike information criterion, or AIC\cite{akaike1974}. The AIC - motivated
from information theory - is more appropriate when none of the models are
considered truly correct, but one is ranking them in terms of their ability
to approximate the truth, which certainly true for heuristic models. Further,
the BIC includes a penalty term which depends on the number of data points,
and this introduces a degree of subjectivity into the model selection
process since it depends on how much one windows the data around each
transit mid-point. For these reasons, we used the AIC to rank these different
models.

In doing so, we find that the 2-spot model is favoured with AIC
improvements versus model i) of 2.6, 8.0 and 4.9 for models ii) to iv)
respectively. For the 2-spot model, the $\chi^2$ score when computed on both
epochs is 702.89, whereas the planet+moon model achieved 712.07. In other
words, the 2-spot model is a better match to the light curve than the
planet+moon model by $\Delta\chi^2 = 9.2$ despite using the same
number of free parameters.

At this point, one could go further and introduce astrophysical spot models,
coupled to the rotational modulations, but the purposes of this work - 
seeking exomoons - this is simply beyond the scope of our objectives. Although
we cannot fully reject the hypothesis of an exomoon, for the reasons described,
there is now sufficient basis to reject this particularly candidate as
a compelling object.

\subsection{Robustness of KIC-7906827.01's moon signature against detrending choices.}
\label{sub:kic790_detrending}

A possible concern with any claimed moon-like signal is that it is sensitive
to the choices of detrending method used to process the data. In this work,
we use the method marginalised light curves, computed as described earlier,
for the model comparison tests. Since this uses the median of eight different
light curve detrendings, it is possible that the signal is present in the
majority of these, but not all. That does not necessarily indicate that the
moon-like signal is spurious, but it would certainly motivate a deeper
investigation as to why this is happening and increases the possibility of a
spurious origin.

We therefore decided to inspect the individual detrendings for evidence that
the signature of the exomoon candidate was not a global feature. This is
complicated by the fact the moon-like signal isn't a single event, but rather
presents itself in both transits through subtle distortions. Although a visual
inspection of the light curve reveals broadly consistent morphologies across
all methods (Extended Data Fig.~\ref{fig:lcdetrendings}), we sought a more
quantitative metric to assess this.

To this end, we took the maximum \textit{a-posteriori} fit of the planet-moon
model conditioned upon the method marginalised light curve as a template, and
compared it to each of the eight detrended light curves. For reference, we also
took the maximum \textit{a-posteriori} planet-only fit. Crucially, we don't
re-fit any of these eight light curves, we simply ask how well these templates
agree with the data in hand. In every case, we find that the planet-moon model
yields superior agreement, indicating that the specific signature of the
hypothesised moon (and not some generic moon) is present in all detrendings. Yet
more, the $\Delta \chi^2$ values obtained are consistent with the value
obtained from the method marginalised light curve ($\Delta\chi^2=23.2$),
yielding $22.2$, $27.1$, $23.4$, $25.0$, $23.9$, $31.4$, $23.8$ and $15.0$.
This list has a median of $23.9$, and a mean of $24.0\pm4.6$ - consistent
with the value obtained from the method marginalised light curve.

On this basis, we conclude that the moon-like signature is robust against
choice of detrending method.

\subsection{Pixel level analysis of KIC-7906827.01.}
\label{sub:kic790_pixels}

We analysed the pixel-level data of \kicname\ to look for anything out of the
ordinary that might suggest that moon-like signal is spurious. To this end,
we largely follow the approach outlined in a previous paper\cite{kepler90g},
where a putative exomoon around Kepler-90g was shown to be a likely
false-positive. This also builds upon the tests already shown for
KIC-8681125.01 discussed previously.

We begin by extracting the individual raw light curves of each pixel within
the postage stamp of the target and for times directly surrounding the
two transits of \kicnameb. Specifically, we extracted light curves
of $\pm 2.5$ transit durations around the two known events. Each light curve
was then detrended using the \local\ method described earlier.

Next, we measure the planet signal-to-noise ratio (SNR) in each pixel by simply
calculating the weighted mean of the detrended pixel light curves
inside/outside the transit region (where we use the duration as determined from
the full planet-only fits found earlier). The standard deviations (divided by
the square root of the number of data points in each section) are used to
compute an error (through quadrature) which then forms the SNR. The result is
illustrated in the middle panel of Extended Data
Fig.~\ref{fig:kic790_pixels}. In comparing to the mean flux counts of each
pixel (shown in the left panel of Extended Data
Fig.~\ref{fig:kic790_pixels}), one sees good agreement between the location
of the highest flux and the location of the highest transit SNR. The planetary
transit thus shows no sign of being dislocated from the target or any other
other strange pixel behaviour.

Turning now to the moon-like signal, we seek to replicate the SNR test but
this is challenged by the fact the moon signature is not a simple box but
rather displays features across the light curve, and in different positions
in each transit. The SNR can instead be measured by asking, in each pixel,
how much better is the maximum \textit{a-posteriori} planet-moon light curve
model template versus that of the planet-only model? Here, ``template'' refers
to the solution obtained by regressing to the method marginalised light curves.
To quantify what we mean by ``better'', we evaluate the $\Delta\chi^2$ between
the two templates, such that positive numbers indicate that the planet-moon
model leads to improved agreement.

As the moon-like signature is inherently much lower SNR than the planetary
signal, the SNR map is correspondingly noisier, but it clearly shows a
concentration of the SNR on top of the target, as expected for a genuine
signal. We highlight that it is precisely this point that the moon candidate of
Kepler-90g failed to pass\cite{kepler90g}. On this basis, we find no evidence
in the pixel-level data to suspect the moon-like signature is i) associated
with a contaminating offset source, ii) is caused by a global dimming of the
detector postage stamp pixels (e.g. due to stray light), or iii) is caused by a
single pixel triggering a false-positive through anomalous behaviour.

\subsection{Centroid analysis of KIC-7906827.01.}
\label{sub:kic790_centroids}


From the \kicname\ {\tt fits} files, we extracted the flux weighted centroid
columns and inspected the time series behaviour of the $X$ and $Y$ positions
within the vicinity of the two transit epochs of \kicnameb. Masking the
transits themselves, and filtering on only data within 6 transit durations of
the eclipses, we fit a series of polynomials through the centroids of
increasing complexity. Scoring with the BIC\cite{schwarz1978}, we identified
the most favourable model for each transit in both $X$ and $Y$ and used this
to remove the long-term trend caused by pointing drift.

We then evaluated the mean position in and out of the transit event,
using the standard deviation to estimate uncertainty, to find that
the centroids exhibit a $\{-0.52\pm0.06,+0.62\pm0.05\}$\,millipixel shift
in the $\{X,Y\}$ directions (Supplementary Fig.~\ref{fig:centroidshift}).
Given the presence of nearby stars observed by \gaia, a centroid shift of some
kind is not surprising but it can also indicate that the transit is not 
associated with the target\cite{bryson2013} - which would open the door to
\kicnameb\ being a false-positive planet.


To investigate the possibility that one of the other known stars was in fact
the host, we created and modelled difference images for the high SNR transit
events in quarters 8 and 16 for \kicname. The results of this very strongly
show that the observed transit signal cannot be due to any stars in the \gaia\
catalog except the target star, \kicname.

We created the difference image as described in previous work\cite{bryson2013}.
Assuming that all flux change is due to the transit event, the difference image
will show a star-like image at the location of the transit signal source. For
each quarter, cadences were chosen in the transit event and the pixel values
were averaged over these cadences, creating an average in-transit image. The
same number of cadences were chosen on both sides of the transit event, and
averaged to create and average out-of-transit image. These observed images are
compared in Supplementary Figs.~\ref{bryson:q8Observed} and
\ref{bryson:q16Observed}. The similarity between the out-of-transit and difference
images very strongly indicate that \kicname\ is the source of the transit.

Even greater confidence in this arrives via modelling of the point response
function (PRF). We modelled the scene using the \kepler\ PRF and stars returned
by a \gaia\ catalog cone search with radius 12.8 arcsec as described in earlier
work\cite{bryson2017}. This search returned 5 stars, as dim as gmag = 21.0. The
\gaia\ proper motion corrected positions of these stars are plotted on all
figures. The stars are placed at pixel locations using \kepler's \radecpix\
code (see \wwwbryson). 

Supplementary Figs.~\ref{bryson:q8Compare} and \ref{bryson:q16Compare} compare
the observed and modeled pixels, demonstrating the quality of the PRF modeling.
Supplementary Figs.~\ref{bryson:q8Simulated} and \ref{bryson:q16Simulated}
compare the observed difference image (top left) with the modeled difference
image assuming that the transit is on each of the five stars in the cone
search. These simulated difference images were created by subtracting simulated
scenes similar to Supplementary Figs.~\ref{bryson:q8Compare} and
\ref{bryson:q16Compare}, with the in-transit scene reducing the flux of the
modelled star by a fitted depth.

It is clear from Supplementary Figs.~\ref{bryson:q8Simulated} and
\ref{bryson:q16Simulated} that a transit on the target star is the only one
that remotely matches the observed difference image. The other stars in the
\gaia\ catalog cannot reproduce the observed signal.


Whilst this analysis excludes the possibility that a different \textit{known}
star hosts the transit signal, it does not address the possibility of an
unknown, blended star with the target. To investigate this, we first measured
the position of the target star by performing a multi-star PRF Markov Chain
Monte Carlo (MCMC) fit to the average out-of-transit image, and the position of
the transit signal source by performing a single-star PRF MCMC fit of the
difference image. These fits computed the posterior distribution of pixel
position and flux for each star consistent with the data, and used a Gaussian
likelihood for each pixel with width given by the propagated per-pixel
uncertainty of the fitted image. These measurements are differenced to give the
distance of the transit source from the target star. Measuring both the target
star source and transit source with PRF fitting mitigates possible bias due to
PRF error, because the same bias likely occurs for both stars.

The blend probability is computed using Equation~(14) of T. Morton's earlier
work\cite{morton2011}, which gives the probability of a blend that can mimic
any planet within 2 arcsec of the target star as a function of the star's
Galactic latitude and \kepler\ magnitude (caution: the columns in Table~1 of
T. Morton's work, which gives the coefficients for Equation~(14), are
reversed). For our star, the probability of a planet-mimicking blend within 2
arcsec is $3.08 \times 10^{-4}$. We compute the $3 \sigma$ radius of the target
star based on the 68th percentile credible interval from the fit to the
difference image, and scale the blend probability by the ratio of the
$3$\,$\sigma$ circle to a 2 arcsec circle.  

The results are summarised in Supplementary Table~3 for
quarters 8 and 16. The transit depth is recovered by taking the ratio of the
fitted fluxes of the difference image to target star from the out-of-transit
image, demonstrating the success of the fit.  The transit source is about 70
milliarcsec from the target star, which is just over $1 \sigma$.  The resulting
blend probability is about $2.6 \times 10^{-6}$.

\subsection{Statistical validation of \kepxb.}
\label{sub:kic790_validation}

Our centroid analysis establishes that the transit signal is associated
with the target star and that a blend is highly improbable given current
observations. This, in isolation, provides a compelling case that \kicnameb\
is a genuine planet. This possibility, often dubbed PRF contamination,
dominates the catalog of known \kepler\ false-positives; for example, 1587
of the 1859 false-positives identified through ephemeris
matching\cite{coughlin2014} to known eclipsing binaries (EBs) originate from
PRF contamination\cite{thompson2018}. However, EBs can also occasionally cause
false-positives without PRF contamination, via column anomalies, cross-talk and
reflections\cite{coughlin2014}. However, we note that \kicnameb\ was already
tested for an EB ephemeris match in the aforementioned work\cite{thompson2018}
and no matches were found, further strengthening the case that \kicnameb\
is a genuine planet. To finalize this, we took the shape of the transit
light curve, in combination with the stellar parameters, to independently
validate \kicnameb.

To this end, we used the \vespa\ package developed for precisely this
task\cite{morton2016}. Here, the shape of the transit light curve is compared
to a suite of models including both planet and false-positive scenarios, to
evaluate the statistical probability of each. The \textit{a-priori} probability
of blending, based on the star's position and fundamental properties
constrained from spectroscopy and \gaia, is used to weigh these scenarios
appropriately in the final evaluation. One additional piece of information
that can be helpful in this task is the existence and upper limit of
an occultation event. A long-period planet like this should not produce a
detectable occultation, and so its existence would put pressure on the
planet hypothesis.

An occultation event is generally expected to be approximately the same
duration as the transit and so we can exploit this feature to provide a
non-parametric means of detrending all of the \kepler\ quarters. Specifically,
we use a median filter where the bandwidth is set to three times the transit
duration of \kicnameb, which essentially acts as a low-cut filter removing all
variability on timescales greater than this threshold. The detrended light
curve was then phase folded onto the ephemeris of the transiting planet modulo
a half-period shift.

For a circular orbit planet, the occultation should occur at precisely a folded
time of zero. However, orbital eccentricity effects cause the occultation to
shift away from zero. Since the eccentricity is unknown, especially if we
remain agnostic about whether the transit signal is truly associated with the
target star, then the shift is also unknown. Accordingly, we created a uniform
grid of possible times across the entire orbit, spaced by one-tenth of the
transit duration.

At each grid point, representing a possible time of occultation, we first
evaluated the standard deviation within an interval equal to the transit
duration. This number was then divided by square root the number of data points
minus one, and thus represents the achievable precision on an occultation
event of similar duration to the transit as a function of orbital phase.
Although this precision score exhibits fluctuations as a result of data gaps
and sampling effects, we find it centres around a value of 62\,ppm. Repeating
using the median deviation as a more robust variance estimator yields 59\,ppm.
If no occultation effect is detected then, one would estimate a 3-$\sigma$
limit of $<180$\,ppm. For much shorter duration occultation events, this would
be overly optimistic though, since the fewer data points would inflate the
uncertainty. Whilst this essentially approaches infinity for infinitesimal
duration events, we adopt an upper limit of $10^{1/2}$ times shorter, which
corresponds to $<330$\,ppm.

The above explicitly assumes no occultation event, which we have to
demonstrate. To this end, we took each grid point and evaluated the SNR of an
occultation at 20 different trial durations (0.1 to 2.0 times the transit
duration in 0.1 steps). From these, we select the highest SNR duration as the
saved solution and continue to move through the grid of possible occultation
times. In this way, a genuine detection would manifest as a high SNR bump
within the grid, where we define the SNR as the mean out-of-occultation minus
the in-occultation intensity divided by the uncertainty on that mean (as
computed using the standard deviation).

For \kepler\ photometry, eclipses generally need to have SNR$>7$ to be
considered significant\cite{christiansen2013}, and we find no values near to
this level. The highest recorded SNR amongst 9225 realised positions with more
than two data points within the interval was 3.0. We thus find no evidence for
an occultation event of \kicnameb. If \kicnameb\ were a real planet, this is
the expected result since its long-period nature means it would be far too
dim to be detected photometrically. From the grid, we can also estimate an
upper limit on the occultation depth in an alternative way. Specifically,
we evaluated $\mathrm{max}[\delta_{\mathrm{occ}},0]+3\sigma_{\mathrm{occ}}$
at each grid point, where $\delta_{\mathrm{occ}}$ is the occultation depth
and $\sigma_{\mathrm{occ}}$ is the uncertainty. We then evaluated the
median of this array and added on 3 times the standard deviation of the
array. This is technically overkill as a 3-$\sigma$ limit since we have
used a 3-$\sigma$ limit twice over, but nevertheless it yields $<350$\,ppm
as an upper limit. This is in good agreement with our $<330$\,ppm value
from earlier and thus we adopt $350$\,ppm as a 3-$\sigma$ upper limit
in what follows.

Using this constraint with the light curve, stellar atmosphere properties
and \gaia\ parallax, we used \vespa\ to calculate the statistical probability
of a false-positive scenario. Eclipsing binary, hierarchical eclipsing binary
and blended eclipsing binary scenarios are all highly disfavoured and
lead to a planet FPP of 1 in 4237, or 0.024\% (see Supplementary Fig.
\ref{fig:vespa}). Combining this with the similar independent conclusion from
the centroid analysis, we conclude that \kicnameb\ is a genuine planet to high
confidence and thus refer to it as \kepxb\ in what follows.

\subsection{Exploring the possibility of alternative astrophysical models for \kepxb-i.}
\label{sub:kic790_badmodel}

The case for an exomoon rests upon the light curve analysis of the \kepler\
photometry. In particular, the Bayes factor of $11.9$ for the planet-moon model
versus the planet-only model drives the exomoon candidacy, as it surpasses
the ``strong evidence'' threshold adopted in this work of $>10$ and recommended
by previous works\cite{kass1995}. Bayes factors are influenced by the likelihood
function and the priors. In this case, the priors do not have arbitrary bounds
but rather well-motivated physical limits (e.g. the longitude of the ascending
node lives on a circle from 0 to $2\pi$ radians). Further, the case for an
exomoon signal remains compelling when viewed in a purely likelihood-based
framework, with a $\Delta\chi^2=23.2$ improved fit, indicating a 4.8-$\sigma$
effect. On this basis, we argue that the likelihood function drives this result
and is the place where we might rightfully apply skeptical interrogation.

The likelihood function can be wrong in two circumstances: 1) the forward
model is wrong 2) the noise model is wrong. We consider each of this in turn
but in this section address the former.

Regarding the forward modelling, the models in question are that of a planet
transiting a limb darkened star versus a planet-moon transiting a limb darkened
star. One might well wonder if some other model is ignored which is truly
responsible. In general, the asymmetric and short-term time-variable nature of
the transit shape is difficult to explain with some other localised
astrophysical effect associated with the planet. For example, a ring
system\cite{barnes2004} would need high obliquity, precession and many times
greater physical extent that Saturn's rings to explain the data. Further, such
an extensive ring system would significantly distort the light curve derived
mean stellar density away from the true value in a manner not observed
here\cite{zuluaga2015}.

We performed an additional check to see if the timing of the two inferred
moments of exomoon transit were suspicious or improbable. Exomoon transits
should be located close to the planetary event, moving back and forth
ostensibly randomly with a range governed by their semi-major axis around the
planet. The probability distribution of times is expected to follow an
arcsin distribution\cite{ose2014}. Although we only have two such times
available, it is possible to evaluate a $p$-value (``surprisingness'' score)
which might indicate tension with our choice of model (i.e. the planet-moon
model). To investigate this, we took our maximum \textit{a-posteriori}
planet-moon and re-generated the light curve but randomising the phase of the
exomoon. Repeating 1000 times, we were able to determine the moon transits
could have occurred up to $\pm0.25$\,days either side of the transit, with a
spread broadly following the arcsin distribution as expected. This may be
compared to the observed times of exomoon transit minima, of $-0.226$\,days and
$+0.136$\,days. Adopting the arcsin distribution, we can evaluate the formal
likelihood of obtaining the two observed times, which was $\log\mathcal{L} = 
1.82$. To put that number in some context, we repeated the above but drew two
random times from the arcsin distribution, evaluated their likelihood, and
built up a distribution of likelihoods under the null hypothesis. The
distribution is shown in Supplementary Fig.~\ref{fig:pvaluetiming}, where one
can see that the real likelihood score sits very close to the centre of the
expected distribution and is thus not remotely surprising. Accordingly, the
timing of the observed moon transits does not appear suspicious or offer
grounds to reject the planet-moon hypothesis.

Aside from a localised effect, the light curve model could be wrong if some
other non-localised phenomena simply coincidentally occurred during the time
of transit of \kepxb. The most obvious example would be a second transiting
planet in the system. Given the local window used of $\pm6.2$\,days, the
probability of this occurring is $\mathrm{min}[12.4/P_c,1]$ (depicted
in Supplementary Fig.~\ref{fig:planetc_prob} by the green dashed line) and
thus improbable for $P_c \gg 12.4$\,days. We note that there no other known
planetary candidates or even threshold crossing events\cite{thompson2018}
reported for this source. Nevertheless, this remains a possibility if the
hypothetical planet were simply too small to have been reliably detected by
the \kepler\ pipeline. Given the depth of the observed deviations, the planet
would need to 2.6\,$R_{\oplus}$ in radius at some unknown period - so how
possible is it that such a planet is hiding in the existing \kepler\ data?

To explore this possibility, we first regressed a two-planet transit model to
the same data used for the planet-only and planet-moon fits. Note that this
data only locally detrends the time series to within $\pm6.2$\,days of the
transit events - which we dub as $\mathcal{D}_{\mathrm{local}}$ in what
follows. We let the second planet have an unknown period with a log-uniform
prior from 10 to 1000\,days and uniform priors for impact parameters, transit
time within the first epoch's window, and ratio-of-radii. Using \multi\ to
explore the parameter space, the best fitting solution yielded a $\chi^2$
substantially lower than the planet-moon model, by $\Delta\chi^2=-16.2$.
Further, the two-planet model is only modestly improved over the one-planet
model despite being a nested model with four additional free parameters (thus
demanding an improved $\chi^2$) with $\Delta\chi^2=+7.0$. Indeed, this leads to
the model having a worse marginal likelihood than the planet-only model with
$\log\mathcal{Z}_{\mathrm{2-planet}} - \log\mathcal{Z}_{\mathrm{2-planet}} =
-0.94$. We note that this is well approximated by evaluating the AIC between
the two models using the $\chi^2$ difference (yielding $-1.05$). Thus, we find
that the local transit photometry, $\mathcal{D}_{\mathrm{local}}$, does not
support for the two-planet hypothesis.

We find that the posterior distribution for $P_c$ almost replicates the prior
of a log-uniform form. Thus, in any given $\log P$ window, we have
approximately the same number of posterior samples. Exploiting this, we group
the posterior into 8 evenly spaced bins in $\log P$ space, with approximately
$4000$ samples in each window. From these, we evaluate the maximum likelihood
solution amongst the subset. Since the AIC well-approximates the marginal
likelihood here, we use it to evaluate the Bayes factor as a function of
$\log P$ at these 8 grid points, which we then spline-interpolate to create a
continuous function. The result is shown in Supplementary
Fig.~\ref{fig:planetc_prob} (red dotted line), where one can see that
long-period solutions are in greatest tension with the
$\mathcal{D}_{\mathrm{local}}$ data.

The above only uses the local photometry to the \kepxb\ transits, but
the broader complete \kepler\ time series would also be expected to exhibit
transit signatures if the signal were caused by an interior transiting planet.

In order to support planet occurrence estimates from the DR25 \kepler\ planet
candidate catalog\cite{thompson2018}, the sensitivity for detecting a planet
of a given period and radius was previously measured in detail\cite{burke2017c,
christiansen2020}. The planet detection sensitivity was measured through
Monte-Carlo transit signal injection and recovery experiments\cite{burke2017b,
christiansen2017}. In previous work\cite{burke2017c}, a model was generated
for planet detection sensitivity that depends on the stellar properties and
noise characteristics of the \kepler\ flux time series based on fits to the
database of transit signal injections. The planet detection sensitivity model
can be calculated for any given \kepler\ target from the data products hosted
by the NASA Exoplanet Archive (see \wwwneaburke) and accompanying 
\keplerports\ \python\ software package (see \wwwkepports). 
Example uses of \keplerports, in the context measuring planet occurrence
rates, have been previously published\cite{burke2015,mulders2018,bryson2020}.

In order to calculate a planet detection contour for \kepx, we use stellar
parameters as updated in this study given in Supplementary
Table~2. We adopt stellar limb darkening parameters for \kepx\
(0.428, 0.4356, -0.1019, -0.0394; four parameter nonlinear limb darkening
parameterisation) by adopting the limb darkening values from the target with
the closest match in stellar properties in the DR25 \kepler\ stellar
catalog\cite{mathur2017}. The photometric noise properties for \kepx\ are
provided as supplemental columns in the DR25 stellar catalog. \kepx\ was
observed for 1459 days with a duty cycle of 87\%. The values of CDPP (Combined
Differential Photometric Precision\cite{christiansen2012}) slope at short and
long durations ($-0.70477$, $-0.3524$, respectively\cite{burke2017c}) indicate
a well-behaved flux time series data series for \kepx\ with minimal amounts of
non-Gaussian noise. The window function and one-sigma depth function
data\cite{burke2017a} for \kepx\ were downloaded from the NASA Exoplanet
Archive.

The resulting planet detection contour from \keplerports\ for \kepx\ is shown
in Supplementary Fig.~\ref{fig:keplerports}. If an additional planet had the
proper inclination to transit, the detection contour provides the probability
that a particular signal of a given period and radius would have been
classified as a planet candidate in the final DR25 \kepler\ planet candidate
catalog. As expected, the detection contour degrades toward small planets as
they have smaller transit depths and at long orbital periods as fewer transits
are available to stack and enhance the detection SNR. The standard version of
\keplerports\ publicly available represents the recoverability of planet
signals due to the \kepler\ pipeline by itself. An additional reduced
sensitivity to planet signals can result from the vetting
procedure\cite{coughlin2016,thompson2018,christiansen2020}. The results shown
in Supplementary Fig.~\ref{fig:keplerports} take into account the additional
loss of sensitivity due to the vetting procedure. The vetting degradation was
measured following a procedure similar to the description in section~4.2 of a
previous analysis\cite{bryson2020}. 

Whilst Supplementary Fig.~\ref{fig:keplerports} depicts the entire range of
possible radii, for this study we are most interested in the radius-slice
corresponding to that necessary to explain to exomoon-like signal, namely
$2.6$\,$R_{\oplus}$. Supplementary Fig.~\ref{fig:planetc_prob} (blue solid
line) shows a slice in the detection contour plane at this radius as a function
of orbital period. The discrete changes in detection probability (at 10, 60,
100, 200, 400, and 700 days) result due to the pipeline detection probability
and vetting recoverability probability are fit independently over orbital
period regions. The detection probability model fits are not required to be
continuous across orbital period region boundaries.

The results described thus far can be combined to evaluate the overall
probability that the observed moon-like deviations were caused by a second,
previously undetected transiting planet. This is evaluated by taking the
product of the three probabilities described thus far: 1) the second planet
coincidentally transits during the local window used to regress \kepxb, 2)
the probability that the two-planet model better explains the local data
versus the one-planet model (in a Bayesian sense), and 3) the probability
that a second planet of the required radius evaded detection from the
\kepler\ pipeline. This combined probability as a function of orbital
period is shown in black solid in the right panel of Supplementary
Fig.~\ref{fig:planetc_prob}. As shown, the probability does not exceed
1\% in the 10-1000\,day region considered. This result is somewhat
over-optimistic in that does assign a prior probability for a such a planet
existing in the first place and thus tacitly it is unity in the above
calculation. On this basis, we find that the probability of a second
transiting planet explaining the observed effects is $\lesssim 1$\%.

\subsection{Noise properties of the detrended \kepxb\ photometry.}
\label{sub:kic790_noise}

As discussed in the previous subsection, the likelihood function could be wrong
if the forward model is wrong or the noise model is wrong. With the former
investigated, we now turn to the latter. The data used to infer the exomoon
candidate spans $\pm6.2$\,days of the two transit events. That data was already
pre-whitened by virtue of the detrending process described earlier -
specifically method marginalised detrending. Since our likelihood function
assumes independent Gaussian noise, the likelihood function adopted (and thus
inferences thereafter) would be technically wrong if the noise were not
described by independent Gaussian noise.

In reality, no observations are ever purely Gaussian. The removal of
instrumental and astrophysical trends cannot ever be a perfect process.
However, we hope to reach a state where the non-Gaussian component is
much smaller than the Gaussian noise, and thus the likelihood function
adopted can be treated as an excellent approximation. In this subsection,
we thus investigate to what extent this appears to be true.

Using the method marginalised time series we excluded the data within
$\pm 0.55$\,days of the times of transit minima, in order to trim the region
where the planet-moon transits occur. The remaining data should now be
described by a normal distribution. One of the simplest tests of this
is to plot a histogram of the normalised intensities (see upper panel
of Supplementary Fig.~\ref{fig:KIC790_noise}). On that same figure, we plot
the probability density function of a normal distribution centred on unity with
a  standard deviation governed by the measurement uncertainties (i.e. this
is not a fit). As expected for Gaussian noise, we find excellent agreement.
Further, a Kolmogorov–Smirnov test of the Gaussianity reports a $p$-value
of 0.10 - consistent with expectation. Finally, the $\chi^2$ of these
data against a flat-line model equals $1073.8$, in close agreement with
the expected value given by the number of data points, $1074$. On this
basis, the data indeed appear normal.

Time-correlated noise can be difficult to identify using the tests
described thus far and a better evaluation comes from looking at the
temporal properties of the noise. Because the jumps between each data
segment, we split the data into four sections; epoch 1 pre-transit,
epoch 1 post-transit, epoch 2 pre-transit, \&
epoch 2 post-transit. The pre-transit moon feature in epoch 1
corresponds to approximately 8 cadences and thus we first tried a
simple autocorrelation test at lag-8 on these four sections, which
finds no significant autocorrelation ($p$-values of 0.54, 0.11, 0.73
and 0.28). We next tried a classic RMS-binning test, where we bin
the data sections into progressively larger bins and evaluate how
the scatter evolves. Supplementary Fig.~\ref{fig:KIC790_noise}
lower-panel shows the results compared to the expected behaviour of
independent Gaussian noise, where again we find no clear evidence for
time-correlation.

On this basis, we conclude that the detrended time series appears
consistent with independent Gaussian noise and thus the adopted
likelihood function is appropriate.

\subsection{False-positive probability of the exomoon signal \kepxb-i.}
\label{sub:kic790_fpp}

An advantage of seeking exomoons is that the null hypothesis is well-defined
and can be injected into real photometric time series. Briefly, we can take the
best-fitting planet-only model parameters, generate a template model, and
inject that into the SAP or PDC photometry as desired, and repeat the same
detection process outlined in this work. This allows us to directly calculate
the false-positive rate (FPP) of detecting an exomoon-like signal.

We emphasise that this is not the same FPP as used earlier when validating
\kepxb. There, the reality of the transit signal was unambiguous, but the
underlying cause was uncertain. Here, the situation is somewhat reversed.
The reality of the exomoon-like signal is unclear - and the FPP in this section
seeks to address that. However, the underlying cause of that signal (assuming
it's true) is not addressed by such a calculation. That has already been
addressed earlier, where we concluded that a second unseen transiting planet
was the most likely astrophysical false-positive, but even this has a
probability of $\lesssim 1$\% of explaining the observations. In what follows,
we focus on the FPP of the signal itself being astrophysical. Although
we refer to this as ``astrophysical'' hereafter, this is technically somewhat
of a misnomer since we are really addressing the probability of time-correlated
noise causing a false-positive, which could in fact be considered astrophysical
if due to stellar activity.

To inject fake planet signals, we took the maximum \textit{a-posteriori}
parameters from the planet-only model fits of \kepxb\ to define a
null-hypothesis template. Quarters 1 through 17 long-cadence data are available
for \kicname\ as possible times to inject the signal into. In some of these
quarters, we observe discontinuities in the SAP time series (for example due to
pointing tweaks) and we went through and located these, saving to a library
file. We then injected two transits of \kepxb, using the template with the only
difference being that $\tau$ (the time of transit minimum) is randomised.
The injection is performed by simplying multiplying the SAP and PDC flux
values by the template model ($=1$ outside of transits and $<1$ inside).
In this way, the time-correlated noise structure of the data is preserved.

There are several cases where the injections were rejected and re-attempted.
For example, if one of the transits is injected into a data gap (e.g. between
quarters) the realisation was rejected and re-tried. Our specific criteria
for a ``good'' injection were:

\begin{enumerate}
\item The injected transits must occurs greater than 3 transit durations away
from the real transit (to avoid signal overlap).
\item The injection has to occur at least half a transit duration after the
beginning of a quarter's start time, and at least half a duration before
the end of a quarter's start time.
\item The trimmed (see below) transit epoch files have to contain at least
530 data points in each (ensuring injections have $\gtrsim 90$\% of the
data volume as the real signals).
\end{enumerate}

The real transits were removed from the photometry, excluding data within
2 transit durations of the best fitting transit times. The real analysis
trimmed the photometry to within 6 transit durations of the central times,
and thus we use the same trimming here. As with the original analysis, if
a discontinuity occurs somewhere within the time series of interest, we only
consider the continuous segments surrounding the transit in question. Finally,
530 points was chosen since the original data has 591 (Q8) and 586 (Q16)
points. We wish to ensure that the injections contain at least 90\% of the
smallest of these two (586), which in principle is 527 data points. However,
we found that out outlier rejection algorithm rejects approximately 2\% of the
time series and by that expectation we need 530 points to ensure the 90\%
threshold.

In total, we created 200 random injections (and thus 400 injected transits).
We next applied the same method marginalised detrending algorithm to all
200, with the only difference being that the GP method was dropped for
computational expediency. The method marginalisation algorithm performs a
final check for the Durbin-Watson statistic and RMS vs bin-size behaviour,
evaluating a $p$-value against bootstrapped experiments. In some rare cases,
this led to an epoch being rejected if none of the methods were able
to produce sufficiently whitened time series. If this occurred, and less than
2 detrended transits were outputted, the realisation was rejected and
restarted with a new random seed.

The detrended light curve were then fit using \multi\ couple to \luna,
as before, trying both the planet-only and planet-moon models with
identical priors (except that $\tau$ is shifted onto the new ephemeris).
If the Bayes factor between the two models exceeded 10, denoting strong
evidence, it was flagged as a possible candidate as with the real analysis,
as shown in Fig.~\ref{fig:moonfpp}. For such cases, of which we found
just three instances, the next step was to perform the negative moon-radius
test. Two of three aforementioned cases (injections \#103 and \#161) pass
this test and we consider these to be ``false-positives'' from
the suite of 200 injections. Their signal shapes are shown in
Supplementary Fig.~\ref{fig:FalseMoons}.

By requiring these signals to be viable moon candidates, in other words
signals that our planet+moon model can explain as being physically sound,
not all dips and bumps in the light curve trigger a false-positive, only
the plausible ones. For example, Fig.~\ref{fig:KIC790_lcs} shows a small
deviation around BJD\,2456341 that was not interpretable by our fits as a
moon signal and thus does not constitute a false-positive by this definition.

On this basis, we conclude that the false-positive rate of the exomoon-like
signal of \kepxb\ is $1.0_{-1.0}^{+0.7}$\% (uncertainty from counting
statistics).

\subsection{Interpreting the FPP.}
\label{sub:kic790_tpp}

Given that we looked at 70 exoplanetary candidates in this survey, one success
from 70 with a 1\% false-positive rate might at first seem to fully explain
this event without invoking an exomoon. Certainly, this is a valid concern, and
one we share. However, that tacitly assumes that the 1\% false-positive rate
computed for \kepxb\ holds for all of the exoplanetary candidates surveyed,
which has neither been demonstrated nor can be reasonably assumed. Each
case will have its own bespoke FPP dependent upon the specific time-correlated
noise properties of each. Further, 45 of the 70 surveyed objects have three or
more transits (unlike \kepxb) and thus have to also pass a TTV test, which
would lower their false-positive rates by virtue of the extra check.

However, let us assume that the FPP rate (as caused by time-correlated noise)
was indeed 1\%, for the sake of making progress. In that case, it's important
to stress that whilst one false-positive is not an unexpected outcome, nor
is zero false-positives. Indeed, the likelihood ratio of the binomial
distribution for $n=70$ samples and $p=0.01$ between $X=1$ false-positives
and $X=0$ false-positives is $0.707$. In other words, it is marginally more
likely that a survey of 70 objects would produce no false-positives than one.
However, even this does not address what we really care about, what is the
probability this is an exomoon given the signal? Let us step back from the
ensemble and look at \kepxb\ in isolation once again.

Let us denote $E$ stands for ``has a \kepxb-i like exomoon'' (and $\bar{E}$
means it does not), and $P$ denotes passes our battery of tests (and $\bar{P}$
does not). With this notation, we can express the probability we seek as

\begin{align}
\mathrm{Pr}(E|P) &=
\frac{ \mathrm{Pr}(P|E) \mathrm{Pr}(E) }{
\mathrm{Pr}(P)
} 
\end{align}

where $\mathrm{Pr}(E)$ is the \textit{a-priori} probability of an
exoplanetary candidate in our sample having a \kepxb-i like exomoon (i.e. the
underlying occurrence rate of such moons in the target sample). The denominator
can be expanded as

\begin{align}
\mathrm{Pr}(P) &= \underbrace{\mathrm{Pr}(P|E)}_{=\mathrm{TPP}} \mathrm{Pr}(E) + \underbrace{\mathrm{Pr}(P|\bar{E})}_{=\mathrm{FPP}} \mathrm{Pr}(\bar{E}),
\end{align}

where we have denoted the false-positive probability (FPP) and true-positive
probabilities (TPP) explicitly. With some re-arranging, one can show

\begin{align}
\frac{\mathrm{Pr}(E|P)}{\mathrm{Pr}(\bar{E}|P)} &=
\frac{ \mathrm{TPP} \times \mathrm{Pr}(E) }{ \mathrm{FPP} \times (1 - \mathrm{Pr}(E)) }
\end{align}

The TPP is essentially the completeness and an accurate assessment is challenged
by defining what we even mean by a ``\kepxb-i like exomoon''. However, given
that the signal had a 4.8-$\sigma$ significance, we should generally expect a
high TPP for such signals (TPP$ \simeq 1$). Put another way, it would be odd if
we missed those. A detailed calculation of the TPP is beyond the scope of this
work and we argue largely unnecessary for the reasons stated above, as well as
the fact $\mathrm{Pr}(E)$ dominates our uncertainty in the calculations. If we
set TPP$ \to 1$, then we have 

\begin{align}
\lim_{\mathrm{TPP}\to1} \frac{\mathrm{Pr}(E|P)}{\mathrm{Pr}(\bar{E}|P)} &=
\frac{1}{\mathrm{FPP}} \Big( \frac{ \mathrm{Pr}(E) }{ 1 - \mathrm{Pr}(E) } \Big) \nonumber\\
\lim_{\mathrm{Pr}(E)\ll1} \lim_{\mathrm{TPP}\to1} \frac{\mathrm{Pr}(E|P)}{\mathrm{Pr}(\bar{E}|P)} &=
\frac{\mathrm{Pr}(E)}{\mathrm{FPP}}.
\end{align}

From this, we estimate that $\mathrm{Pr}(E|P) > \mathrm{Pr}(\bar{E}|P)$ if
$\mathrm{Pr}(E) > $FPP. Here, then, if 1\% or more of our sample host
\kepxb-i like exomoon, we should expect the detected signal is most likely
a real exomoon rather than a false-positive. This calculation reveals
the catch-22 conundrum facing the interpretation of this detection. In
isolation, it is not possible to reliably assess the odds that it is
real since we don't know the underlying occurrence rate of similar sized
moons around cold Jupiters.

Zooming back out to the ensemble, the total number of detections of
\kepxb-i like exomoons should be

\begin{align}
\sum_{i=1}^{70} \mathrm{TPP}_i \mathrm{Pr}(E) + \mathrm{FPP}_i.
\end{align}

In principle, one could define a likelihood function from this to infer
$\mathrm{Pr}(E)$ based on our one success and see if it's consistent with zero
- which would favour \kepxb-i being a false-positive. However, assuming
$\mathrm{TPP}_i \simeq 1$ for all 70 is not well-motivated here due to the
different noise properties of each source, and similarly the FPPs will be
distinct, as already discussed.

As this section establishes then, an accurate calculation of the probability of
\kepxb-i being genuine is marred with challenges, stemming from the unknown
occurrence rate of exomoons and the individual target FPP/TPP properties. This
also extends to considerations of specific parameters of our retrieved fit
versus false-positive scenarios. In principle, the FPPs and TPPs could be
determined with a far more extensive computational runs than done here,
although we highlight that this study already took several years to complete
and leveraged supercomputing time throughout (although not continuously). The
enormous computational challenge, human time, and CO2 production associated
with such an endeavour has to weighed against the benefits, or the simple act
of just re-observing \kepxb\ in the future to more straight-forwardly (and less
ambiguously) address this question.

In conclusion, in considering the exomoon-like signal associated with \kepxb,
we can find no firm grounds to reject it as a candidate at this time.
Future supporting evidence could be found by detecting TTVs, predicted in the
main text to have an amplitude between 1.2 to 77.0 minutes (95\% confidence).
In isolation, this would not be sufficient to confirm the moon due to the
possibility of perturbing planets. In practice, we argue that the only real
way to confirm/deny the existence of the moon convincingly would be 
high-precision transits of several future epochs, with the next event due
24th March 2023 (BJD\,2460027.86).

\end{methods}


\clearpage
\begin{addendum}
\item[Data Availability]
The data that support the plots within this paper and other findings of this
study are available at \wwwcoolworlds; or from the corresponding author
upon reasonable request.

\item[Code Availability]
The \multi\ regression algorithm\cite{feroz2009} is publicly available at
\href{https://github.com/farhanferoz/MultiNest}{https://github.com/farhanferoz/MultiNest}.
The \vespa\ software package\cite{morton2011,morton2016} is publicly available at
\href{https://github.com/timothydmorton/VESPA}{https://github.com/timothydmorton/VESPA}.
The \isochrones\ software package\cite{morton2015} is publicly available at
\href{https://github.com/timothydmorton/isochrones}{https://github.com/timothydmorton/isochrones}.
The \keplerports\ software package\cite{burke2017a,burke2017b,burke2017c} is
publicly available at
\href{https://github.com/nasa/KeplerPORTs}{https://github.com/nasa/KeplerPORTs}.
The \radecpix\ software package is publicly available at
\href{https://github.com/stevepur/Kepler-RaDex2Pix}{https://github.com/stevepur/Kepler-RaDex2Pix}.

\end{addendum}



\newpage
\begin{efigure}
  \centering
  \includegraphics[angle=0, width=16.0cm]{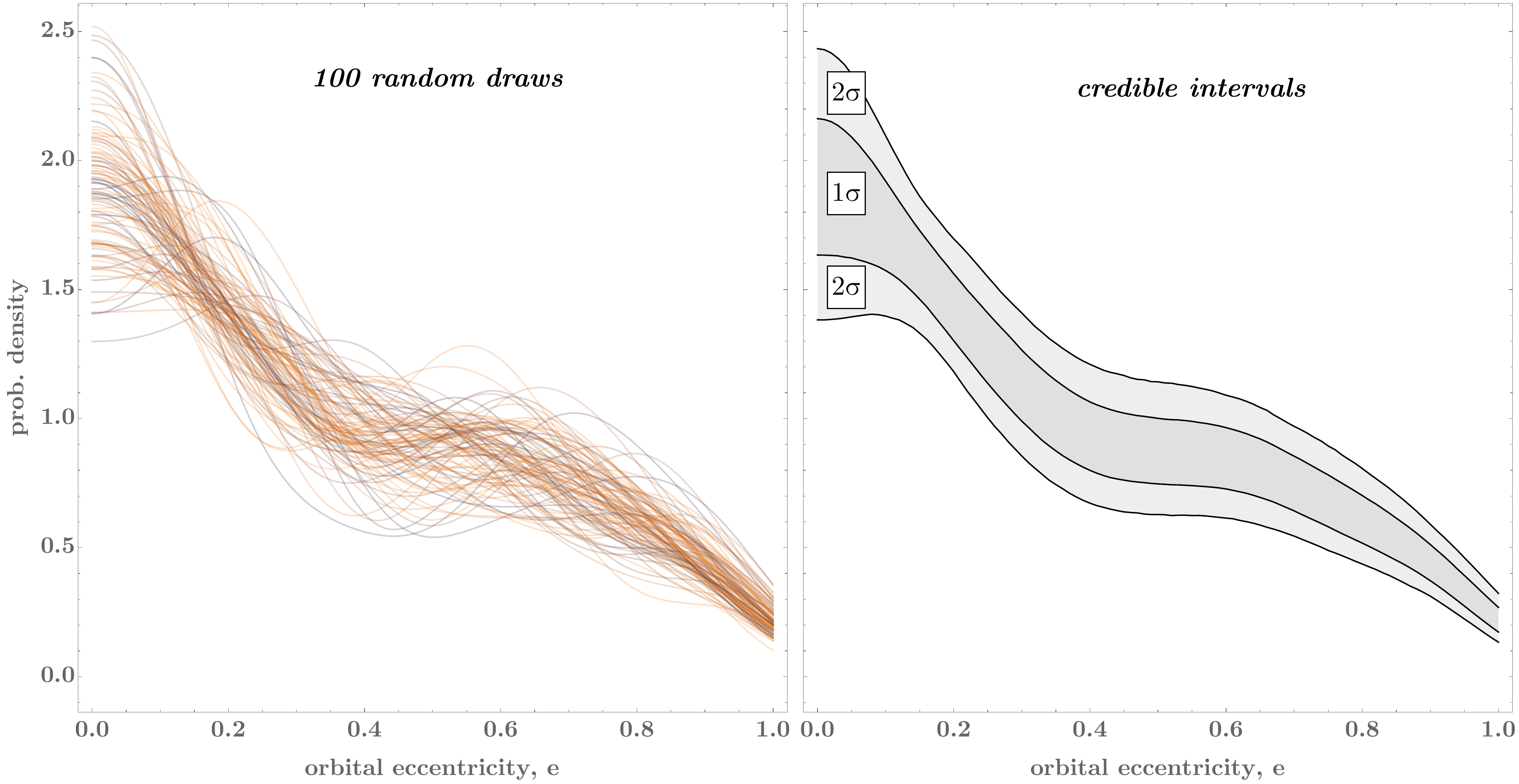}
  \caption{\label{fig:e_distribution}
  \textbf{Probability distribution of the cool giant's eccentricities.}
  Left: We extract a random draw from the eccentricity posterior distribution
  of each planet and apply a smooth kernel density estimator (KDE) to the
  sample with a Gaussian kernel. Each line represents 1 of 100 such realisations.
  Right: Credible intervals evaluated using $10^5$ such samples as computed in the
  left panel. 
  }
\end{efigure}

\newpage
\begin{efigure}
  \centering
  \includegraphics[angle=0, width=16.0cm]{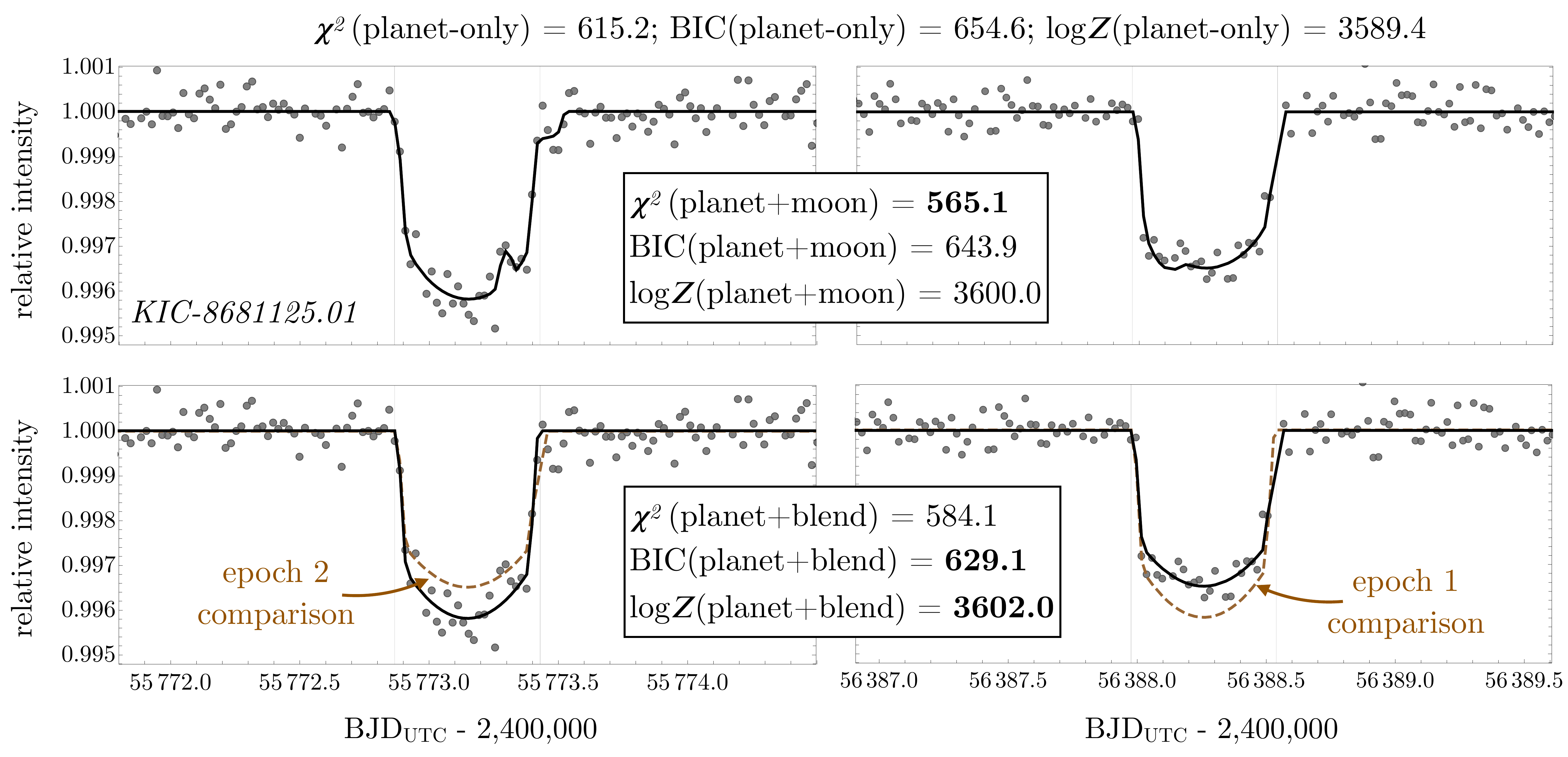}
  \caption{\label{fig:KIC868_lcs}
  \textbf{Transit light curves of KIC-8681125.01 for the first (left) and
  second (right) epochs.}
  Top: Each panel shows the method marginalised detrended photometry centred
  on the times of transit, with the maximum likelihood planet-moon fit overlaid
  in solid black. Model comparison statistics are provided within the inset
  box.
  Bottom: Same as above but for a model with a single planet and variable
  blend factor between the two epochs. This model substantially outperforms the
  planet-moon model.
  }
\end{efigure}

\newpage
\begin{efigure}
  \centering
  \includegraphics[angle=0, width=16.0cm]{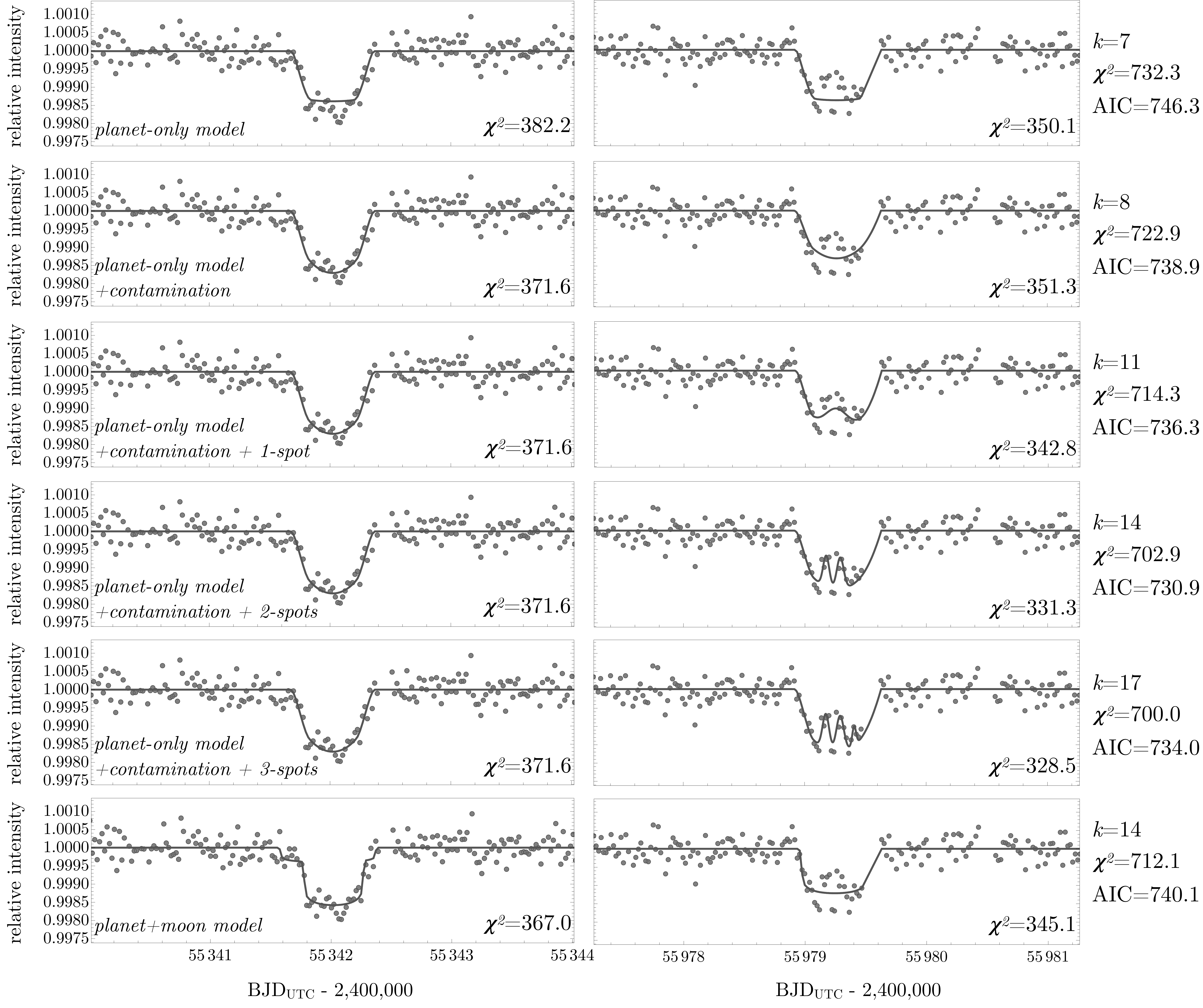}
  \caption{\label{fig:KIC535_lcs}
  \textbf{Transit light curves of KIC-5351250.06/Kepler-150f for the first (left)
	and second (right) epochs.}
  Each row shows a different model fit to the same data. Whilst the planet-moon
	model is clearly a better fit than the planet-only model, a 2-spot model
	is able to out-perform either and is well-motivated from the activity
	levels observed in the out-of-transit light curve.
  }
\end{efigure}

\newpage
\begin{efigure}
  \centering
  \includegraphics[angle=0, width=16.0cm]{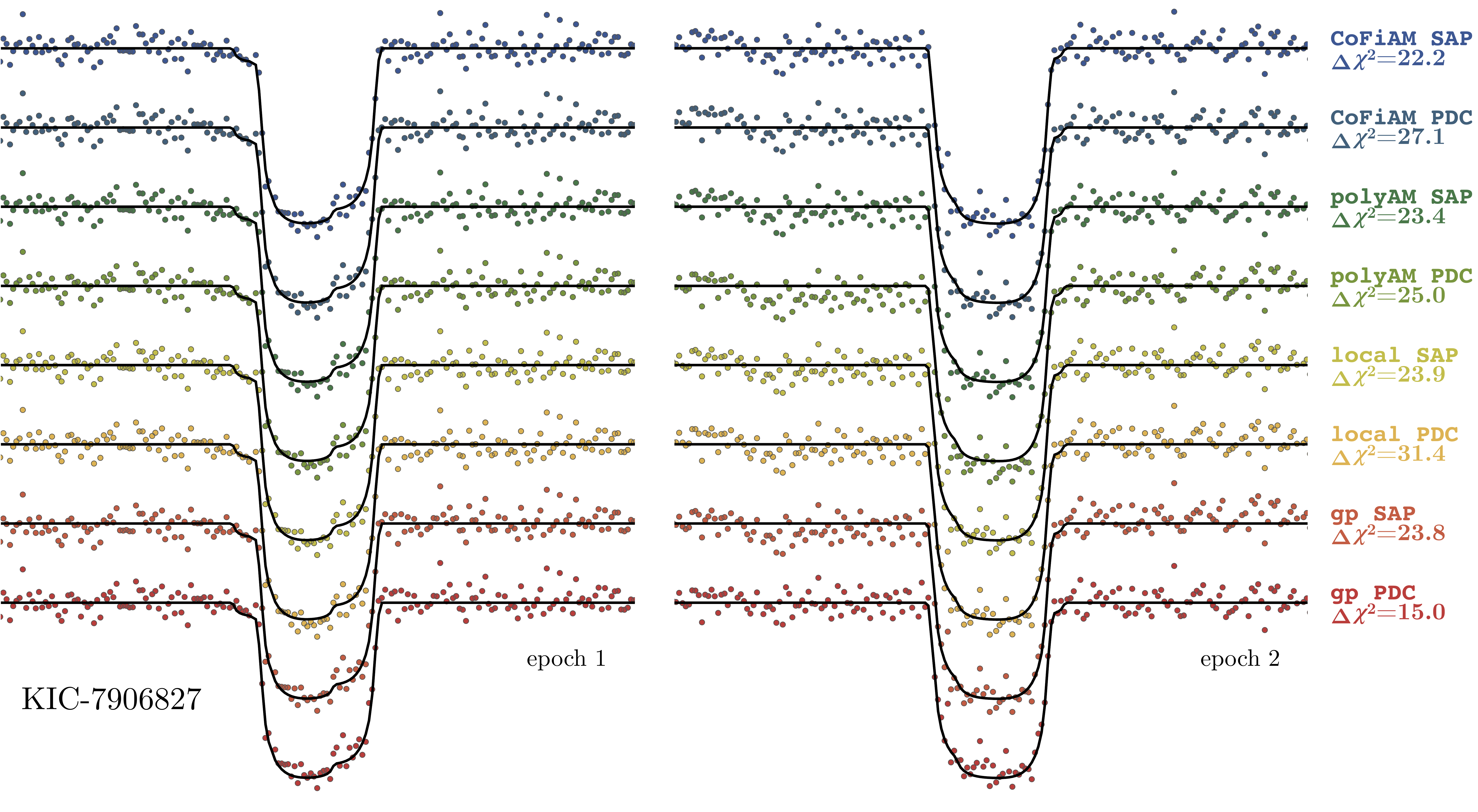}
  \caption{\label{fig:lcdetrendings}
  \textbf{Detrended transit light curves of \kicnameb\ for the first (left)
	and second (right) epochs.}
    Each row shows a different combination of light curve detrending method
	and input data, which are combined to build the method marginalised product.
	For each, we overlay the maximum \textit{a-posteriori} planet-moon
	model as conditioned upon the method marginalised light curve, and a comparison
	of how much better it matches the data versus the planet-only model, in a
	$\chi^2$-sense.
  }
\end{efigure}

\newpage
\begin{efigure}
  \centering
  \includegraphics[angle=0, width=16.0cm]{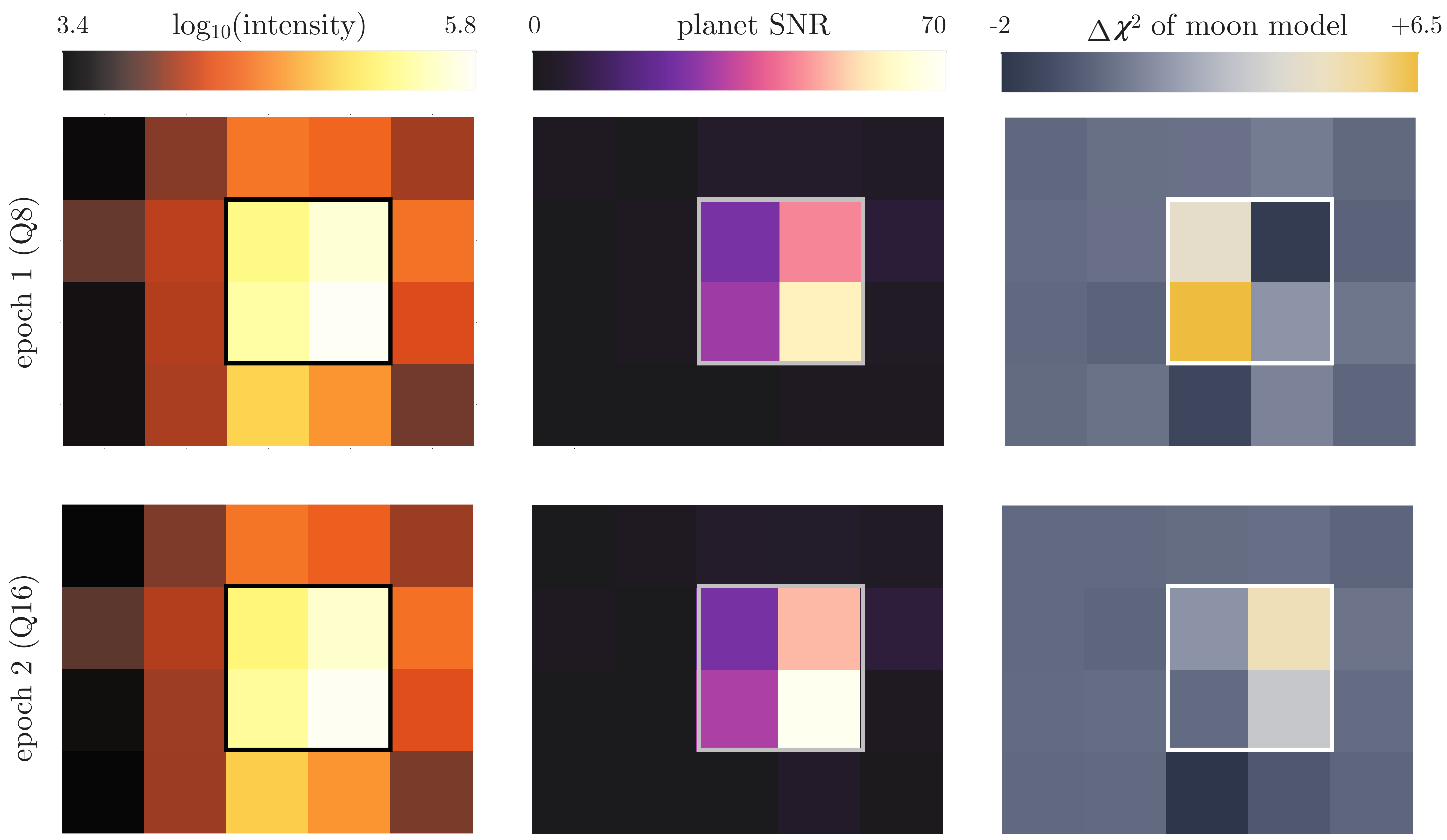}
  \caption{\label{fig:kic790_pixels}
  \textbf{Pixel-level comparison the two transits of \kicnameb.}
  Left: Pixel log-intensity is shown for the postage stamp downloaded for
  KIC-7906827 from the \kepler\ spacecraft, for epochs 1 (top) and 2 (bottom).
  The black solid outline shows the optimal aperture selected by the \kepler\
  pipeline.
  Middle: Same as the left, except we show the signal to noise ratio (SNR)
  of the planetary transit signal in each pixel. As expected, the transit signal
  is co-located with the brightest source in view.
  Right: Same as the middle, except we show the $\chi^2$ improvement of the
  planet-moon model over the planet-only model in each pixel light curve
  after \local\ detrending. As expected, the moon signal appears co-located with
  the target.
  }
\end{efigure}

\clearpage


\newpage
\begingroup
\captionsetup{width=16.8cm,font=normalsize}
\begin{longtable}{lcccccc}
\hline
KIC & Circular? & TTVs? & Moon favoured? & Signif. mass? & Signif. radius? & + radius? \\
\hline
3239945.02	 & 	\cmark	 & 	N/A	 & 	\xmark \\
3345675.01	 & 	\xmark	 & 	\xmark	 & 	\cmark \\
3534076.01	 & 	\xmark	 & 	\xmark	 & 	\cmark \\
3634051.01	 & 	\cmark	 & 	\xmark	 & 	\cmark \\
3756801.01	 & 	\cmark	 & 	N/A	 & 	\cmark	 & N/A & 1.9 [\xmark] & 2.5 [\xmark] \\
4346339.01	 & 	\xmark	 & 	\xmark	 & 	\cmark \\
4820550.01	 & 	\xmark	 & 	\xmark	 & 	\xmark \\
5010054.01	 & 	\cmark	 & 	N/A	 & 	\cmark	 & N/A & 0.5 [\xmark] & 2.1 [\xmark] \\
5094412.01	 & 	\cmark	 & 	\xmark	 & 	\xmark \\
5110453.01	 & 	\xmark$^{\dagger}$	 & 	N/A	 & 	\cmark \\
5115688.01	 & 	\xmark	 & 	\xmark	 & 	\cmark \\
5181299.01	 & 	\xmark	 & 	\xmark	 & 	\cmark \\
5184479.01	 & 	\cmark	 & 	N/A	 & 	\xmark \\
5351250.06	 & 	\cmark	 & 	N/A	 & 	\cmark	 & N/A & 2.8 [\cmark] & 74.8 [\cmark]\\
5437945.01	 & 	\cmark	 & 	\xmark	 & 	\cmark \\
5437945.02	 & 	\cmark	 & 	\xmark	 & 	\cmark \\
5732155.01	 & 	\cmark	 & 	N/A	 & 	\xmark \\
5792202.04	 & 	\cmark	 & 	\xmark	 & 	\xmark \\
6191521.02	 & 	\cmark	 & 	N/A	 & 	\xmark \\
6372194.01	 & 	\xmark	 & 	\xmark	 & 	\cmark \\
6443093.01	 & 	\cmark	 & 	\xmark	 & 	\xmark \\
6517255.01	 & 	\xmark	 & 	\xmark	 & 	\cmark \\
6867155.01	 & 	\xmark	 & 	\xmark	 & 	\cmark \\
6878240.01	 & 	\cmark	 & 	\cmark	 & 	\cmark	 & $>100$ [\cmark] & 1.4 [\xmark] & 0.0 [\xmark] \\
7198587.01	 & 	\xmark	 & 	\xmark	 & 	\cmark \\
7282470.01	 & 	\xmark	 & 	\xmark	 & 	\cmark \\
7363829.01	 & 	\xmark	 & 	N/A	 & 	\xmark \\
7383840.01	 & 	\xmark	 & 	N/A	 & 	\cmark \\
7619236.01	 & 	\cmark	 & 	\xmark	 & 	\cmark	 & 0.6 [\xmark] & $>100$ [\cmark] & $>100$ [\cmark]\\
7630229.01	 & 	\xmark	 & 	\xmark	 & 	\xmark \\
7731281.01	 & 	\cmark	 & 	\xmark	 & 	\cmark \\
7811397.01	 & 	\cmark	 & 	\cmark	 & 	\cmark	 & $>100$ [\cmark] & 1.5 [\xmark] & 78.4 [\cmark] \\
7906827.01	 & 	\cmark	 & 	N/A	 & 	\cmark	 & N/A & 2.54 [\cmark] & 63.3 [\cmark] \\
7917068.01	 & 	\cmark	 & 	\xmark	 & 	\cmark \\
8012732.01	 & 	\cmark	 & 	N/A	 & 	\xmark \\
8168509.01	 & 	\cmark	 & 	\xmark	 & 	\cmark \\
8240617.01	 & 	\cmark	 & 	\xmark	 & 	\cmark \\
8410697.01	 & 	\cmark	 & 	N/A	 & 	\xmark \\
8508736.01	 & 	\xmark	 & 	N/A	 & 	\xmark \\
8681125.01	 & 	\cmark	 & 	N/A	 & 	\cmark	 & N/A & 11.7 [\cmark] & 5.2 [\cmark] \\
8800954.01	 & 	\xmark	 & 	N/A	 & 	\xmark \\
9011955.01	 & 	\cmark	 & 	N/A	 & 	\cmark	 & N/A & $>100$ [\cmark] & 0.0 [\xmark] \\
9079767.01	 & 	\cmark	 & 	\xmark	 & 	\cmark \\
9147029.01	 & 	\xmark	 & 	\xmark	 & 	\xmark \\
9214713.01	 & 	\xmark	 & 	N/A	 & 	\cmark \\
9363944.01	 & 	\xmark	 & 	\xmark	 & 	\cmark \\
9413313.01	 & 	\xmark	 & 	\xmark	 & 	\cmark \\
9425139.01	 & 	\xmark	 & 	\xmark	 & 	\xmark \\
9512981.01	 & 	\xmark	 & 	\xmark	 & 	\xmark \\
9662267.01	 & 	\cmark	 & 	N/A	 & 	\xmark \\
9663113.02	 & 	\cmark	 & 	N/A	 & 	\xmark \\
9772531.01	 & 	\xmark	 & 	\xmark	 & 	\cmark \\
10272858.01	 & 	\xmark	 & 	\xmark	 & 	\cmark \\
10403228.02	 & 	\xmark$^{\dagger}$	 & 	\xmark	 & 	\cmark \\
10460629.01	 & 	\xmark	 & 	N/A	 & 	\cmark \\
10525077.02	 & 	\xmark	 & 	\xmark	 & 	\cmark \\
10552151.01	 & 	\xmark$^{\dagger}$	 & 	\xmark	 & 	\cmark \\
10850327.01	 & 	\xmark$^{\dagger}$	 & 	\xmark	 & 	\cmark \\
10937029.02	 & 	\xmark	 & 	\xmark	 & 	\cmark \\
11075279.01	 & 	\cmark	 & 	\xmark	 & 	\xmark \\
11442793.01	 & 	\cmark	 & 	\cmark	 & 	\cmark	 & 35.0 [\cmark] & 0.0 [\xmark] & 1.1 [\xmark] \\
11465813.01	 & 	\xmark	 & 	\xmark	 & 	\cmark \\
11805075.01	 & 	\xmark	 & 	\xmark	 & 	\cmark \\
11853130.01	 & 	\xmark	 & 	\xmark	 & 	\cmark \\
12266600.01	 & 	\cmark	 & 	N/A	 & 	\xmark \\
12356617.01	 & 	\xmark	 & 	N/A	 & 	\cmark \\
12416987.01	 & 	\xmark$^{\dagger}$	 & 	\xmark	 & 	\cmark \\
12454613.01	 & 	\cmark	 & 	N/A	 & 	\cmark	 & 	N/A & 1.0 [\xmark] & 9.5 [\cmark] \\
12647757.01	 & 	\xmark	 & 	N/A	 & 	\cmark \\
12735740.01	 & 	\xmark	 & 	\xmark	 & 	\cmark \\ [1ex]
\hline 
\caption*{\textbf{Supplementary Table 1 | Initial (columns 2-4) and secondary (columns 5-7) exomoon candidacy
tests applied to the 70 cool giants (column 1) in our survey}. For each, we simply
mark whether the test was passed/failed with a \cmark/\xmark. The $^{\dagger}$
symbol denotes that the circularity test was only failed for planet-moon
model.
}
\label{tab:targets} 
\end{longtable}

\newpage
\begingroup
\captionsetup{width=16.8cm,font=normalsize}
\renewcommand*{\arraystretch}{0.5}
\begin{longtable}{lcccccc}
\hline
KIC &
$M_{\star}$\,[$M_{\odot}$] &
$R_{\star}$\,[$R_{\odot}$] &
$\log_{10}(\mathrm{A}\,[\mathrm{yr}])$] & 
$\log_{10}(L_{\star}\,[L_{\odot}])$] & 
$d$\,[pc] & 
$\log_{10}(\rho_{\star}\,[\mathrm{g}\,\mathrm{cm}^{-3}])$] \\ [0.5ex] 
\hline 
\multirow{ 2}{*}{3239945} & $0.774$ & $0.735$ & $9.78$ & $-0.584$ & $345.3$ & $0.440$ \\
 & $\pm0.030$ & $\pm0.019$ & $\pm0.34$ & $\pm0.040$ & $\pm2.0$ & $\pm0.028$ \\
\multirow{ 2}{*}{3345675} & $0.680$ & $0.658$ & $9.87$ & $-0.871$ & $475.3$ & $0.528$ \\
 & $\pm0.029$ & $\pm0.021$ & $\pm0.34$ & $\pm0.058$ & $\pm8.3$ & $\pm0.030$ \\
\multirow{ 2}{*}{3534076} & $1.019$ & $1.015$ & $9.62$ & $0.022$ & $1085.1$ & $0.138$ \\
 & $\pm0.039$ & $\pm0.066$ & $\pm0.30$ & $\pm0.062$ & $\pm108.6$ & $\pm0.082$ \\
\multirow{ 2}{*}{3634051} & $1.081$ & $1.384$ & $9.82$ & $0.309$ & $672.9$ & $-0.239$ \\
 & $\pm0.047$ & $\pm0.141$ & $\pm0.12$ & $\pm0.089$ & $\pm6.9$ & $\pm0.129$ \\
\multirow{ 2}{*}{3756801} & $1.472$ & $2.213$ & $9.47$ & $0.756$ & $1607.3$ & $-0.718$ \\
 & $\pm0.059$ & $\pm0.147$ & $\pm0.05$ & $\pm0.055$ & $\pm45.3$	 & $\pm0.077$ \\
\multirow{ 2}{*}{4346339} & $1.074$ & $5.043$ & $9.86$ & $1.169$ & $1193.0$ & $-1.931$ \\
 & $\pm0.112$ & $\pm0.983$ & $\pm0.16$ & $\pm0.169$ & $\pm38.9$	 & $\pm0.390$ \\
\multirow{ 2}{*}{4820550} & $0.943$ & $0.950$ & $9.83$ & $-0.117$ & $596.4$ & $0.188$ \\
 & $\pm0.037$ & $\pm0.055$ & $\pm0.30$ & $\pm0.055$ & $\pm5.4$ & $\pm0.078$ \\
\multirow{ 2}{*}{5010054} & $1.185$ & $1.526$ & $9.65$ & $0.444$ & $1197.8$ & $-0.332$ \\
 & $\pm0.071$ & $\pm0.159$ & $\pm0.16$ & $\pm0.077$ & $\pm22.5$ & $\pm0.13$ \\
\multirow{ 2}{*}{5094412} & $0.883$ & $0.878$ & $9.90$ & $-0.252$ & $1103.$ & $0.264$ \\
 & $\pm0.035$ & $\pm0.044$ & $\pm0.31$ & $\pm0.054$ & $\pm39.0$ & $\pm0.068$ \\
\multirow{ 2}{*}{5110453} & $0.974$ & $4.328$ & $10.01$ & $1.013$ & $1664.6$ & $-1.770$ \\
 & $\pm0.080$ & $\pm0.281$ & $\pm0.11$ & $\pm0.059$ & $\pm49.6$ & $\pm0.105$ \\
\multirow{ 2}{*}{5115688} & $1.012$ & $3.805$ & $9.96$ & $0.950$ & $1836.7$ & $-1.586$ \\
 & $\pm0.085$ & $\pm0.223$ & $\pm0.11$ & $\pm0.056$ & $\pm70.0$ & $\pm0.083$ \\
\multirow{ 2}{*}{5181299} & $1.065$ & $3.645$ & $9.88$ & $0.923$ & $1612.8$ & $-1.504$ \\
 & $\pm0.134$ & $\pm0.732$ & $\pm0.18$ & $\pm0.141$ & $\pm43.1$ & $\pm0.341$ \\
\multirow{ 2}{*}{5184479} & $1.178$ & $1.502$ & $9.68$ & $0.415$ & $1226.7$ & $-0.312$ \\
 & $\pm0.062$ & $\pm0.147$ & $\pm0.09$ & $\pm0.082$ & $\pm31.6$ & $\pm0.112$ \\
\multirow{ 2}{*}{5351250} & $0.928$ & $0.923$ & $9.79$ & $-0.136$ & $914.8$ & $0.219$ \\
 & $\pm0.041$ & $\pm0.055$ & $\pm0.32$ & $\pm0.062$ & $\pm17.5$ & $\pm0.079$ \\
\multirow{ 2}{*}{5437945} & $1.313$ & $1.707$ & $9.47$ & $0.637$ & $1322.2$ & $-0.439$ \\
 & $\pm0.098$ & $\pm0.212$ & $\pm0.14$ & $\pm0.091$ & $\pm30.0$ & $\pm0.145$ \\
\multirow{ 2}{*}{5732155} & $1.355$ & $1.818$ & $9.48$ & $0.664$ & $2811.5$ & $-0.501$ \\
 & $\pm0.076$ & $\pm0.179$ & $\pm0.10$ & $\pm0.076$ & $\pm129.7$ & $\pm0.117$ \\
\multirow{ 2}{*}{5792202} & $0.880$ & $0.805$ & $9.81$ & $-0.310$ & $1077.3$ & $0.306$ \\
 & $\pm0.036$ & $\pm0.033$ & $\pm0.32$ & $\pm0.051$ & $\pm41.0$ & $\pm0.050$ \\
\multirow{ 2}{*}{6191521} & $0.968$ & $1.294$ & $10.05$ & $0.158$ & $1538.3$ & $-0.202$ \\
 & $\pm0.040$ & $\pm0.121$ & $\pm0.09$ & $\pm0.075$ & $\pm59.3$ & $\pm0.114$ \\
\multirow{ 2}{*}{6372194} & $0.886$ & $0.852$ & $9.69$ & $-0.242$ & $1284.2$ & $0.307$ \\
 & $\pm0.051$ & $\pm0.042$ & $\pm0.36$ & $\pm0.062$ & $\pm64.9$ & $\pm0.059$ \\
\multirow{ 2}{*}{6443093} & $1.029$ & $1.184$ & $9.83$ & $0.161$ & $1146.9$ & $-0.063$ \\
 & $\pm0.057$ & $\pm0.113$ & $\pm0.23$ & $\pm0.074$ & $\pm27.3$ & $\pm0.121$ \\
\multirow{ 2}{*}{6517255} & $1.124$ & $6.923$ & $9.79$ & $1.410$ & $2716.9$ & $-2.324$ \\
 & $\pm0.101$ & $\pm1.104$ & $\pm0.12$ & $\pm0.130$ & $\pm126.4$ & $\pm0.258$ \\
\multirow{ 2}{*}{6867155} & $0.687$ & $0.662$ & $9.85$ & $-0.864$ & $407.3$ & $0.523$ \\
 & $\pm0.020$ & $\pm0.013$ & $\pm0.31$ & $\pm0.032$ & $\pm4.1$ & $\pm0.022$ \\
\multirow{ 2}{*}{6878240} & $0.826$ & $0.790$ & $9.73$ & $-0.352$ & $1072.2$ & $0.375$ \\
 & $\pm0.045$ & $\pm0.036$ & $\pm0.35$ & $\pm0.058$ & $\pm135.4$ & $\pm0.051$ \\
\multirow{ 2}{*}{7198587} & $0.976$ & $4.355$ & $10.00$ & $1.018$ & $1253.4$ & $-1.776$ \\
 & $\pm0.087$ & $\pm0.318$ & $\pm0.11$ & $\pm0.067$ & $\pm93.0$ & $\pm0.112$ \\
\multirow{ 2}{*}{7282470} & $1.122$ & $4.372$ & $9.80$ & $1.099$ & $1118.1$ & $-1.725$ \\
 & $\pm0.119$ & $\pm0.723$ & $\pm0.14$ & $\pm0.127$ & $\pm30.8$ & $\pm0.291$ \\
\multirow{ 2}{*}{7363829} & $1.106$ & $1.773$ & $9.84$ & $0.476$ & $2121.5$ & $-0.548$ \\
 & $\pm0.073$ & $\pm0.178$ & $\pm0.09$ & $\pm0.082$ & $\pm98.2$ & $\pm0.122$ \\
\multirow{ 2}{*}{7383840} & $0.910$ & $0.918$ & $9.86$ & $-0.157$ & $955.9$ & $0.219$ \\
 & $\pm0.058$ & $\pm0.036$ & $\pm0.39$ & $\pm0.059$ & $\pm32.7$ & $\pm0.059$ \\
\multirow{ 2}{*}{7619236} & $0.989$ & $1.086$ & $9.91$ & $0.011$ & $662.2$ & $0.032$ \\
 & $\pm0.046$ & $\pm0.098$ & $\pm0.26$ & $\pm0.080$ & $\pm7.7$ & $\pm0.116$ \\
\multirow{ 2}{*}{7630229} & $1.030$ & $1.089$ & $9.75$ & $0.082$ & $688.5$ & $0.047$ \\
 & $\pm0.044$ & $\pm0.082$ & $\pm0.26$ & $\pm0.068$ & $\pm7.5$ & $\pm0.095$ \\
\multirow{ 2}{*}{7731281} & $0.581$ & $0.563$ & $9.88$ & $-1.187$ & $421.4$ & $0.662$ \\
 & $\pm0.032$ & $\pm0.030$ & $\pm0.36$ & $\pm0.085$ & $\pm71.2$ & $\pm0.048$ \\
\multirow{ 2}{*}{7811397} & $0.876$ & $0.833$ & $9.73$ & $-0.318$ & $1072.6$ & $0.331$ \\
 & $\pm0.033$ & $\pm0.029$ & $\pm0.32$ & $\pm0.043$ & $\pm37.4$ & $\pm0.044$ \\
\multirow{ 2}{*}{7906827} & $1.056$ & $1.098$ & $9.61$ & $0.140$ & $1753.9$ & $0.048$ \\
 & $\pm0.067$ & $\pm0.095$ & $\pm0.33$ & $\pm0.087$ & $\pm100.3$ & $\pm0.105$ \\
\multirow{ 2}{*}{7917068} & $1.402$ & $2.990$ & $9.47$ & $0.974$ & $1339.5$ & $-1.138$ \\
 & $\pm0.127$ & $\pm0.422$ & $\pm0.10$ & $\pm0.094$ & $\pm41.6$ & $\pm0.200$ \\
\multirow{ 2}{*}{8012732} & $1.195$ & $1.341$ & $9.53$ & $0.377$ & $1032.5$ & $-0.160$ \\
 & $\pm0.069$ & $\pm0.140$ & $\pm0.26$ & $\pm0.088$ & $\pm18.0$ & $\pm0.125$ \\
\multirow{ 2}{*}{8168509} & $1.415$ & $1.560$ & $9.17$ & $0.705$ & $1428.4$ & $-0.283$ \\
 & $\pm0.087$ & $\pm0.154$ & $\pm0.22$ & $\pm0.088$ & $\pm35.5$ & $\pm0.115$ \\
\multirow{ 2}{*}{8240617} & $0.887$ & $0.863$ & $9.79$ & $-0.248$ & $540.7$ & $0.290$ \\
 & $\pm0.043$ & $\pm0.046$ & $\pm0.34$ & $\pm0.077$ & $\pm75.2$ & $\pm0.069$ \\
\multirow{ 2}{*}{8410697} & $0.998$ & $1.035$ & $9.69$ & $0.071$ & $577.6$ & $0.097$ \\
 & $\pm0.070$ & $\pm0.087$ & $\pm0.34$ & $\pm0.078$ & $\pm3.9$ & $\pm0.104$ \\
\multirow{ 2}{*}{8508736} & $0.846$ & $0.819$ & $9.82$ & $-0.352$ & $1073.6$ & $0.339$ \\
 & $\pm0.054$ & $\pm0.055$ & $\pm0.37$ & $\pm0.112$ & $\pm176.6$ & $\pm0.078$ \\
\multirow{ 2}{*}{8681125} & $1.030$ & $1.066$ & $9.68$ & $0.075$ & $1097.4$ & $0.079$ \\
 & $\pm0.060$ & $\pm0.088$ & $\pm0.33$ & $\pm0.084$ & $\pm27.9$ & $\pm0.104$ \\
\multirow{ 2}{*}{8800954} & $0.840$ & $0.802$ & $9.73$ & $-0.337$ & $352.1$ & $0.361$ \\
 & $\pm0.041$ & $\pm0.031$ & $\pm0.33$ & $\pm0.043$ & $\pm2.1$ & $\pm0.046$ \\
\multirow{ 2}{*}{9011955} & $1.275$ & $1.516$ & $9.47$ & $0.521$ & $2364.2$ & $-0.293$ \\
 & $\pm0.075$ & $\pm0.162$ & $\pm0.19$ & $\pm0.086$ & $\pm135.9$ & $\pm0.127$ \\
\multirow{ 2}{*}{9079767} & $1.169$ & $1.253$ & $9.47$ & $0.335$ & $949.7$ & $-0.082$ \\
 & $\pm0.069$ & $\pm0.109$ & $\pm0.28$ & $\pm0.080$ & $\pm39.4$ & $\pm0.103$ \\
\multirow{ 2}{*}{9147029} & $1.076$ & $1.323$ & $9.82$ & $0.268$ & $1710.9$ & $-0.184$ \\
 & $\pm0.049$ & $\pm0.136$ & $\pm0.11$ & $\pm0.087$ & $\pm73.3$ & $\pm0.122$ \\
\multirow{ 2}{*}{9214713} & $1.189$ & $1.350$ & $9.52$ & $0.413$ & $1641.4$ & $-0.175$ \\
 & $\pm0.074$ & $\pm0.130$ & $\pm0.23$ & $\pm0.077$ & $\pm54.2$ & $\pm0.116$ \\
\multirow{ 2}{*}{9363944} & $0.992$ & $3.812$ & $9.98$ & $0.914$ & $1418.$ & $-1.587$ \\
 & $\pm0.127$ & $\pm0.638$ & $\pm0.16$ & $\pm0.129$ & $\pm42.4$ & $\pm0.277$ \\
\multirow{ 2}{*}{9413313} & $0.874$ & $0.855$ & $9.82$ & $-0.270$ & $476.7$ & $0.297$ \\
 & $\pm0.047$ & $\pm0.051$ & $\pm0.38$ & $\pm0.073$ & $\pm4.6$ & $\pm0.075$ \\
\multirow{ 2}{*}{9425139} & $1.055$ & $1.113$ & $9.74$ & $0.071$ & $552.$ & $0.028$ \\
 & $\pm0.047$ & $\pm0.092$ & $\pm0.27$ & $\pm0.074$ & $\pm4.8$ & $\pm0.102$ \\
\multirow{ 2}{*}{9512981} & $0.783$ & $0.755$ & $9.90$ & $-0.555$ & $894.4$ & $0.409$ \\
 & $\pm0.027$ & $\pm0.019$ & $\pm0.29$ & $\pm0.036$ & $\pm27.3$ & $\pm0.031$ \\
\multirow{ 2}{*}{9662267} & $1.059$ & $1.103$ & $9.61$ & $0.145$ & $1183.7$ & $0.045$ \\
 & $\pm0.066$ & $\pm0.095$ & $\pm0.33$ & $\pm0.083$ & $\pm28.7$ & $\pm0.105$ \\
\multirow{ 2}{*}{9663113} & $1.367$ & $1.825$ & $9.46$ & $0.678$ & $1648.2$ & $-0.501$ \\
 & $\pm0.052$ & $\pm0.133$ & $\pm0.05$ & $\pm0.059$ & $\pm46.2$ & $\pm0.084$ \\
\multirow{ 2}{*}{9772531} & $0.511$ & $0.494$ & $9.77$ & $-1.324$ & $304.4$ & $0.776$ \\
 & $\pm0.016$ & $\pm0.015$ & $\pm0.38$ & $\pm0.033$ & $\pm3.5$ & $\pm0.028$ \\
\multirow{ 2}{*}{10272858} & $1.090$ & $3.255$ & $9.85$ & $0.869$ & $568.9$ & $-1.353$ \\
 & $\pm0.124$ & $\pm0.209$ & $\pm0.15$ & $\pm0.063$ & $\pm16.5$ & $\pm0.071$ \\
\multirow{ 2}{*}{10403228} & $0.482$ & $0.460$ & $9.75$ & $-1.494$ & $299.5$ & $0.844$ \\
 & $\pm0.012$ & $\pm0.011$ & $\pm0.35$ & $\pm0.029$ & $\pm3.6$ & $\pm0.023$ \\
\multirow{ 2}{*}{10460629} & $1.425$ & $1.896$ & $9.38$ & $0.770$ & $1796.1$ & $-0.534$ \\
 & $\pm0.080$ & $\pm0.177$ & $\pm0.08$ & $\pm0.070$ & $\pm48.9$ & $\pm0.109$ \\
\multirow{ 2}{*}{10525077} & $1.075$ & $1.132$ & $9.59$ & $0.183$ & $1530.6$ & $0.018$ \\
 & $\pm0.069$ & $\pm0.102$ & $\pm0.32$ & $\pm0.087$ & $\pm58.2$ & $\pm0.109$ \\
\multirow{ 2}{*}{10552151} & $1.042$ & $3.055$ & $9.91$ & $0.804$ & $2116.7$ & $-1.285$ \\
 & $\pm0.101$ & $\pm0.406$ & $\pm0.12$ & $\pm0.110$ & $\pm88.8$ & $\pm0.210$ \\
\multirow{ 2}{*}{10850327} & $1.183$ & $1.391$ & $9.55$ & $0.438$ & $719.8$ & $-0.217$ \\
 & $\pm0.085$ & $\pm0.154$ & $\pm0.21$ & $\pm0.088$ & $\pm13.0$ & $\pm0.131$ \\
\multirow{ 2}{*}{10937029} & $0.883$ & $0.868$ & $9.88$ & $-0.292$ & $743.8$ & $0.279$ \\
 & $\pm0.036$ & $\pm0.038$ & $\pm0.31$ & $\pm0.053$ & $\pm12.7$ & $\pm0.058$ \\
\multirow{ 2}{*}{11075279} & $0.981$ & $0.977$ & $9.73$ & $-0.075$ & $478.4$ & $0.171$ \\
 & $\pm0.044$ & $\pm0.057$ & $\pm0.32$ & $\pm0.062$ & $\pm3.4$ & $\pm0.076$ \\
\multirow{ 2}{*}{11442793} & $1.103$ & $1.230$ & $9.68$ & $0.238$ & $869.2$ & $-0.081$ \\
 & $\pm0.048$ & $\pm0.100$ & $\pm0.19$ & $\pm0.069$ & $\pm11.2$ & $\pm0.100$ \\
\multirow{ 2}{*}{11465813} & $1.011$ & $1.017$ & $9.73$ & $-0.065$ & $1106.$ & $0.129$ \\
 & $\pm0.038$ & $\pm0.065$ & $\pm0.30$ & $\pm0.065$ & $\pm27.2$ & $\pm0.082$ \\
\multirow{ 2}{*}{11805075} & $0.956$ & $1.046$ & $9.94$ & $-0.017$ & $1017.3$ & $0.068$ \\
 & $\pm0.050$ & $\pm0.082$ & $\pm0.28$ & $\pm0.066$ & $\pm19.9$ & $\pm0.104$ \\
\multirow{ 2}{*}{11853130} & $0.592$ & $0.571$ & $9.77$ & $-1.153$ & $435.6$ & $0.651$ \\
 & $\pm0.015$ & $\pm0.014$ & $\pm0.37$ & $\pm0.030$ & $\pm7.1$ & $\pm0.024$ \\
\multirow{ 2}{*}{12266600} & $0.880$ & $0.856$ & $9.75$ & $-0.243$ & $835.7$ & $0.3$ \\
 & $\pm0.051$ & $\pm0.048$ & $\pm0.37$ & $\pm0.069$ & $\pm17.7$ & $\pm0.069$ \\
\multirow{ 2}{*}{12356617} & $1.184$ & $1.525$ & $9.71$ & $0.395$ & $855.1$ & $-0.333$ \\
 & $\pm0.063$ & $\pm0.120$ & $\pm0.10$ & $\pm0.063$ & $\pm9.6$ & $\pm0.091$ \\
\multirow{ 2}{*}{12416987} & $0.758$ & $0.718$ & $9.71$ & $-0.637$ & $798.7$ & $0.462$ \\
 & $\pm0.031$ & $\pm0.020$ & $\pm0.35$ & $\pm0.039$ & $\pm20.9$ & $\pm0.028$ \\
\multirow{ 2}{*}{12454613} & $0.919$ & $0.918$ & $9.81$ & $-0.153$ & $450.6$ & $0.224$ \\
 & $\pm0.050$ & $\pm0.064$ & $\pm0.36$ & $\pm0.075$ & $\pm2.6$ & $\pm0.089$ \\
\multirow{ 2}{*}{12647757} & $1.011$ & $3.994$ & $9.95$ & $0.983$ & $1327.6$ & $-1.65$ \\
 & $\pm0.100$ & $\pm0.379$ & $\pm0.16$ & $\pm0.077$ & $\pm27.7$ & $\pm0.181$ \\
\multirow{ 2}{*}{12735740} & $0.977$ & $0.967$ & $9.67$ & $-0.047$ & $347.4$ & $0.181$ \\
 & $\pm0.044$ & $\pm0.057$ & $\pm0.32$ & $\pm0.061$ & $\pm3.6$ & $\pm0.075$ \\ [1ex]
\hline 
\caption*{\textbf{Supplementary Table 2 | Fundamental stellar parameters inferred for the cool giant host stars in our
sample using an isochrone analysis.} Values quoted define the median and surrounding
68.3\% confidence interval of the posterior distributions.}
\label{tab:stellar} 
\end{longtable}
\endgroup

\newpage
\begin{sfigure}
  \centering
  \includegraphics[angle=0, width=16.0cm]{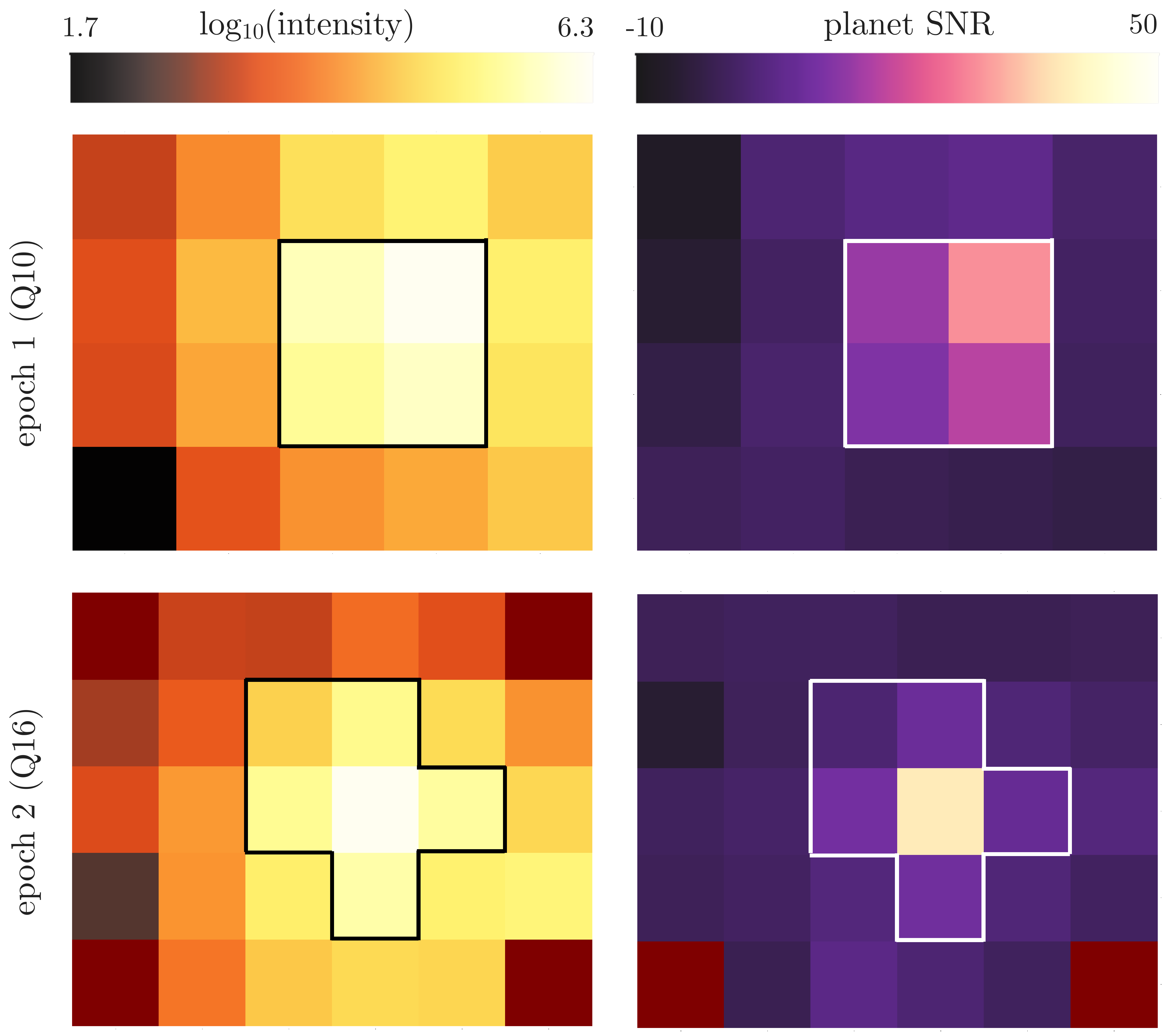}
  \caption{\label{fig:KIC868_pixels}
  \textbf{Comparison of the aperture used between the two transit epochs of
  KIC-8681125.01.}
  Left: Pixel log-intensity is shown for the postage stamp downloaded for
  KIC-8681125 from the \kepler\ spacecraft, for epochs 1 (top) and 2 (bottom).
  The black solid outline shows the optimal aperture selected by the \kepler\
  pipeline.
  Right: Same as the left, except we show the signal to noise ratio (SNR)
  of the transit signal in each pixel. As expected, the transit signal
  is co-located with the brightest source in view.
  }
\end{sfigure}

\newpage
\begin{sfigure}
  \centering
  \includegraphics[angle=0, width=16.0cm]{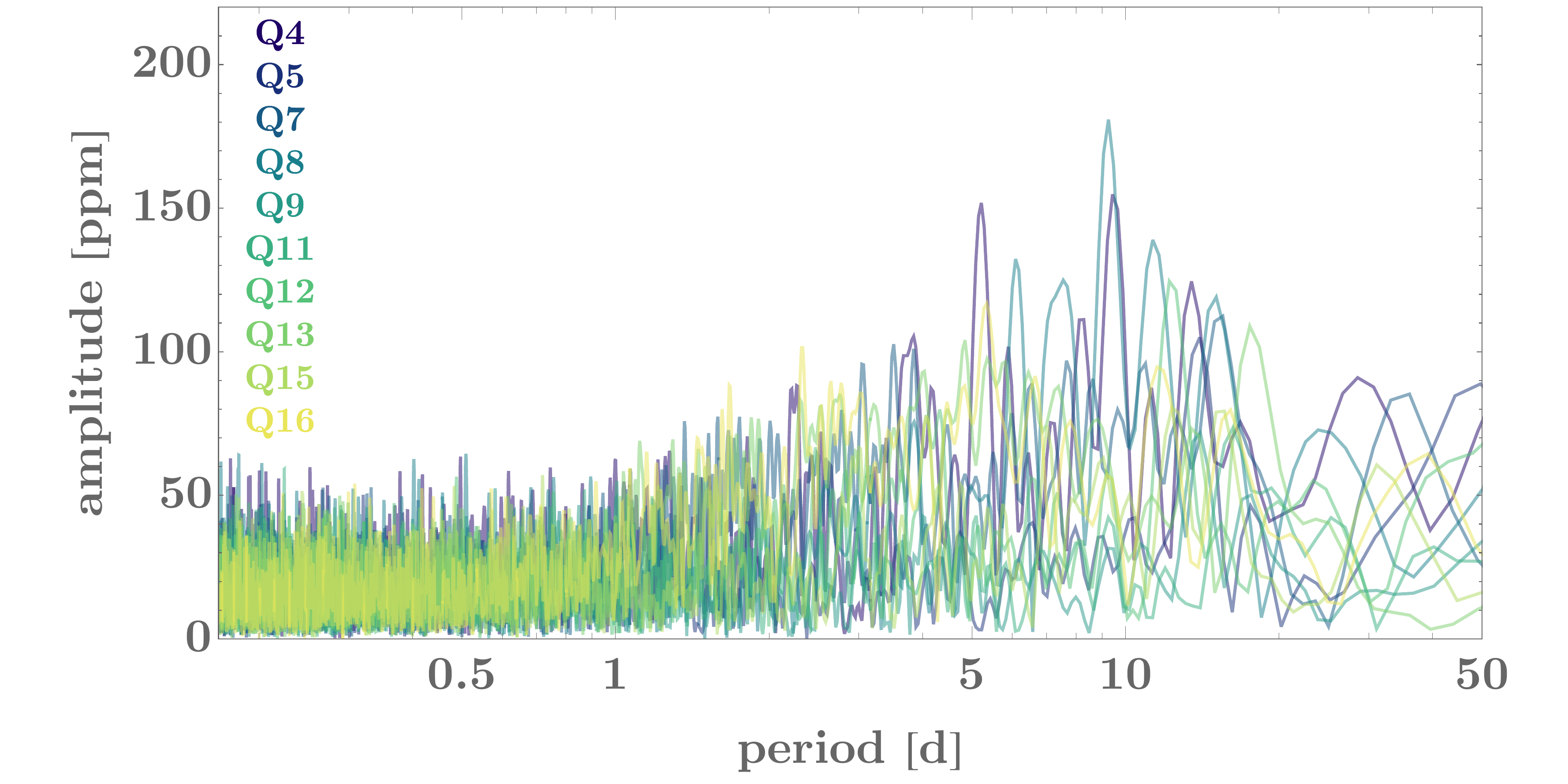}
  \caption{\label{fig:KIC868_activity}
  \textbf{Lomb-Scargle periodogram of each available \kepler\ quarter for
  KIC-8681125.}
  Colours delineate each quarter, as denoted by the legend. The amplitude
  appears bound to be less than 200\,ppm for all quarters and thus
  relatively quiet.
  }
\end{sfigure}

\newpage
\begin{sfigure}
  \centering
  \includegraphics[angle=0, width=16.0cm]{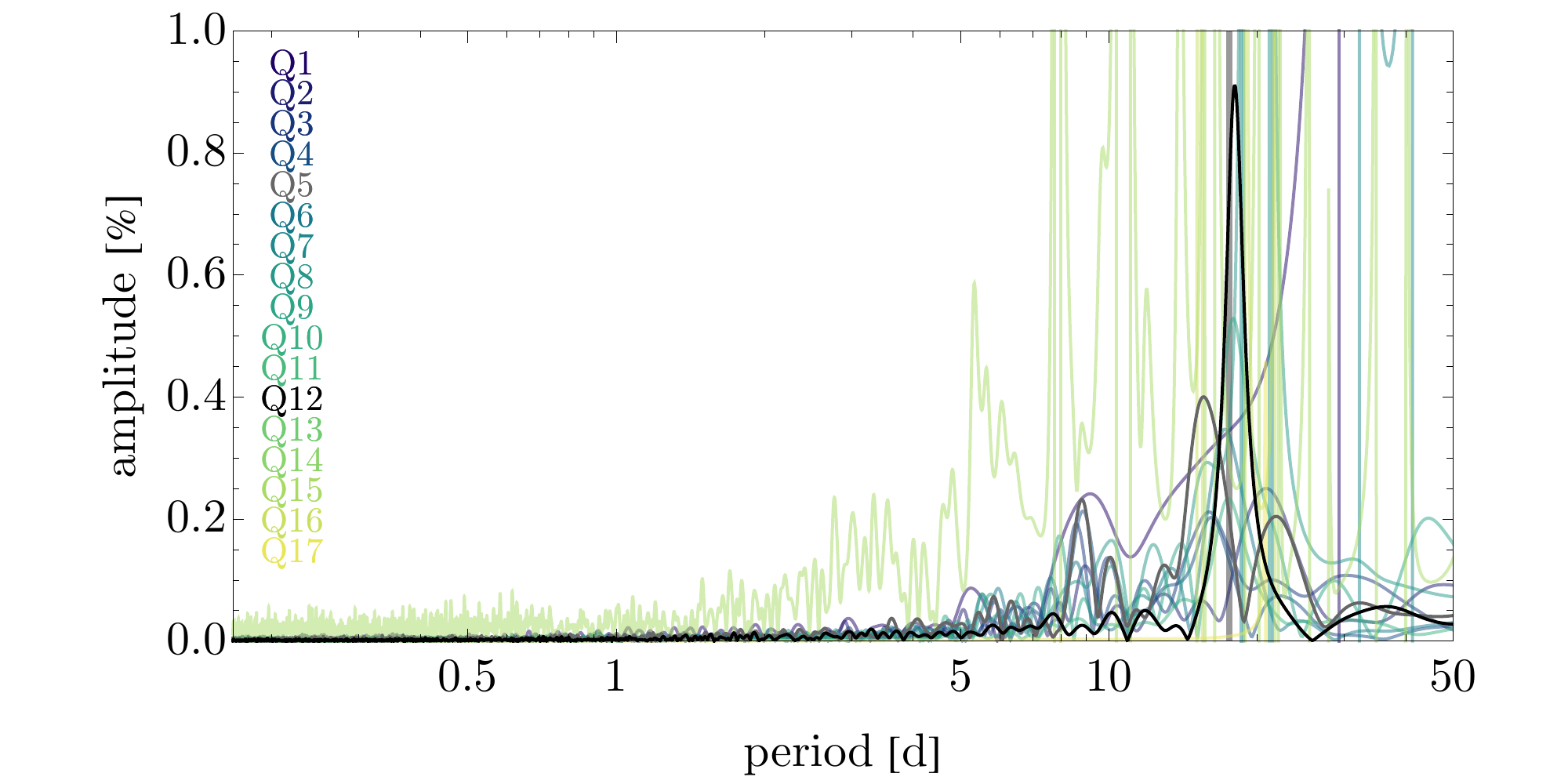}
  \caption{\label{fig:KIC535_activity}
  \textbf{Lomb-Scargle periodogram of each available \kepler\ quarter for
  KIC-5351250/Kepler-150.}
  Colours delineate each quarter, as denoted by the legend. The amplitude
  reaches up to 1\%, and is particularly active in Q12 (highlighted in black) -
  corresponding to second epoch of Kepler-150f indicating that spots are more
  likely to observed then.
  }
\end{sfigure}

\newpage
\begin{sfigure}
  \centering
  \includegraphics[angle=0, width=16.0cm]{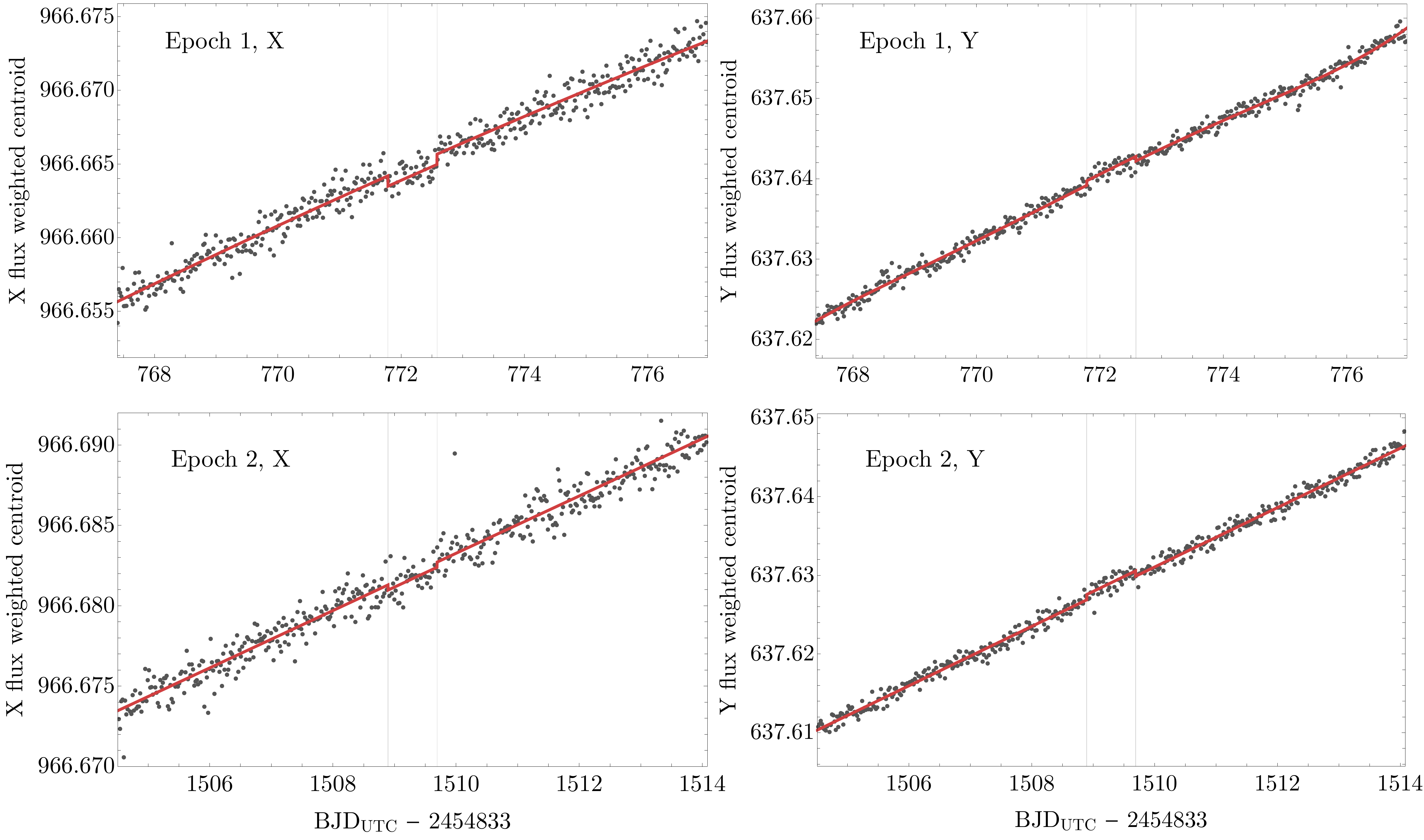}
  \caption{\label{fig:centroidshift}
  \textbf{Flux weighted centroid time series of \kicnameb.}
  As visible from the plots, which are labeled in the top-left corner of
  each panel, the centroids exhibit a small shift during the time of the
  transits of \kicnameb.
  }
\end{sfigure}

\newpage
\begin{sfigure}
  \centering
  \includegraphics[width=0.48\linewidth]{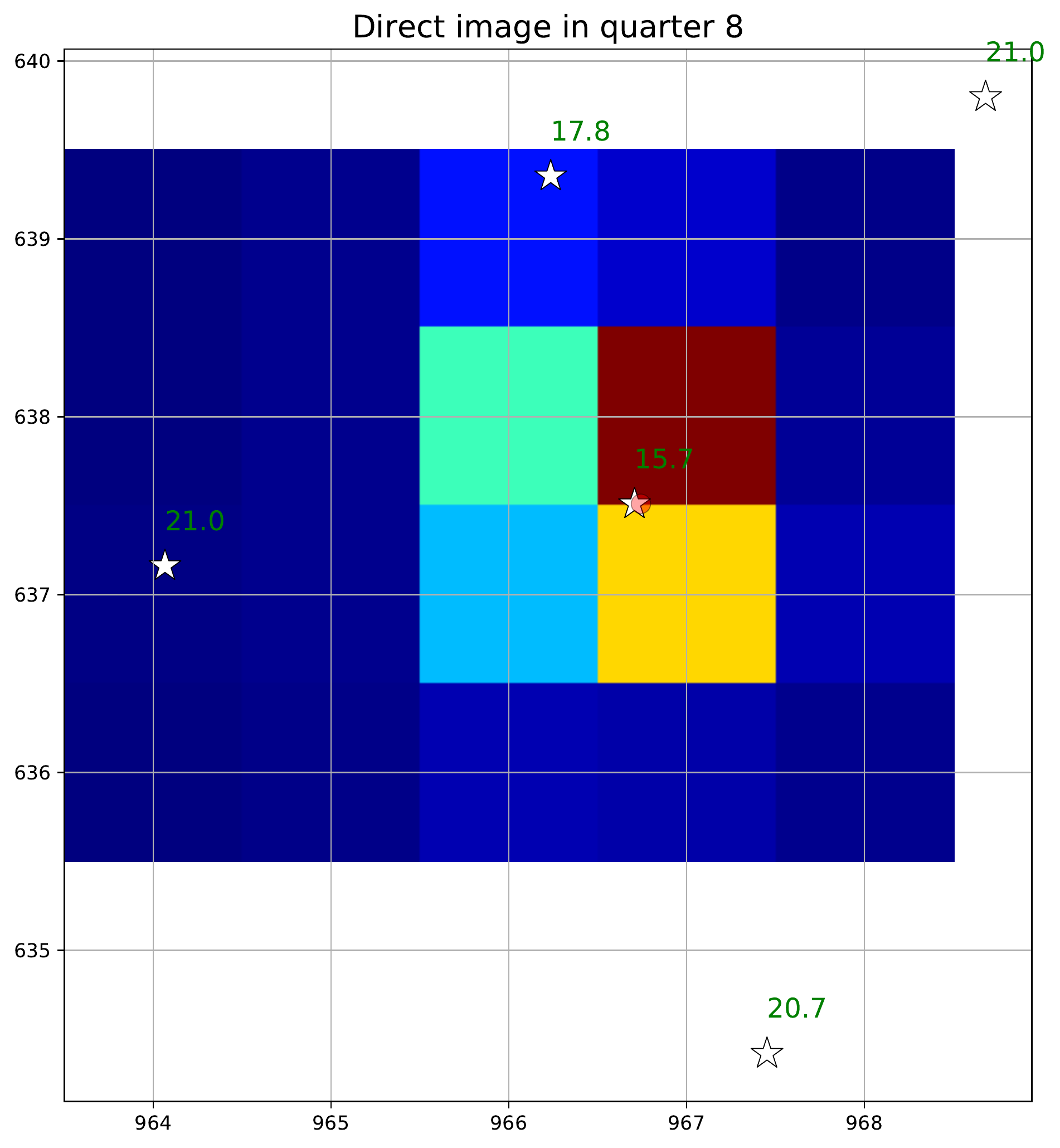} 
  \includegraphics[width=0.48\linewidth]{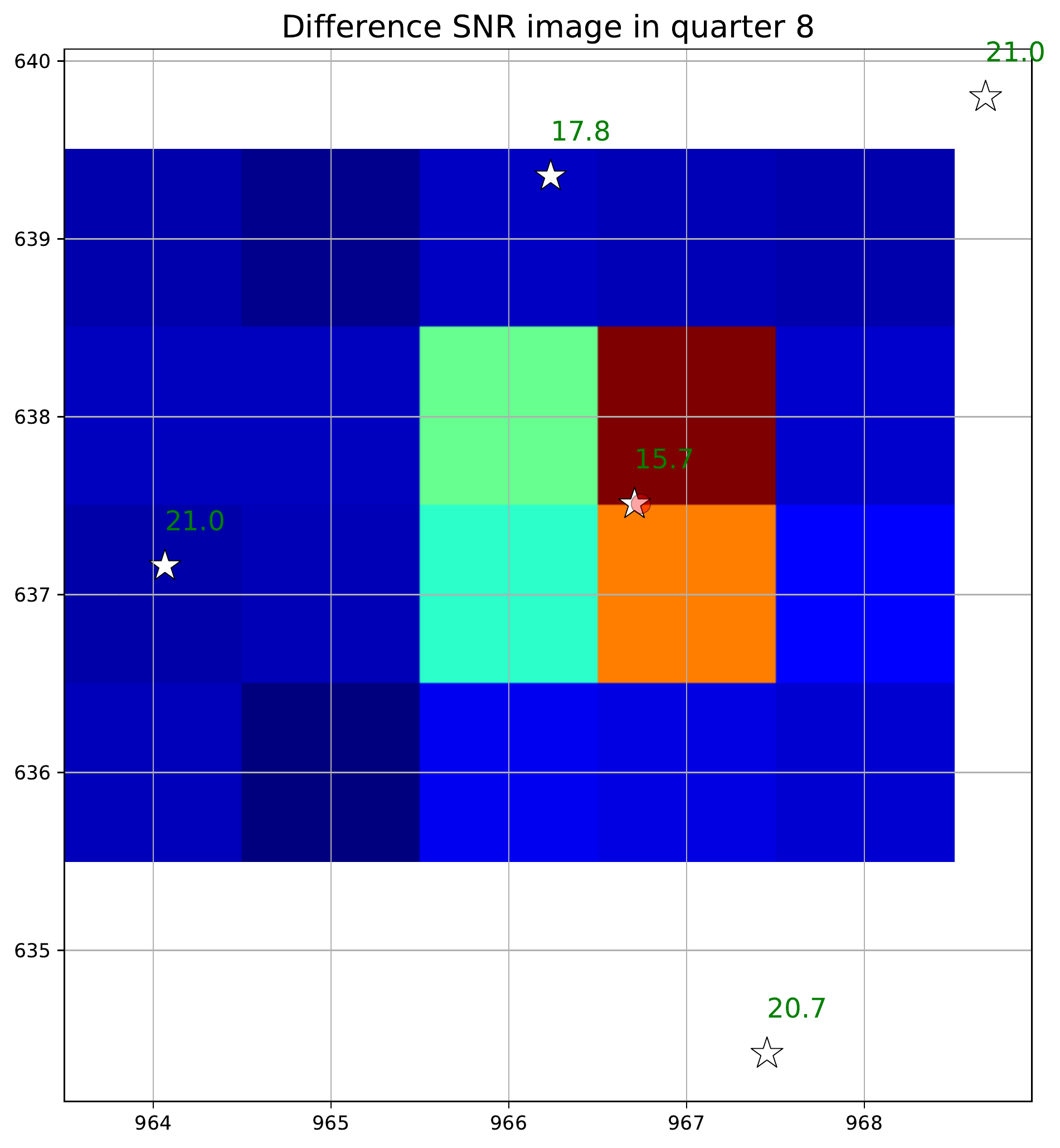} 
\caption{
\textbf{Observed \kepler\ image of \kicname\ during quarter 8.}
Left: Observed average out-of-transit image. Right: Observed difference image
normalised by pixel-by-pixel uncertainty. The star symbols are the
proper-motion-corrected \gaia\ star positions, and the semi-transparent red circle is the non-proper-motion-corrected \gaia\ position of the target star.
}
\label{bryson:q8Observed}
\end{sfigure}

\begin{sfigure}
  \centering
  \includegraphics[width=0.48\linewidth]{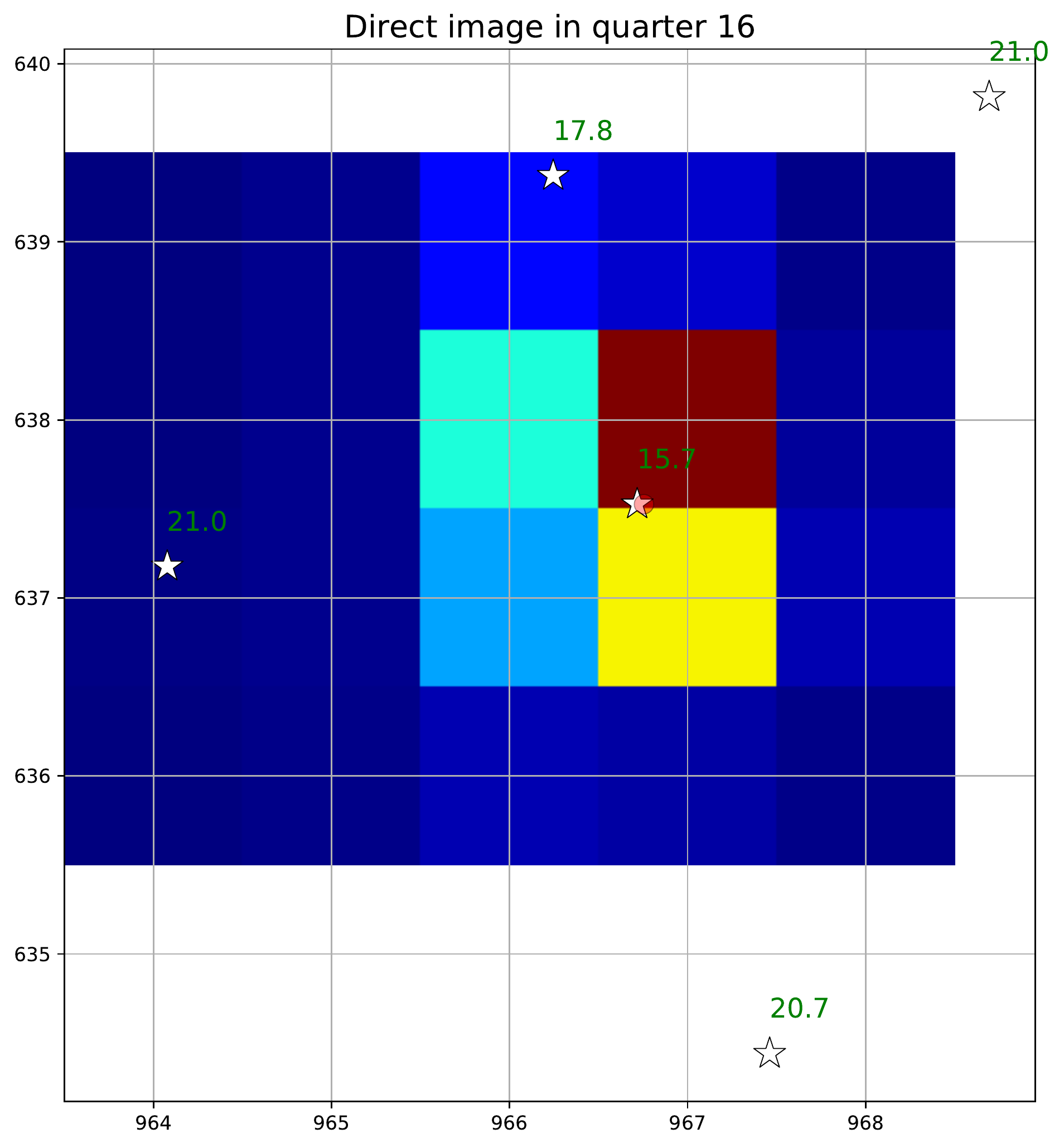} 
  \includegraphics[width=0.48\linewidth]{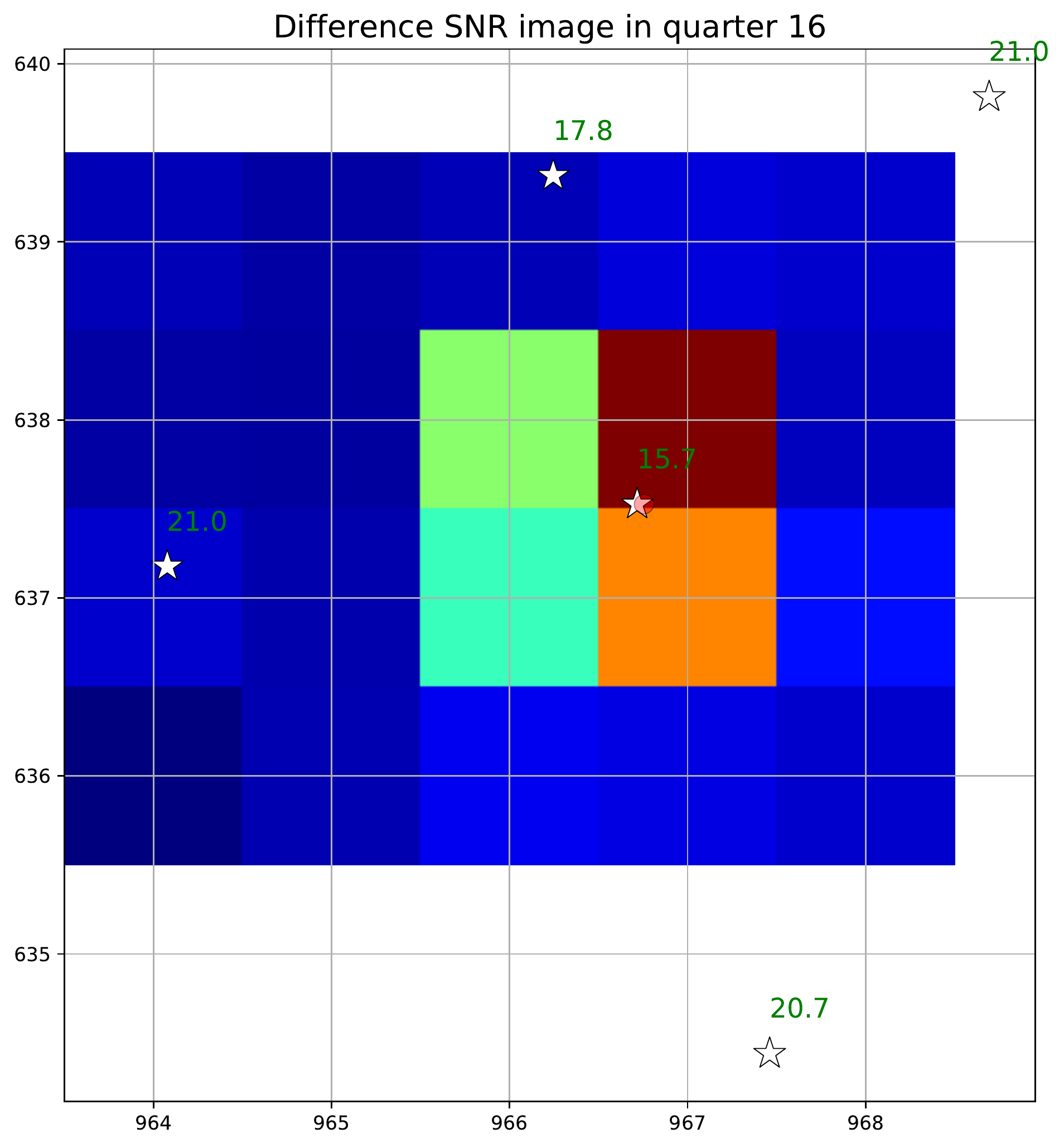} 
\caption{
\textbf{Observed \kepler\ image of \kicname\ during quarter 16.}
Left: Observed average out-of-transit image. Right: Observed difference image.
The star symbols are the proper-motion-corrected \gaia\ star positions, and the
red circle is the non-proper-motion-corrected \gaia\ position of the target
star.
}
\label{bryson:q16Observed}
\end{sfigure}

\newpage
\begin{sfigure}
  \centering
  \includegraphics[width=0.48\linewidth]{bryson_figures/observed_OOT_image_q8.pdf} 
  \includegraphics[width=0.48\linewidth]{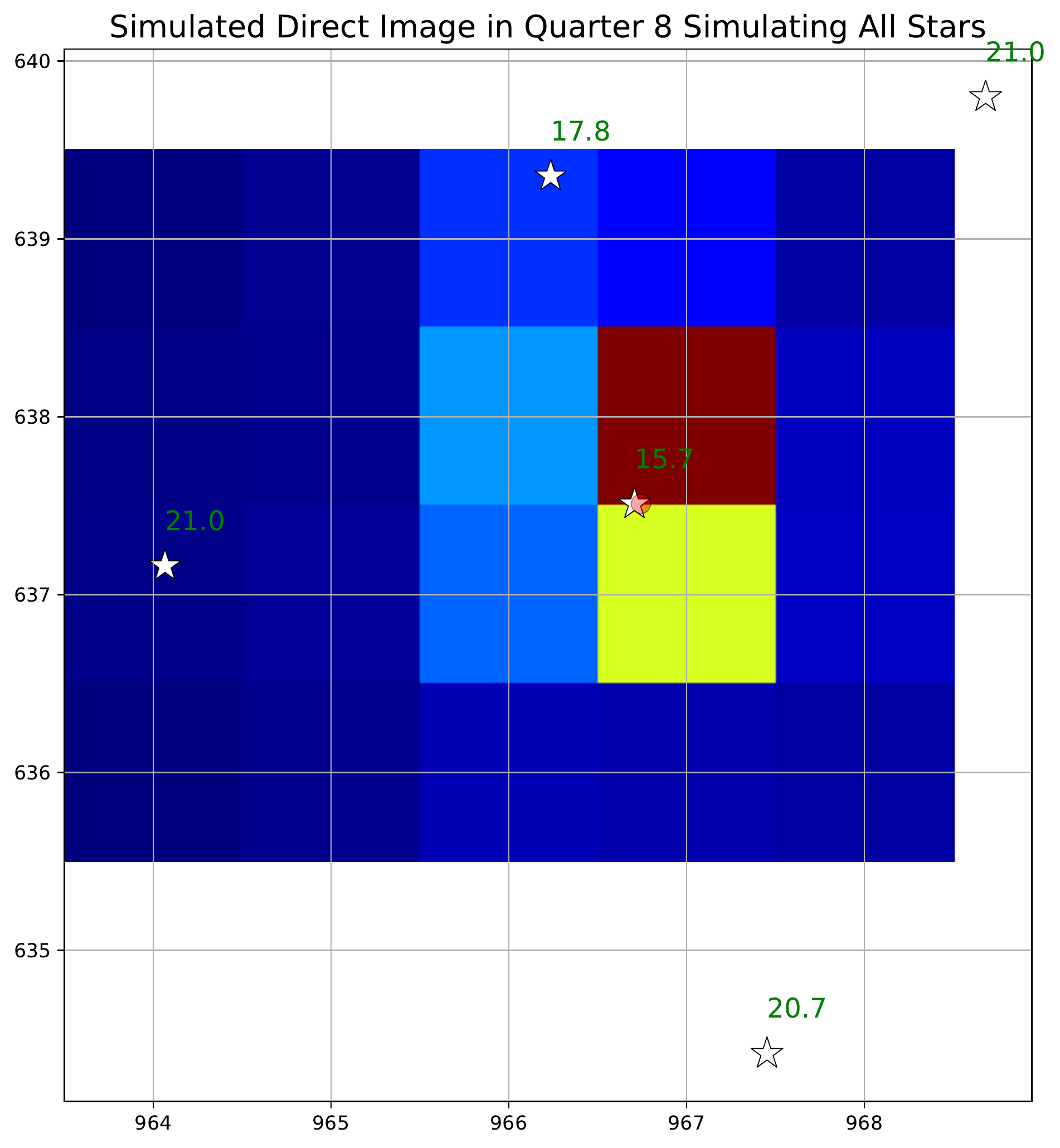} 
\caption{
\textbf{Comparison of the observed and simulated \kepler\ images of \kicname\
during quarter 8.}
Left: Observed average out-of-transit image.
Right: Simulated average out-of-transit image.
}
\label{bryson:q8Compare}
\end{sfigure}

\newpage
\begin{sfigure}
  \centering
  \includegraphics[width=0.48\linewidth]{bryson_figures/observed_OOT_image_q16.pdf} 
  \includegraphics[width=0.48\linewidth]{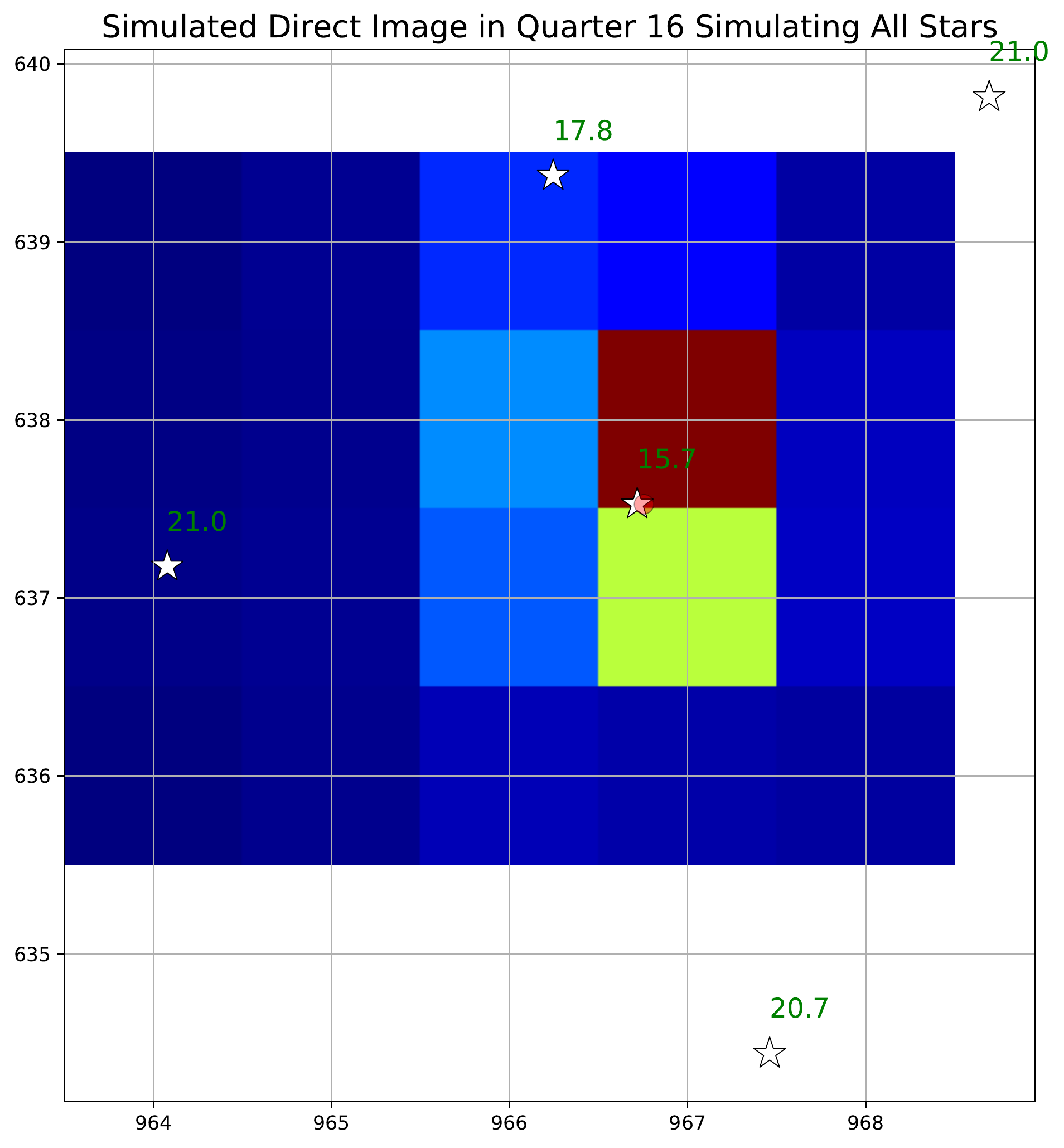} 
\caption{
\textbf{Comparison of the observed and simulated \kepler\ images of \kicname\
during quarter 16.}
Left: Observed average out-of-transit image.
Right: Simulated average out-of-transit image.
}
\label{bryson:q16Compare}
\end{sfigure}

\newpage
\begin{sfigure}
  \centering
  \includegraphics[width=0.33\linewidth]{bryson_figures/observed_diff_SNR_image_q8.pdf} 
  \includegraphics[width=0.33\linewidth]{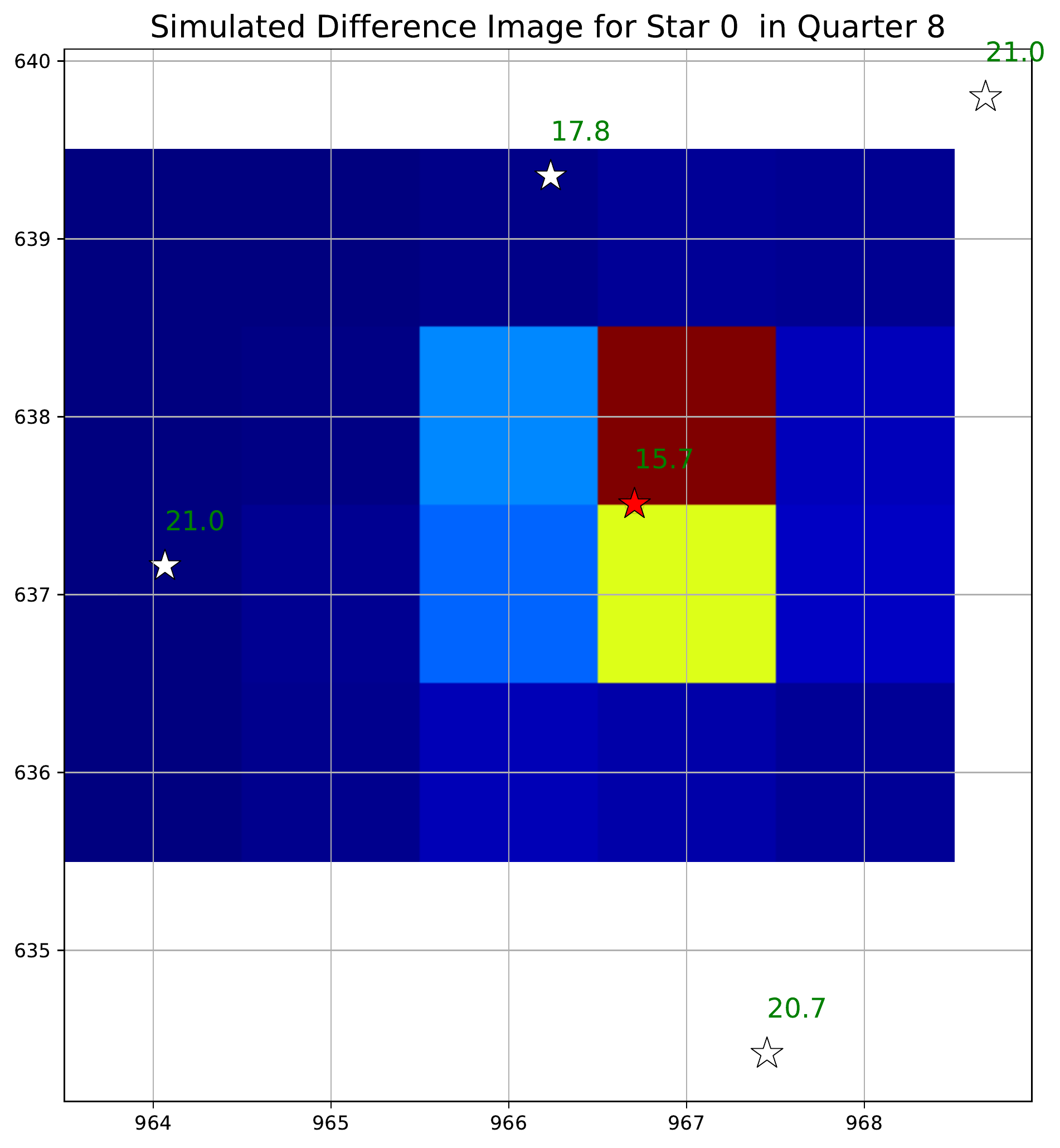} \\
  \includegraphics[width=0.33\linewidth]{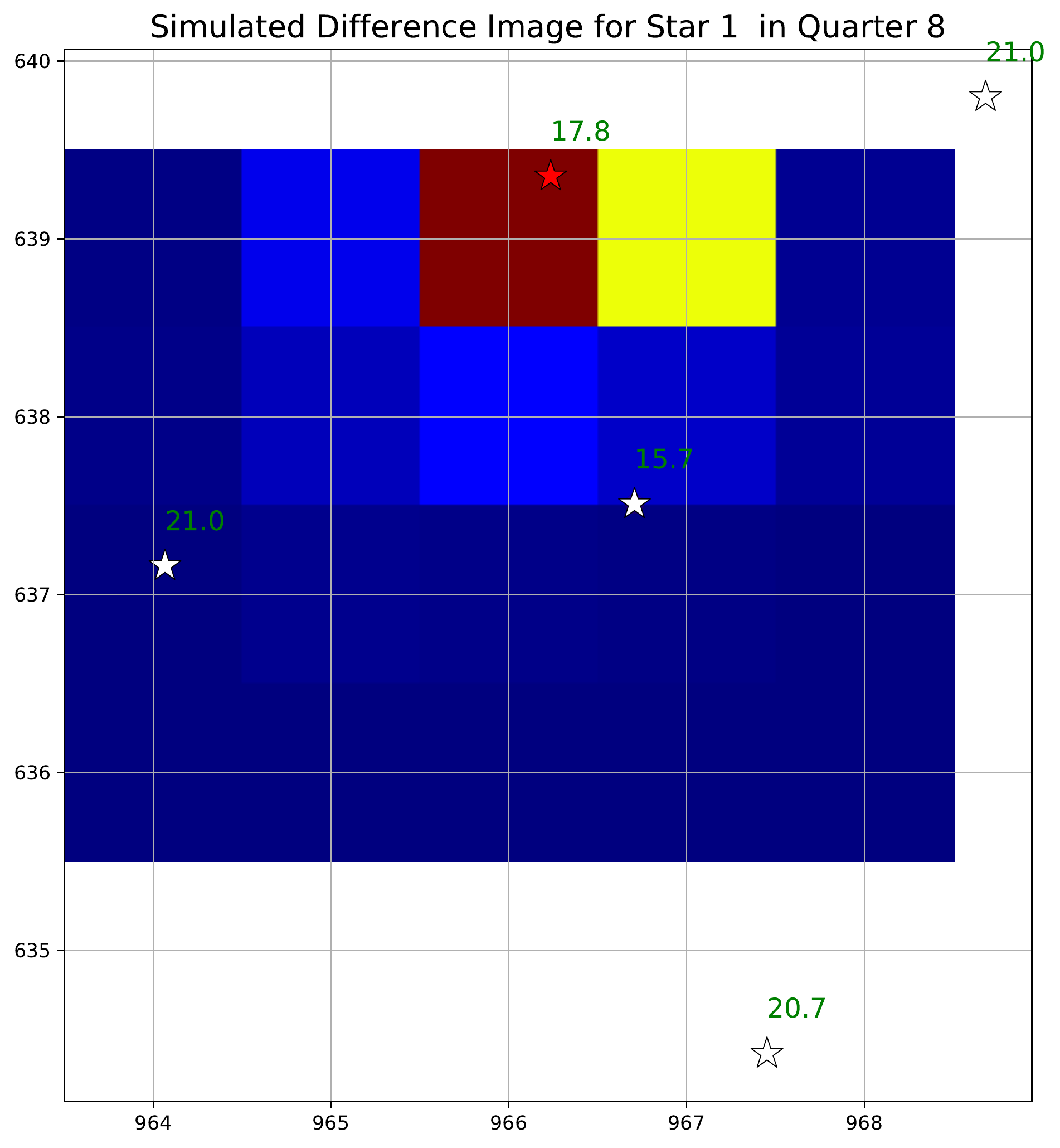} 
  \includegraphics[width=0.33\linewidth]{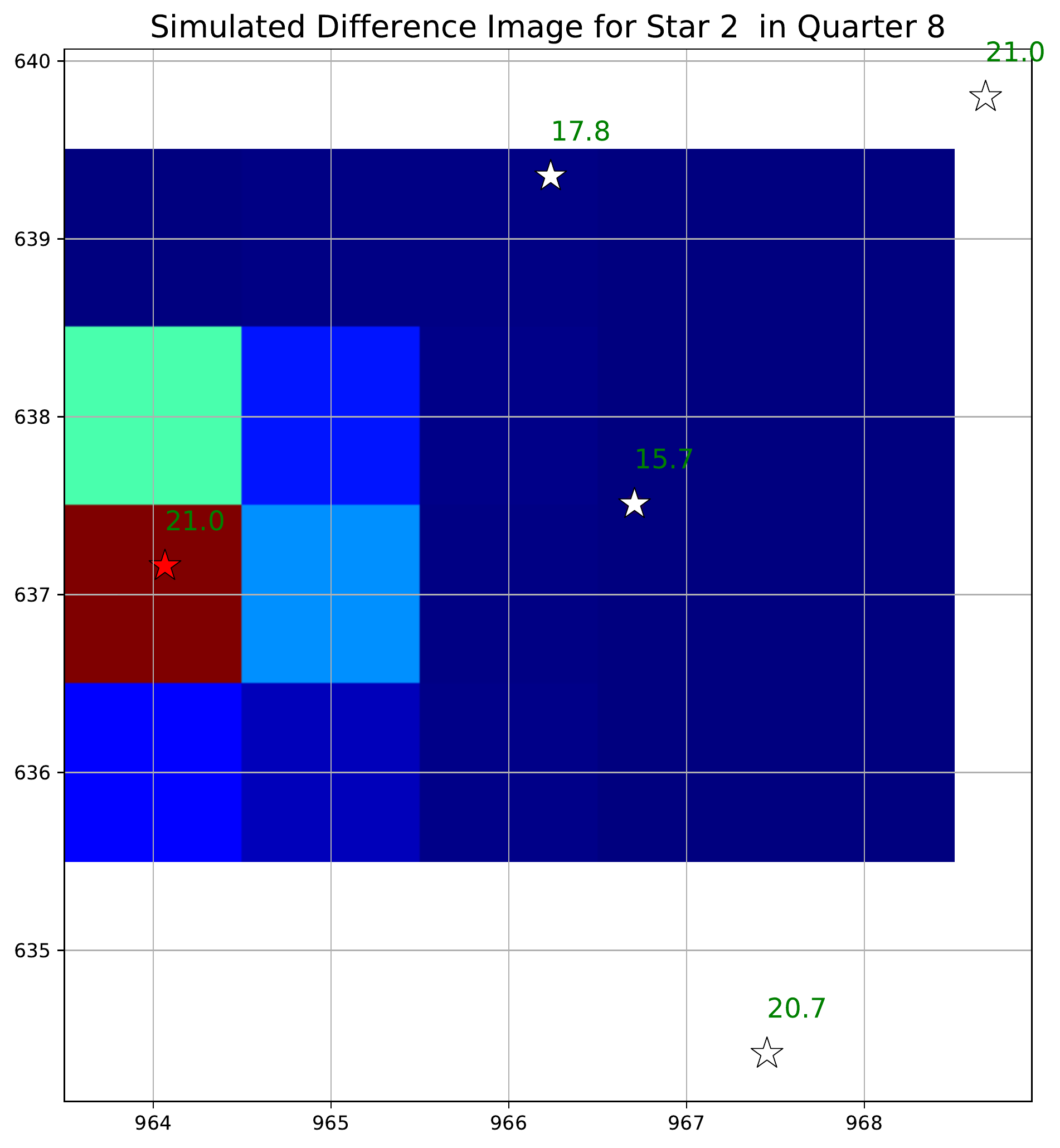} \\
  \includegraphics[width=0.33\linewidth]{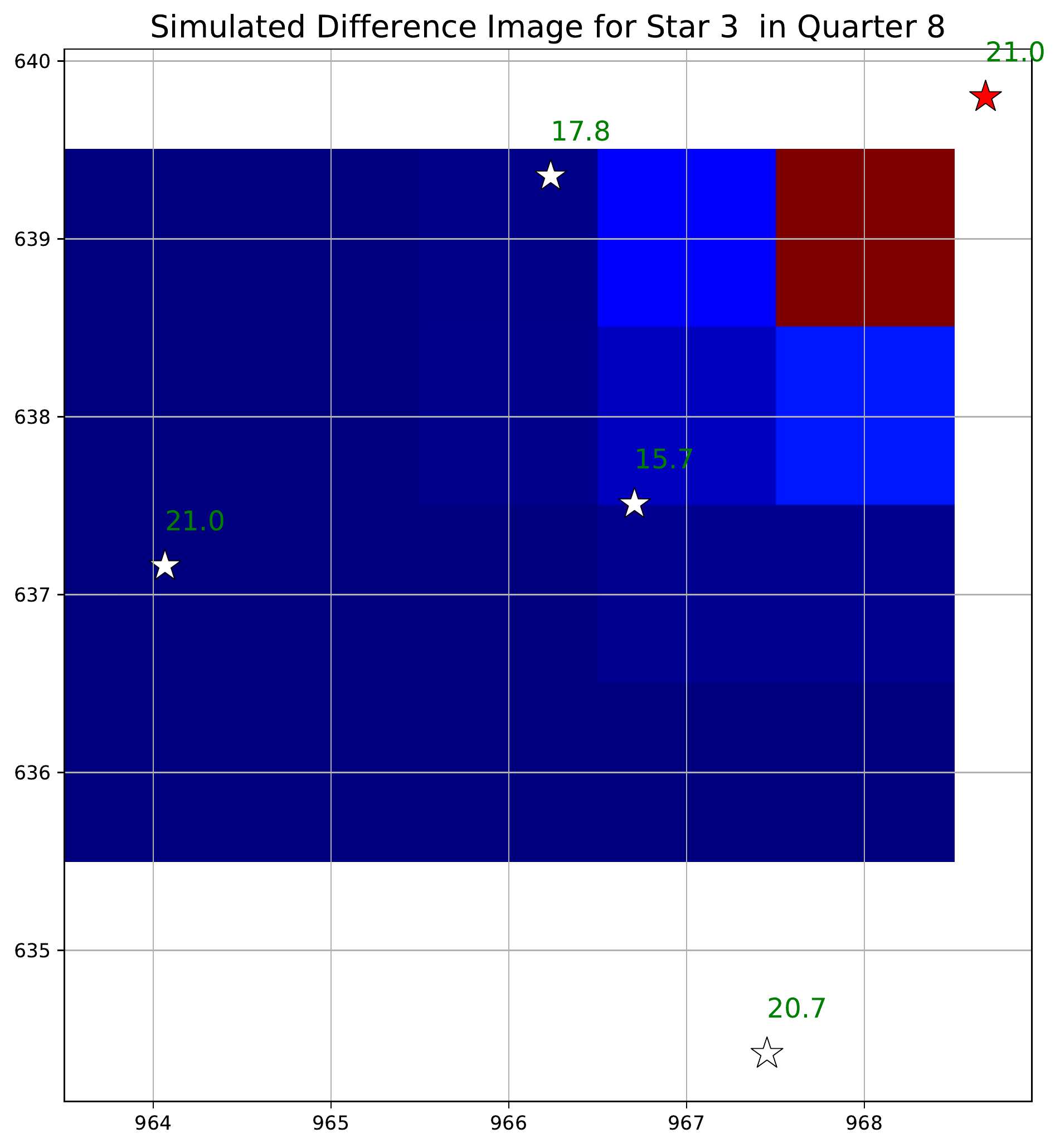} 
  \includegraphics[width=0.33\linewidth]{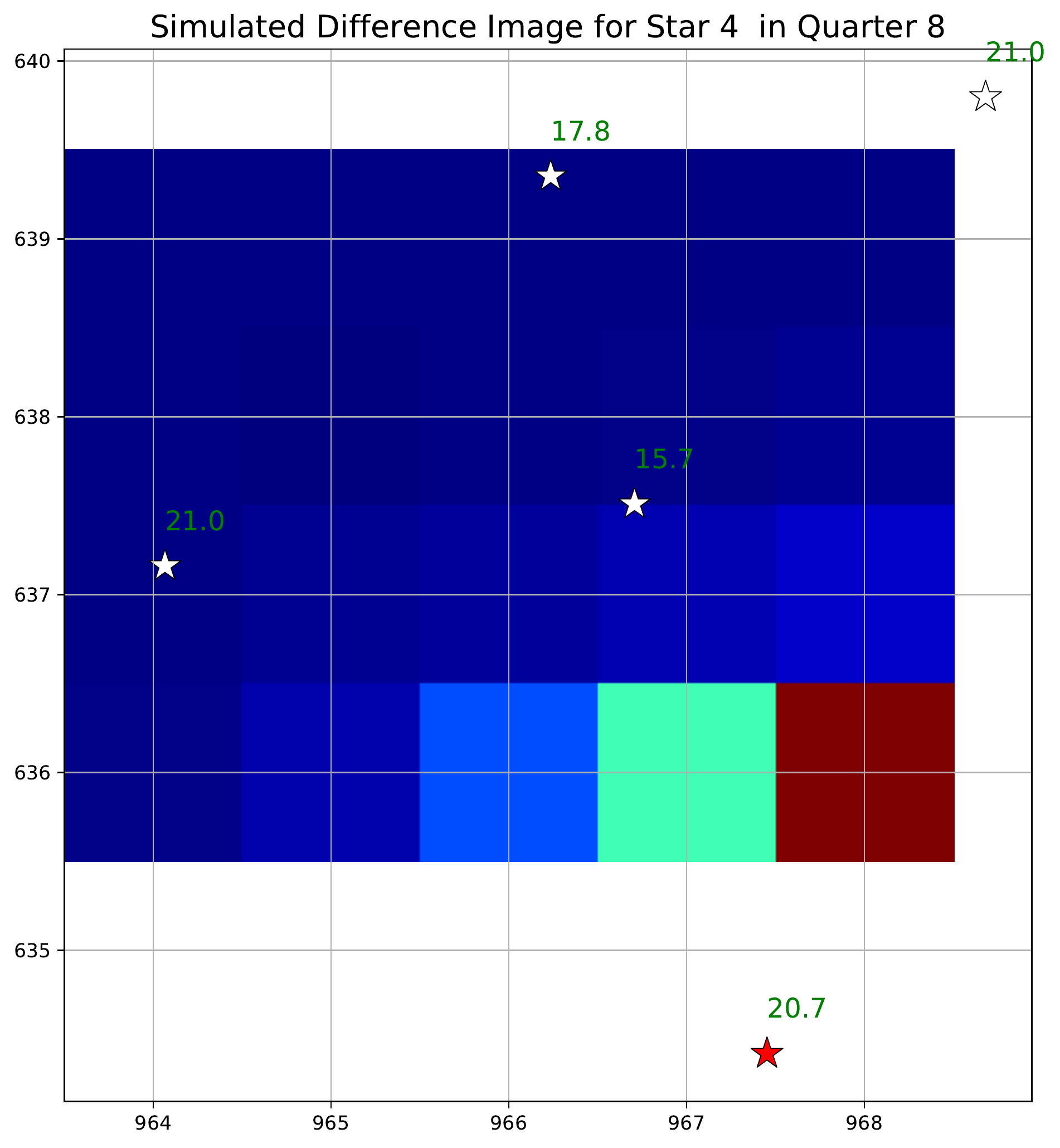} 
\caption{
\textbf{Simulated difference images of the \kicname\ postage stamp during
quarter 8.}
Simulated difference images for the various stars compared with the observed
difference image. The simulated star is shown with the red star symbol.
}
\label{bryson:q8Simulated}
\end{sfigure}

\newpage
\begin{sfigure}
  \centering
  \includegraphics[width=0.33\linewidth]{bryson_figures/observed_diff_SNR_image_q16.pdf} 
  \includegraphics[width=0.33\linewidth]{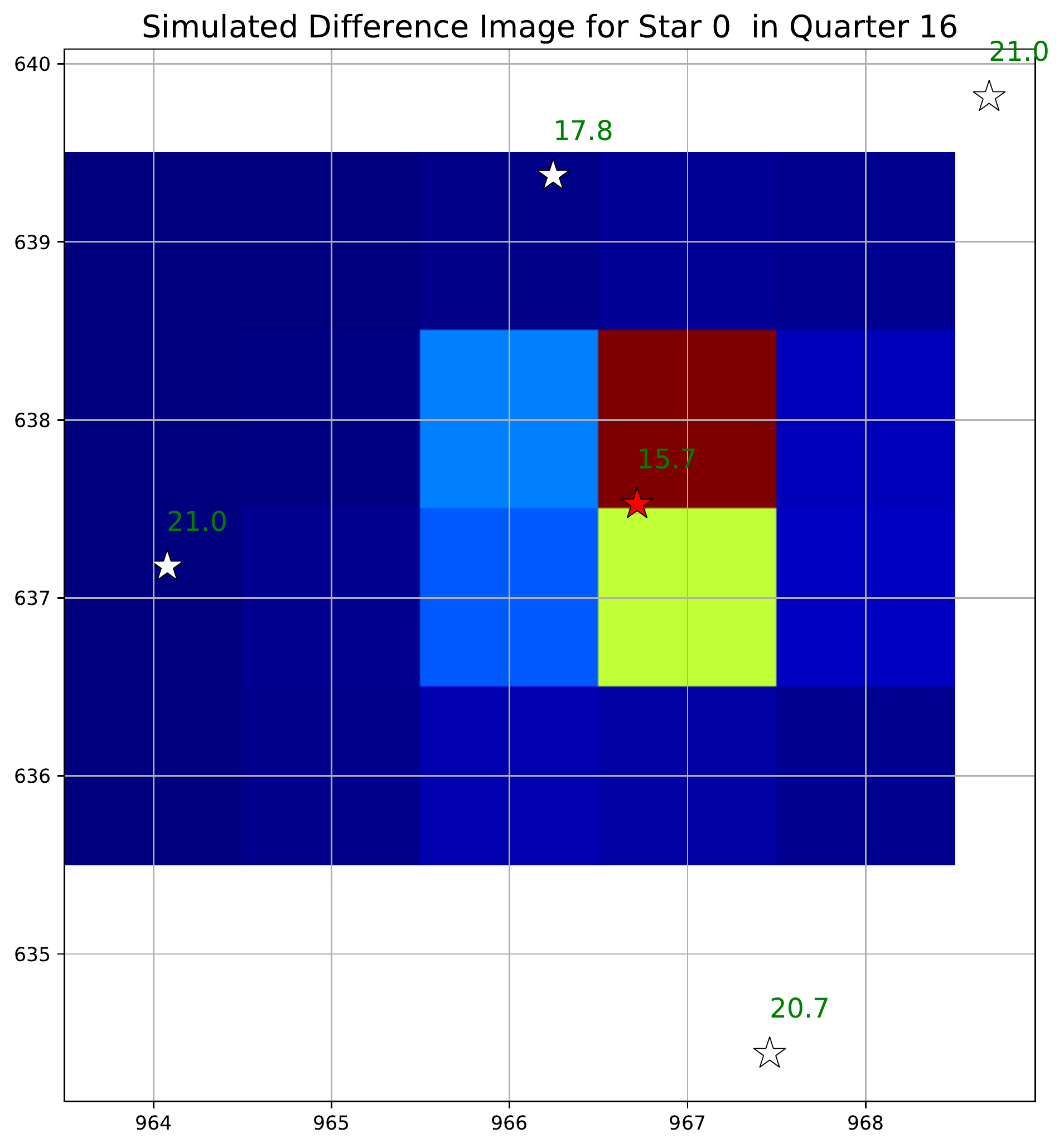} \\
  \includegraphics[width=0.33\linewidth]{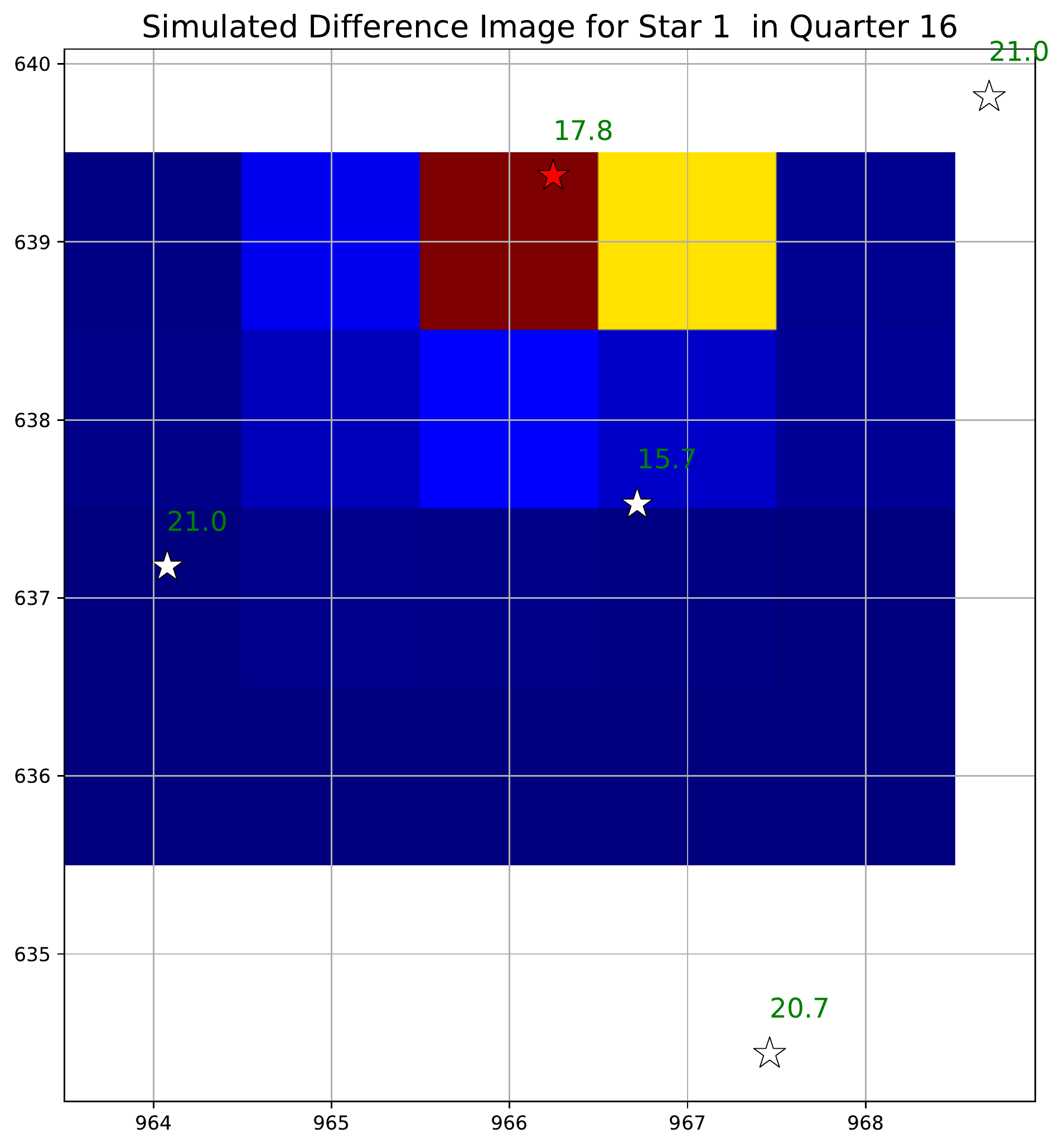} 
  \includegraphics[width=0.33\linewidth]{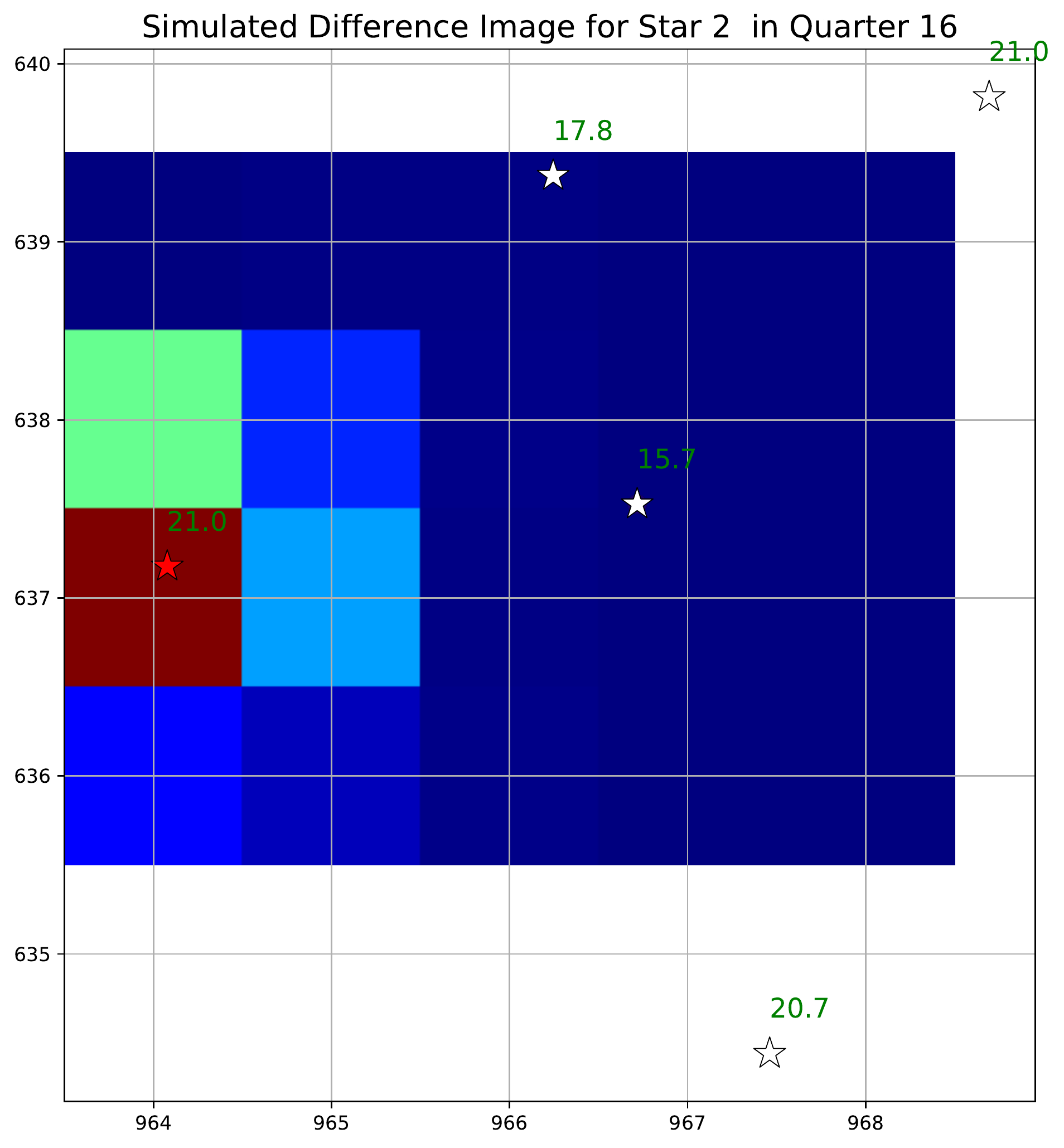} \\
  \includegraphics[width=0.33\linewidth]{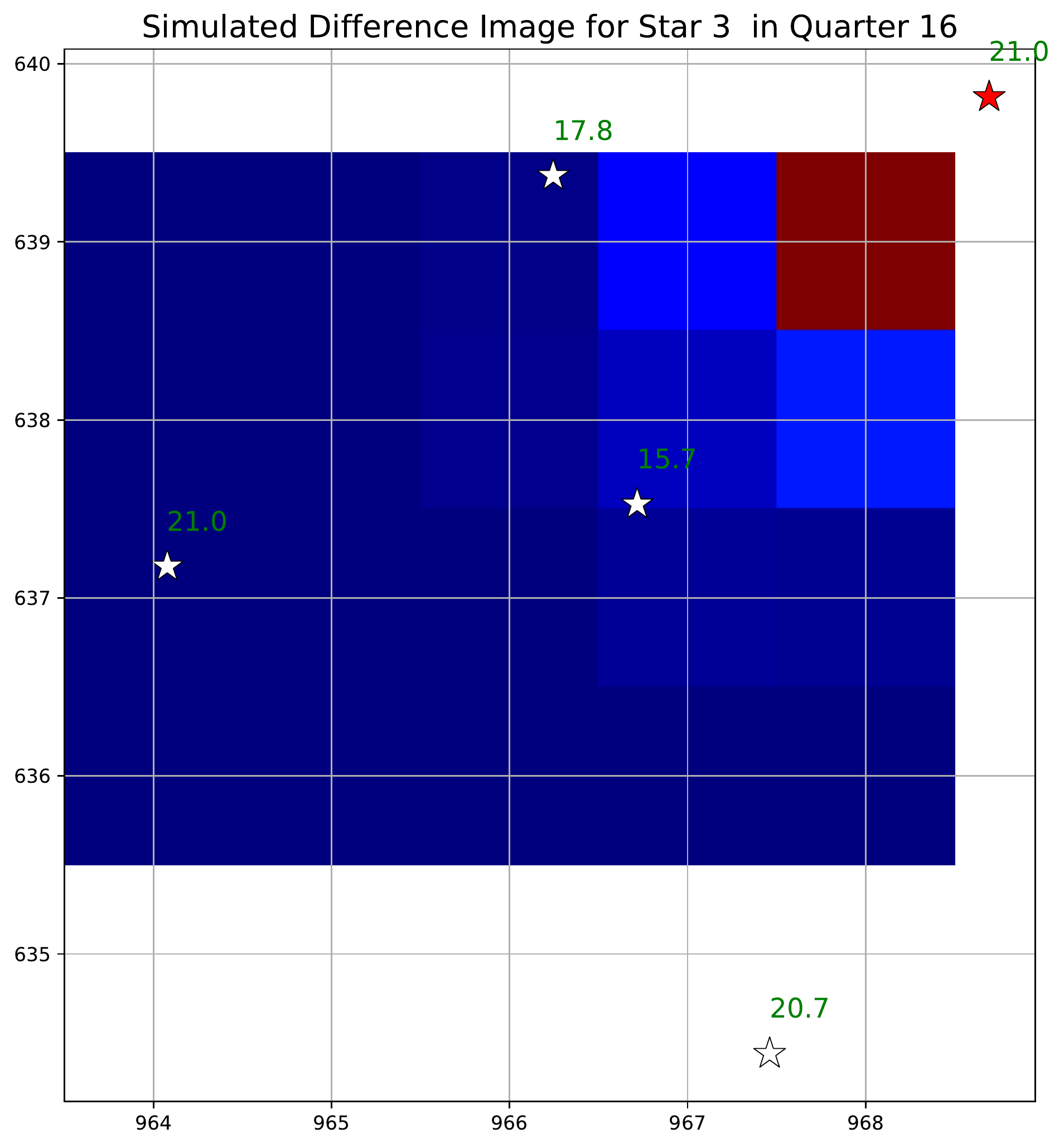} 
  \includegraphics[width=0.33\linewidth]{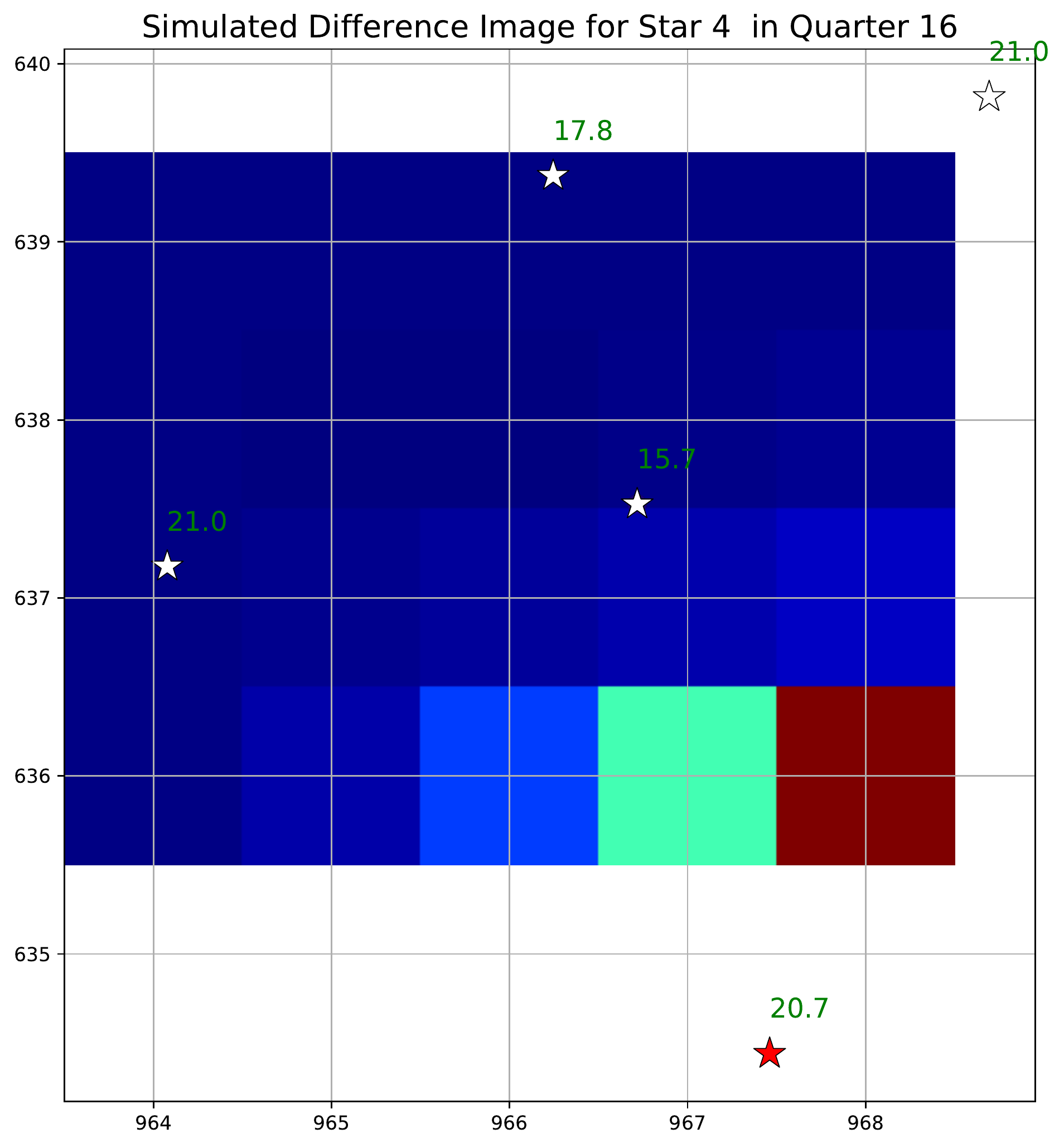} 
\caption{
\textbf{Simulated difference images of the \kicname\ postage stamp during
quarter 16.}
Simulated difference images for the various stars compared with the observed
difference image. The simulated star is shown with the red star symbol.
}
\label{bryson:q16Simulated}
\end{sfigure}

\newpage
\setcounter{stable}{2}
\begin{stable}[ht]
\centering
\begin{tabular}{ r c c }
 & Quarter 8 & Quarter 16
\\
\hline
Recovered transit depth & $7.78 \times 10^{-3}$ & $7.68 \times 10^{-3}$ \\
Distance from target star (pixels) & $0.0146 \pm 0.0154$ & $0.0192 \pm 0.0153$ \\
Distance from target star (arcsec) & $0.058 \pm 0.061$ & $0.076 \pm 0.061$ \\
$3\sigma$ circle area (square arcsec) & 0.107 & 0.104 \\
Blend probability & $2.62 \times 10^{-6}$ & $2.56 \times 10^{-6}$ \\
\end{tabular}
\caption{
\textbf{Summary of the results from our MCMC PRF blend analysis.}
}
\label{bryson:blend}
\end{stable}

\newpage
\begin{sfigure}
  \centering
  \includegraphics[angle=0, width=16.0cm]{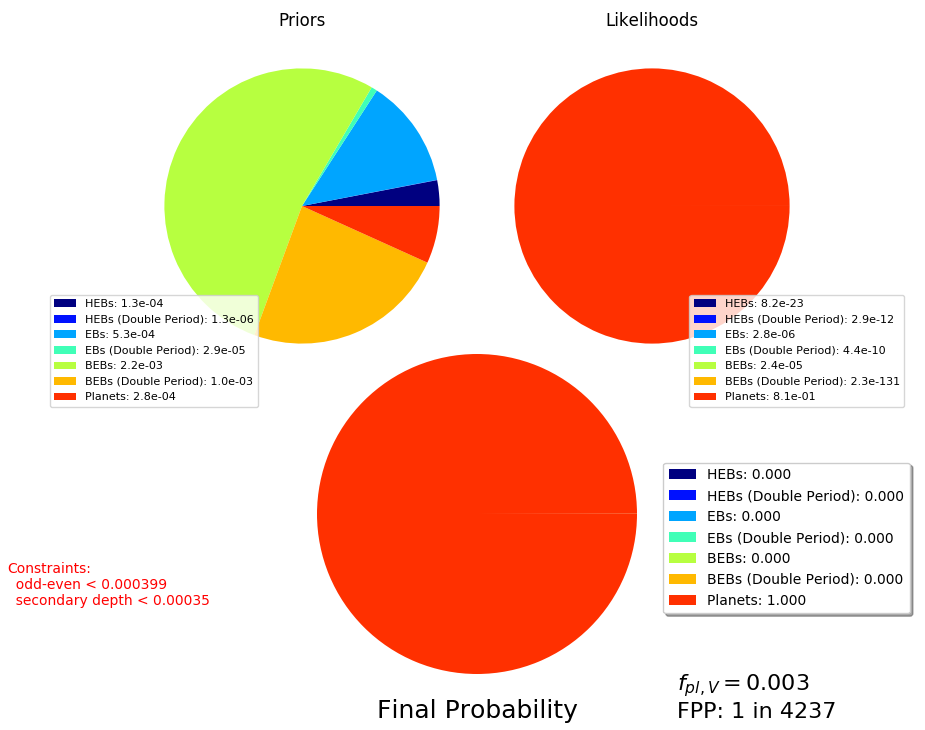}
  \caption{\label{fig:vespa}
  \textbf{Auto-generated output summary figure from \vespa\ for \kicnameb.}
    Pie-charts showing the relative odds of various astrophysical models for the
	observed transits of \kicname. Top-left shows the adopted priors, based
	on the stellar properties and position. Top-left shows the likelihood
	governed by the transit light curve morphology. Bottom shows the posteriors
	odds, from which we conclude high confidence the planetary nature of \kicnameb\ (henceforth \kepxb).
  }
\end{sfigure}

\newpage
\begin{sfigure}
  \centering
  \includegraphics[angle=0, width=16.0cm]{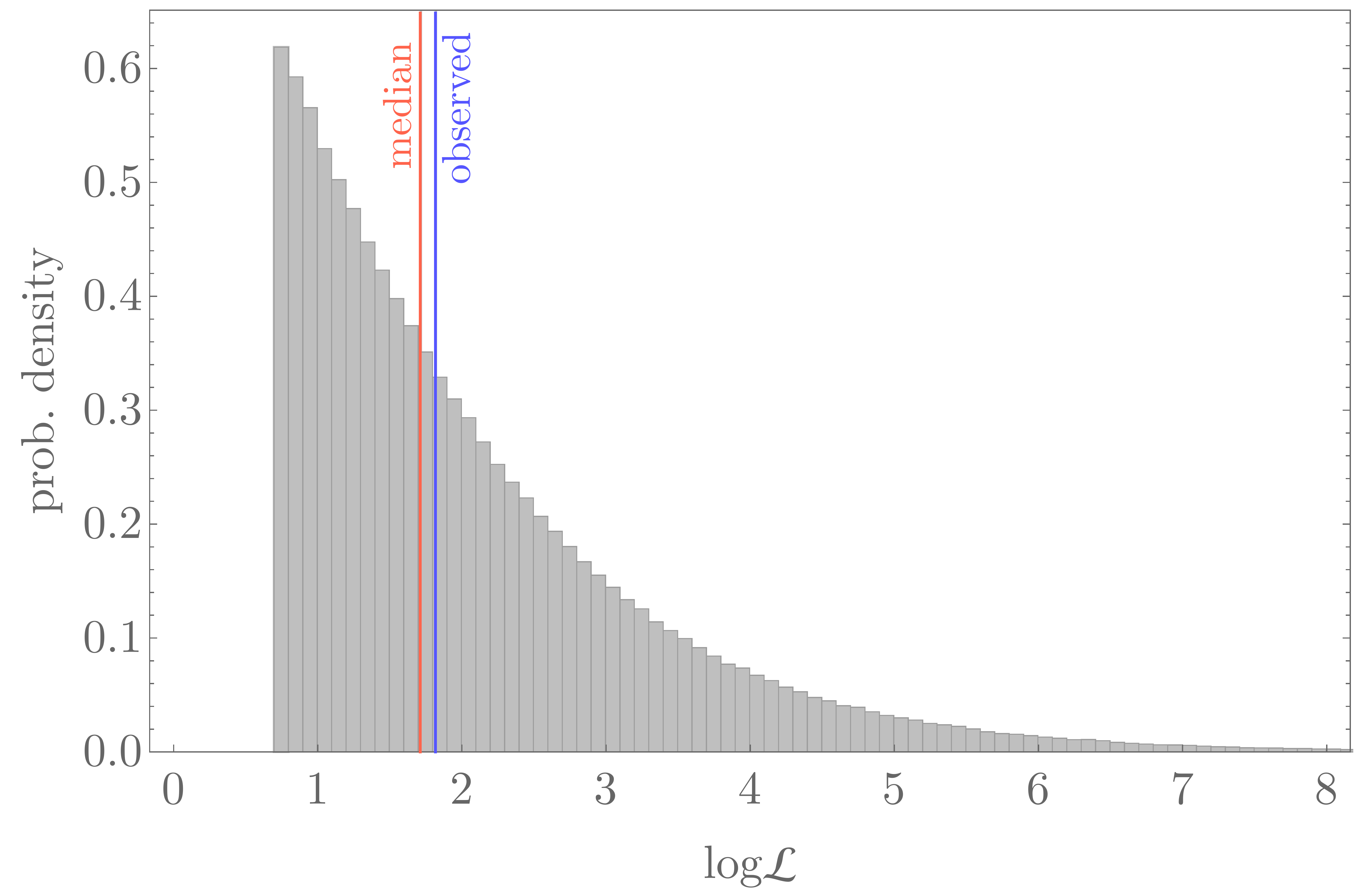}
  \caption{\label{fig:pvaluetiming}
  \textbf{Log-likelihood distribution of two randomly sampled exomoon transit
  times.}
    Through Monte Carlo simulation and geometric arguments\cite{ose2014},
	the distribution of exomoon transit times is expected to follow an arcsin
	distribution. Here, we evaluate the log-likelihood of observing the two
	transit times of \kepxb-i under this assumption (blue line), which is
	compared the distribution expected for random samplings. This reveals that
	the observed times are fully consistent with the expected behaviour.
  }
\end{sfigure}

\newpage
\begin{sfigure}
  \centering
  \includegraphics[angle=0, width=16.0cm]{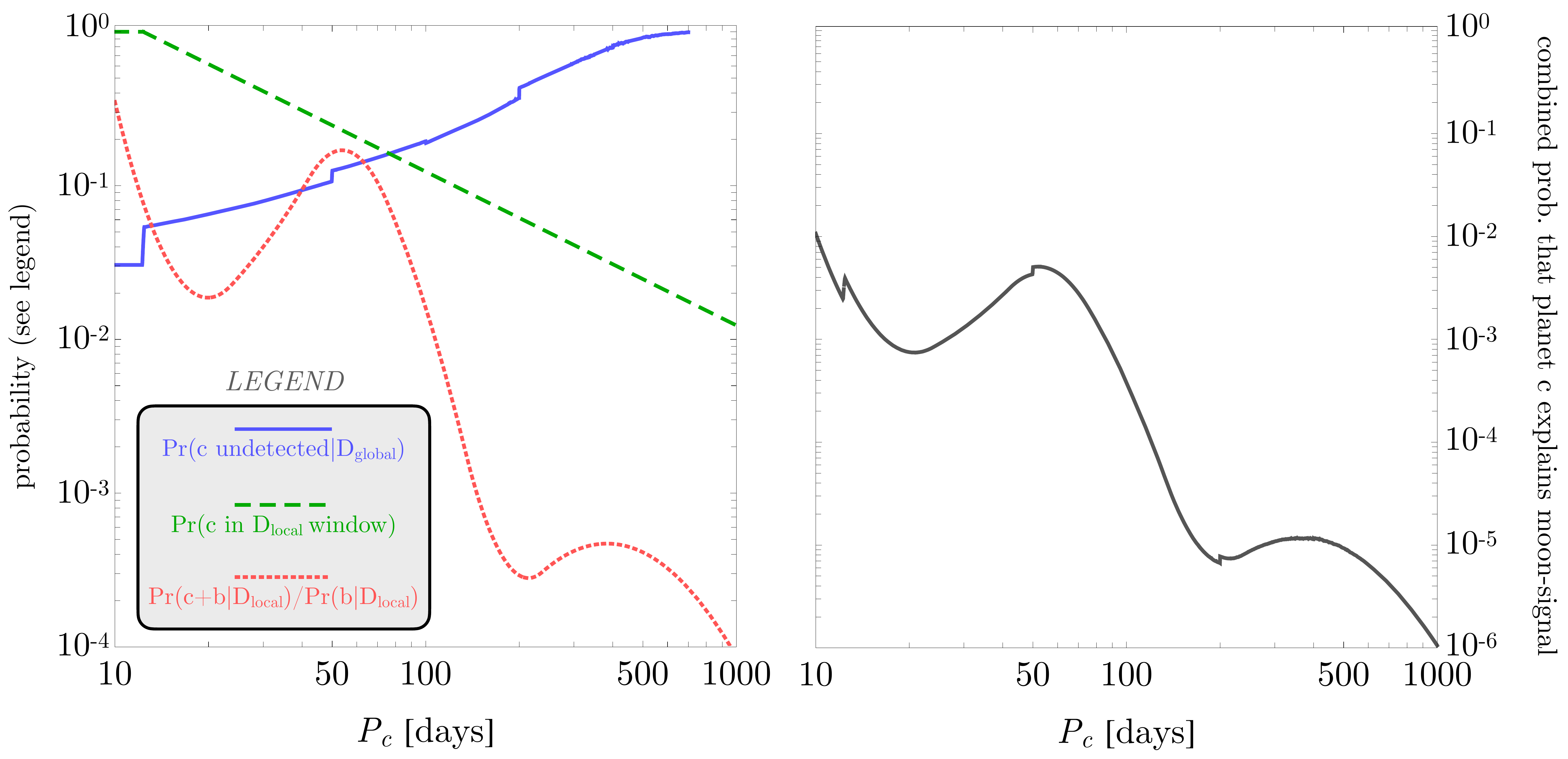}
  \caption{\label{fig:planetc_prob}
  \textbf{Probability of a second transiting planet explaining the moon-like
  deviations observed for \kepxb.}
    Left: The probability of the three independent necessary criteria:
    1) planet c evades detection by \kepler\ (blue solid).
	2) planet c transited during the transit window of \kepxb\ (green dashed).
	3) planet b+c model is statistically favoured over planet b alone (red dotted).
	Right: Combined probability using all three effects.
  }
\end{sfigure}

\newpage
\begin{sfigure}
  \centering
  \includegraphics[angle=0, width=16.0cm]{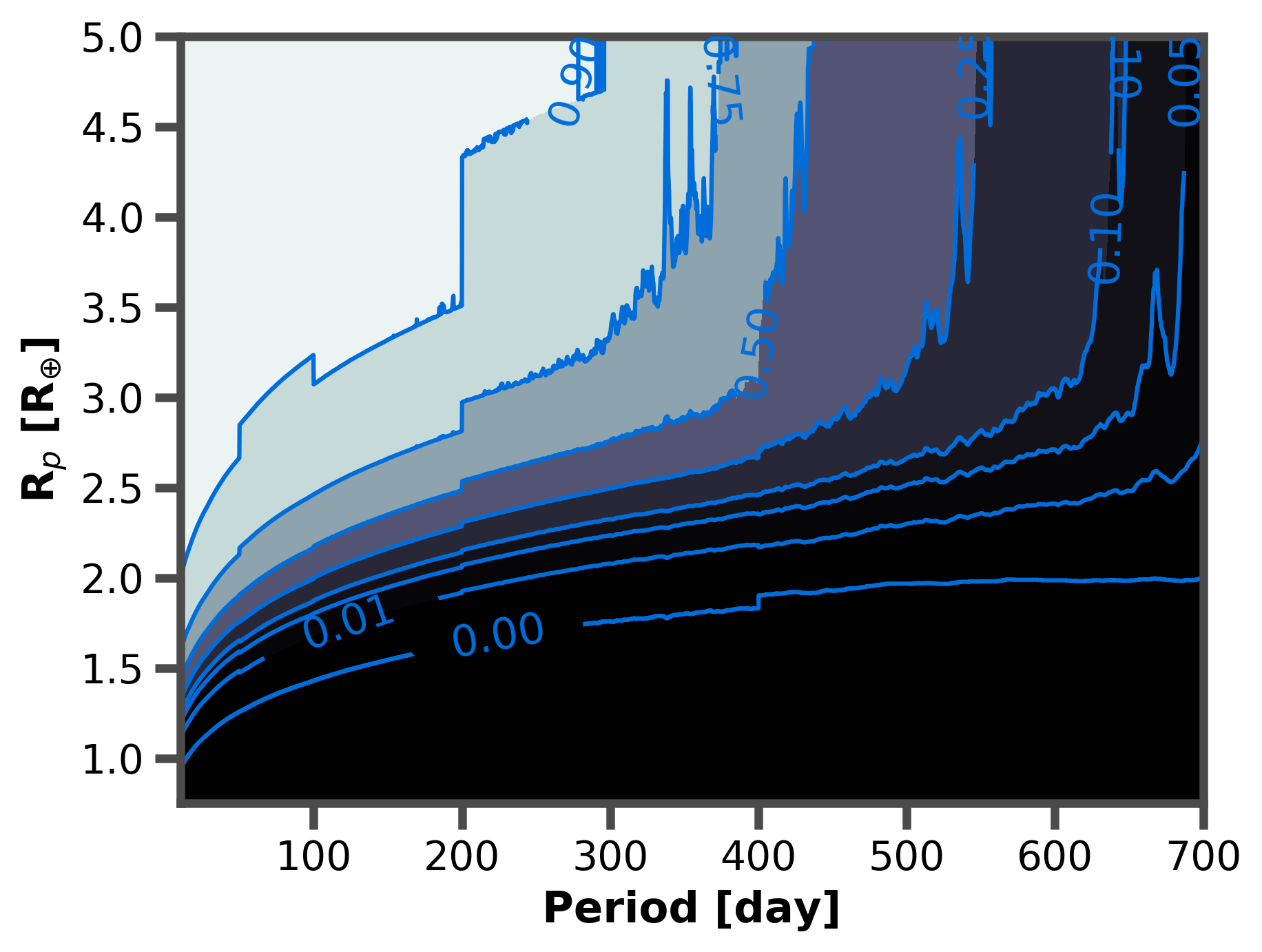}
  \caption{\label{fig:keplerports}
  \textbf{Detection probabilities for a second transiting planet orbiting \kepx.}
    Detection probability contour for a planet of a given orbital period and
	planetary radius ($R_P$) for the target \kepx. Results are calculated for the
	DR25 \kepler\ planet candidate catalog and include the effects of the vetting
	procedure.
  }
\end{sfigure}

\begin{sfigure}
  \centering
  \includegraphics[angle=0, width=14.0cm]{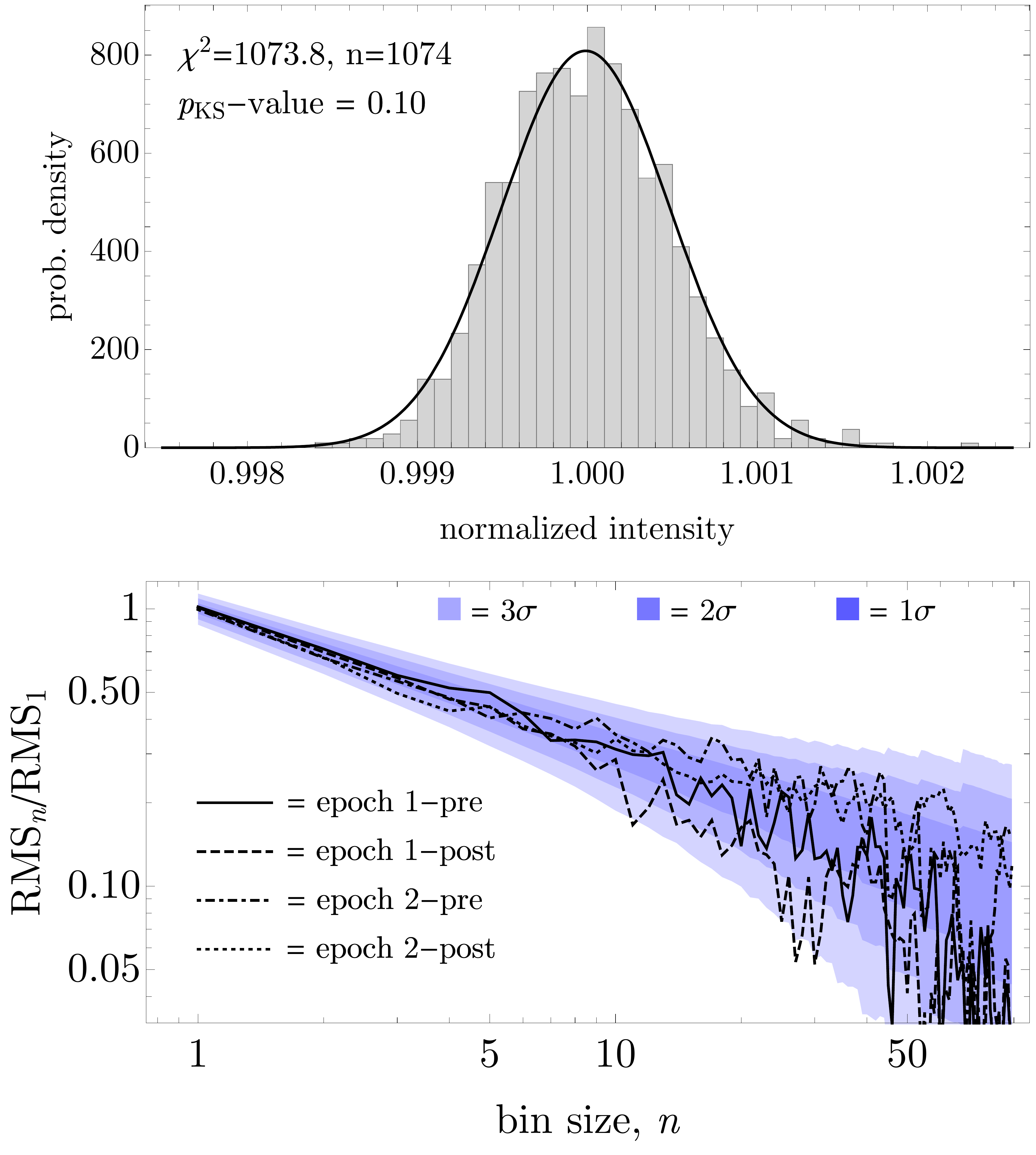}
  \caption{\label{fig:KIC790_noise}
  \textbf{Tests for the Gaussianity on the local out-of-transit photometry of \kepxb.}
    Top: Histogram of the method marginalised detrended photometry (excluding
	the transits) of \kepx. The plotted function is not a fit but the expected
	form if the data were normally distributed and governed by the
	measurement uncertainties. Bottom: RMS vs bin test of the data, showing the
	four relevant regions with separate black lines and the expected range for
	Gaussian statistics in blue.
  }
\end{sfigure}

\newpage
\begin{sfigure}
  \centering
  \includegraphics[angle=0, width=16.0cm]{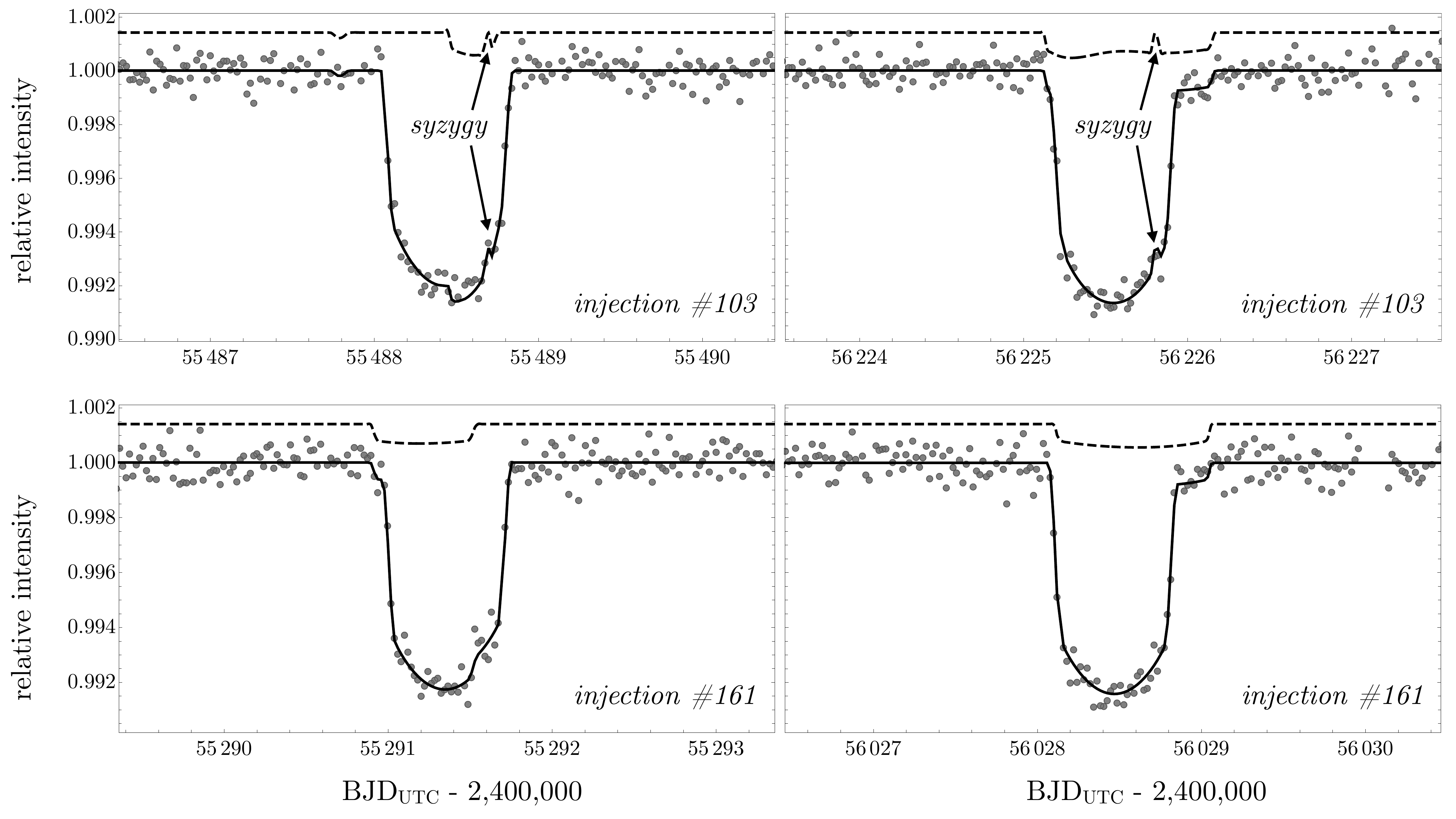}
  \caption{\label{fig:FalseMoons}
  \textbf{Transit light curves of the two false-positives found in our
  injection-recovery exercise for \kepxb-i, with the moon components
  shown as dashed lines above.}
    Top shows injection \#103 and bottom shows injections \#161, both
	of which include two planetary transits (two columns). Both cases
	correspond to positive radii and have ``strong evidence'' via the
	Bayes factor tests. The spurious moon of injection \#103
	requires an inclined ($\sim45^{\circ}$) moon (causing the slope in its
	trough due to limb darkening effects) and a short-period of 36\,hours
	(leading to syzygies highlighted).
  }
\end{sfigure}

\end{document}